\documentclass[a4]{article}

\PassOptionsToPackage{numbers, sort}{natbib}
\usepackage[preprint]{neurips_2019}

\usepackage[utf8]{inputenc} 
\usepackage[T1]{fontenc}    
\usepackage{hyperref}       
\usepackage{url}            
\usepackage{booktabs}       
\usepackage{amsfonts}       
\usepackage{nicefrac}       
\usepackage{microtype}      
\usepackage{graphicx}
\usepackage{amsmath}
\usepackage{pdfpages}
\usepackage{makecell}
\usepackage{todonotes}
\usepackage{color}
\usepackage{caption}
\usepackage{tabularx}
\usepackage{cleveref}
\usepackage[numbers]{natbib}
\usepackage{paralist}
\usepackage{relsize}
\usepackage{bbm}
\usepackage{lipsum}  
\usepackage{float}
\usepackage{bibentry}
\usepackage{subfig}

\usepackage{lipsum}
\usepackage{graphicx}
\usepackage{caption}
\usepackage{adjustbox}
\usepackage{pdfpages}
\usepackage{gensymb}
\usepackage{amsfonts}
\usepackage{setspace}
\usepackage{pdflscape}
\usepackage{lscape}
\usepackage{fancyhdr}
\usepackage{relsize}
\usepackage{tabularx}
\usepackage{paralist}

\usepackage{enumitem}
\setitemize{noitemsep,topsep=2pt,parsep=2pt,partopsep=2pt}
\setenumerate{noitemsep,topsep=2pt,parsep=2pt,partopsep=2pt}

\usepackage{bbm}

\usepackage{color}

\usepackage{tikz}
\usetikzlibrary{shapes,arrows}

\usepackage{amsmath}

\usepackage{algorithm}
\usepackage[noend]{algpseudocode}
\makeatletter
\def\BState{\State\hskip-\ALG@thistlm}
\makeatother
\usepackage{multirow}
\DeclareMathOperator*{\argmax}{\arg\!\max}

\newcommand{\cef}{\emph{Caenorhabditis elegans}}
\newcommand{\ce}{\emph{C. elegans}}

\title{The Virtual Patch Clamp: Imputing {\ce} Membrane Potentials from Calcium Imaging}

\author{
Andrew Warrington \\
Department of Engineering Science \\
University of Oxford \\
OX1 3PJ, United Kingdom \\
\texttt{andrew.warrington@keble.ox.ac.uk} \\
\And
Arthur Spencer \\
School of Physiology, \\ Pharmacology \& Neuroscience \\
University of Bristol \\
BS8 1TD, United Kingdom \\
\And
Frank Wood \\
Department of Computer Science \\
University of British Columbia \\
V6T 1Z4, Canada}%

\begin{document}

\maketitle


\begin{abstract} 
We develop a stochastic whole-brain and body simulator of the nematode roundworm {\cef} ({\ce}) and show that it is sufficiently regularizing to allow imputation of latent membrane potentials from partial calcium fluorescence imaging observations.
This is the first attempt we know of to ``complete the circle,'' where an anatomically grounded whole-connectome simulator is used to impute a time-varying ``brain'' state at single-cell fidelity from covariates that are measurable in practice.  
The sequential Monte Carlo (SMC) method we employ not only enables imputation of said latent states but also presents a strategy for learning simulator parameters via variational optimization of the noisy model evidence approximation provided by SMC.  Our imputation and parameter estimation experiments were conducted on distributed  systems using novel implementations of the aforementioned techniques applied to synthetic data of dimension and type representative of that which are measured in laboratories currently.

\end{abstract}

\section{Introduction}
\label{sec:intro}

One of the goals of artificial intelligence, neuroscience and connectomics~\cite{seung2011neuroscience} is to understand how sentience emerges from the interactions of the atomic units of the brain~\cite{kristan2006form, doty1975consciousness}, to be able to probe these mechanisms on the deepest level in living organisms, and to be able to simulate this interaction \emph{ad infinitum}~\cite{sarma2018openworm}.  
Models imbued with anatomically correct structure exploring the nature of these interactions have been developed for the widely studied nematode roundworm {\cef} ({\ce}).
These models span from whole-body locomotion~\cite{palyanov2016sibernetic, boyle2012gait} to whole-connectome neuron membrane potential and ion dynamics~\cite{gleeson2018c302, kunert2014low}.
However, anatomically grounded {\ce} connectome models have never been conditioned on data for brain-wide latent state imputation and parameter estimation.
We create a novel simulator of the whole {\ce} connectome and body by combining existing models~\cite{sce, boyle2012gait,rahmati2016inferring} and present methods for performing inference in the latent space of this simulator, conditioned on the type of data that non-invasive measurement techniques are currently capable of capturing~\cite{kato2015global, nguyen2016whole}.


\begin{figure}[tp]																								  %
\setcounter{figure}{1}

\setcounter{subfigure}{0}

\begin{minipage}[b!]{0.48\textwidth}
\subfloat[][]{
	\centering
	\includegraphics[width=\textwidth]{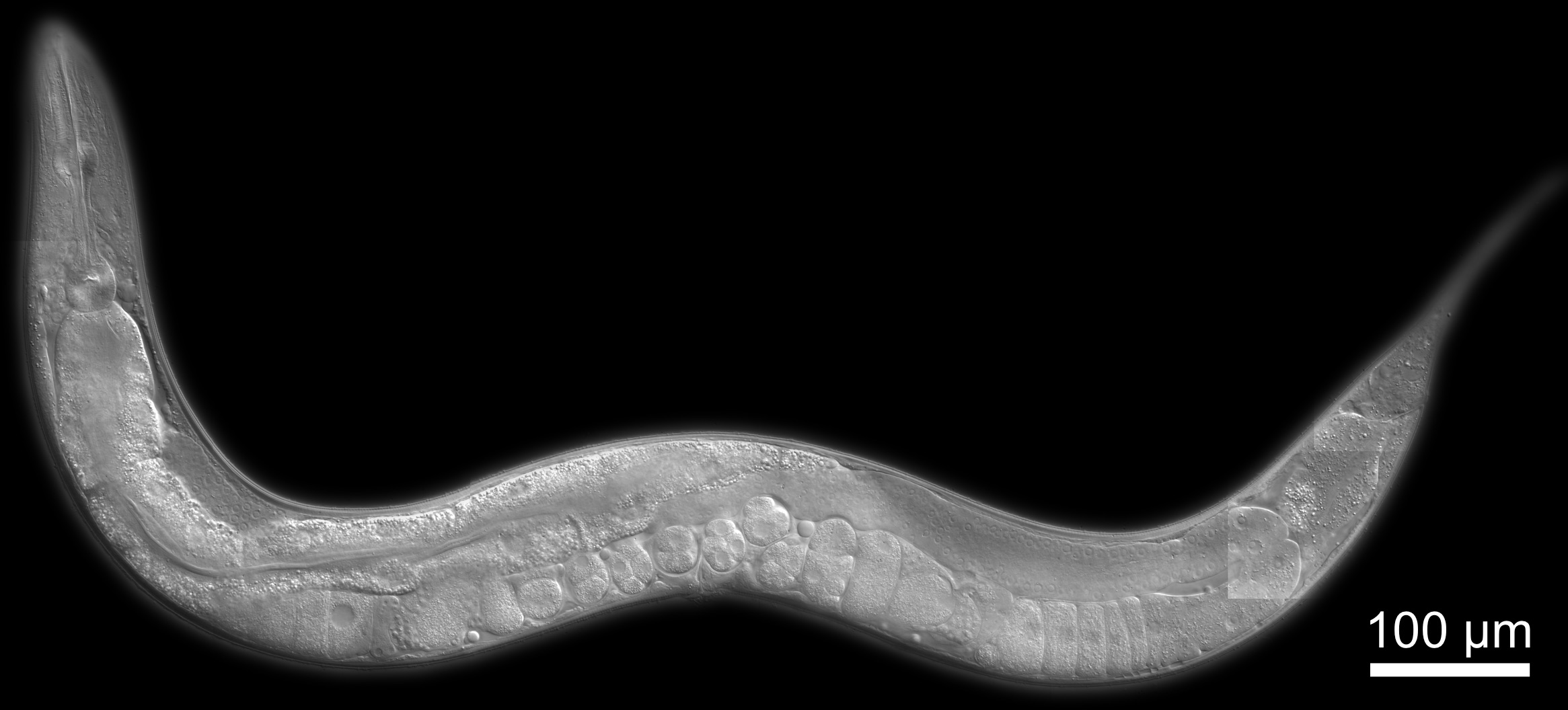}
	\label{fig:ce_img}
}

\setcounter{subfigure}{1}
\vspace*{-0.2cm}
\subfloat[][]{
	\centering
	\includegraphics[width=\textwidth]{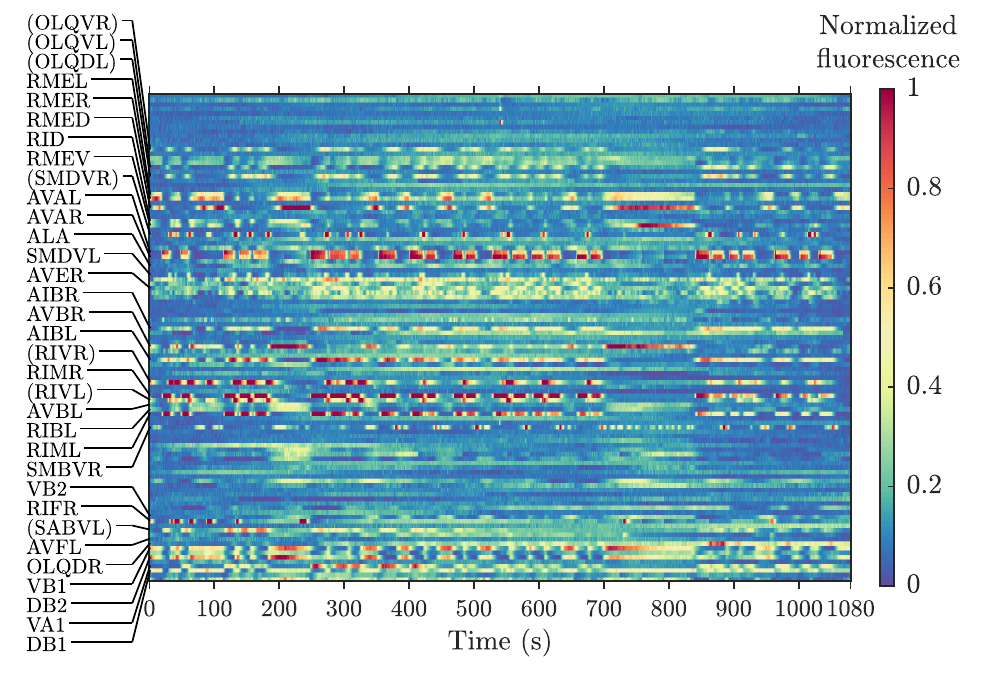}
	\label{fig:kato_calcium}
}

\setcounter{subfigure}{4}
\subfloat[][]{
	\centering%
	\begin{minipage}[b!]{0.32\textwidth}%
	\centering%
	\includegraphics[width=1.\textwidth]{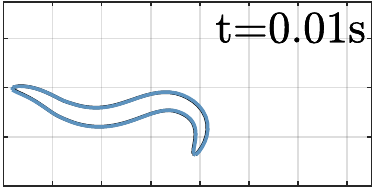}
	\includegraphics[width=1.\textwidth]{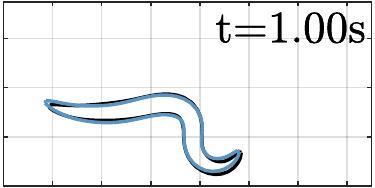}
	\end{minipage}%
	\hfill%
	\begin{minipage}[b!]{0.32\textwidth}%
	\centering%
	\includegraphics[width=1.\textwidth]{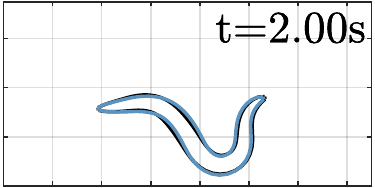}
	\includegraphics[width=1.\textwidth]{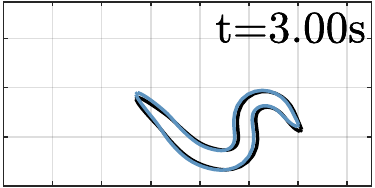}
	\end{minipage}%
	\hfill%
	\begin{minipage}[b!]{0.32\textwidth}%
	\centering%
	\includegraphics[width=1.\textwidth]{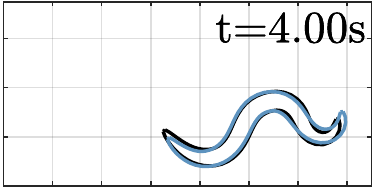}
	\includegraphics[width=1.\textwidth]{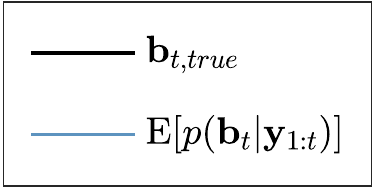}
	\end{minipage}%
	\label{fig:sce:wormsim}%
}
\end{minipage}
\hfill%
\begin{minipage}[b!]{0.48\textwidth}
\centering

\setcounter{subfigure}{2}
\vspace*{-0.2cm}

\subfloat[][]{
	\centering
	\includegraphics[width=\textwidth]{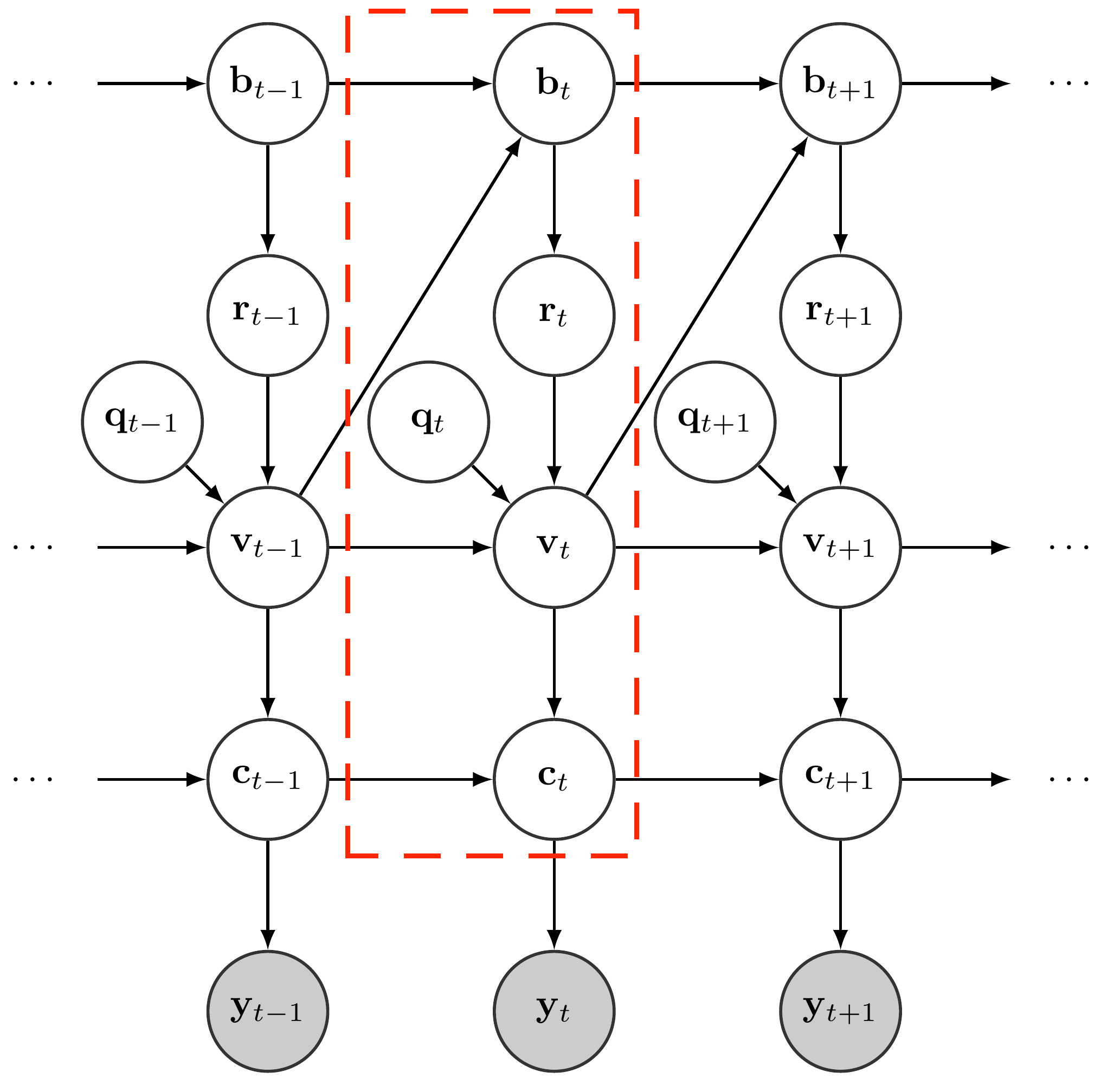}
	\label{fig:hmm_full}
}

\setcounter{subfigure}{3}

\subfloat[][]{
	\centering%
	\includegraphics[width=\textwidth]{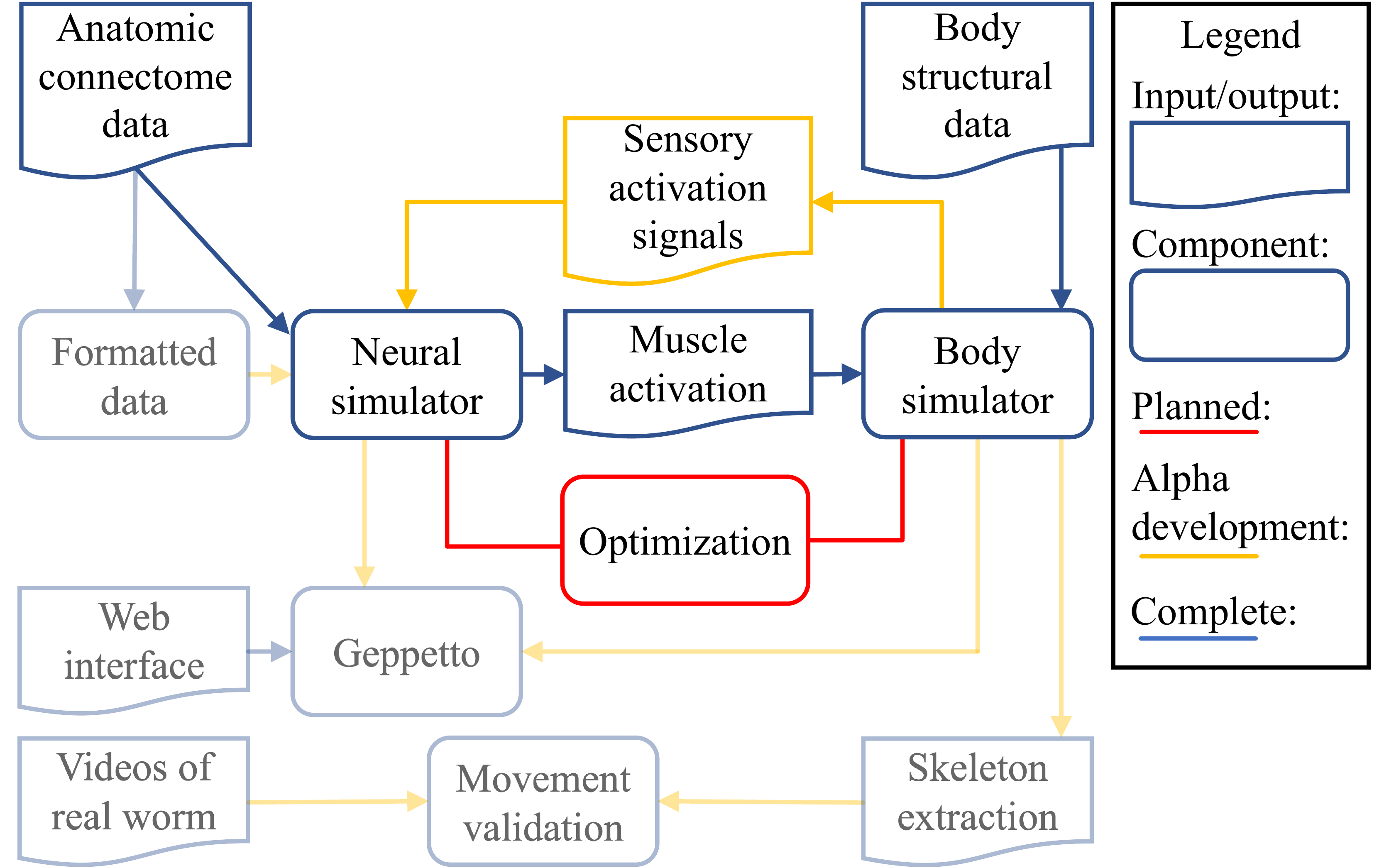}
	\label{fig:openworm}
}
\end{minipage}

\setcounter{subfigure}{5}
\vspace*{-0.2cm}
\subfloat[][]{
	\centering
	\includegraphics[width=\textwidth]{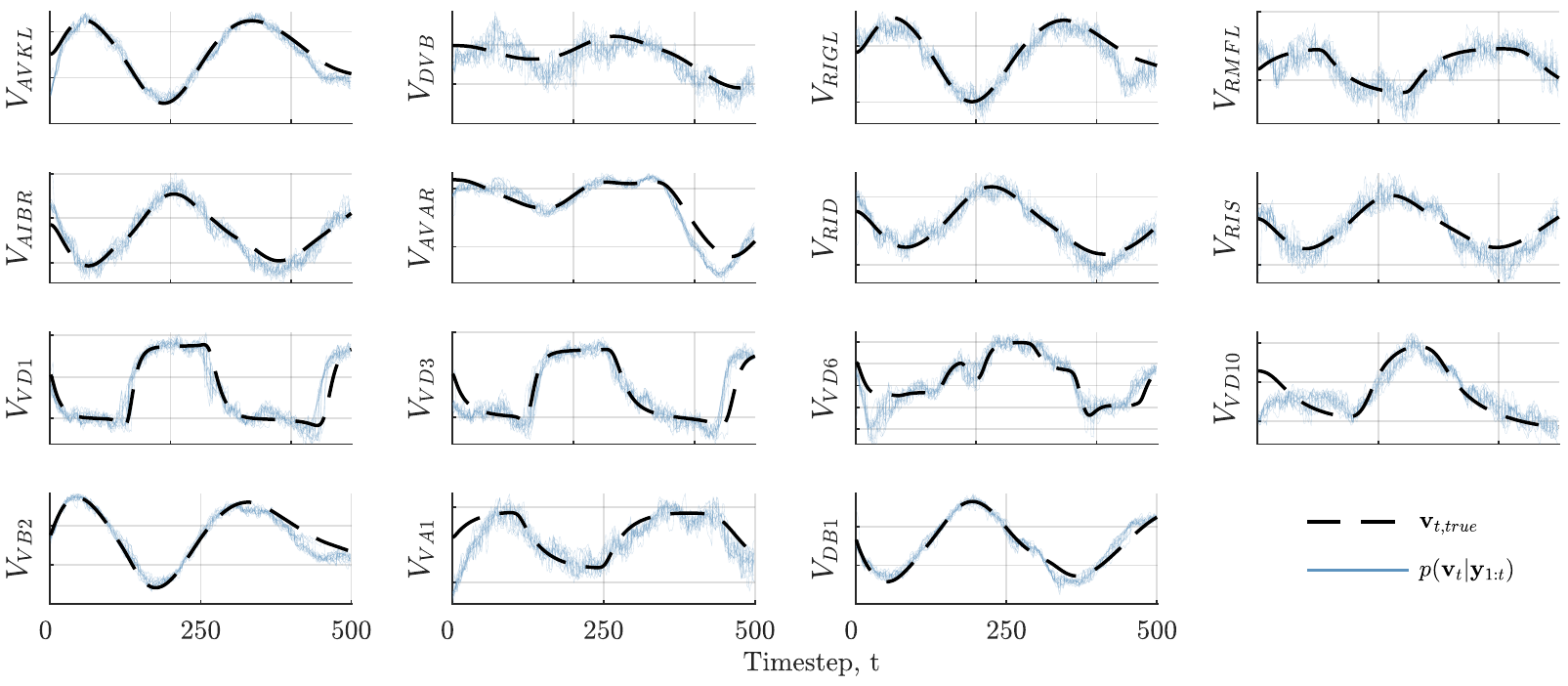}  
	\label{fig:sce:reconstruction}
}

\setcounter{figure}{0}
\captionof{figure}
{
\protect\subref{fig:ce_img}: The {\ce} roundworm, reproduced with permission from \emph{anon}. 
\protect\subref{fig:kato_calcium}: Typical {\em in vivo} fluorescence data on which we propose to condition (from \citet{kato2015global}). 
\protect\subref{fig:hmm_full}: Graphical model of our {\ce} simulator; variables defined in Section \ref{sec:simulator}. 
The dashed box denotes the $994$ dimensional latent state of the worm at each time step.
\protect\subref{fig:openworm}: Diagram adapted from \citet{sarma2018openworm} reflecting the community planned development pipeline for {\ce} simulation. Greyed out components are not considered in this paper.  The status of components are as categorized by OpenWorm \cite{sarma2018openworm, szigeti2014openworm}.
\protect\subref{fig:sce:wormsim} and \protect\subref{fig:sce:reconstruction}: Results of the virtual patch clamp experiment introduced in Section \ref{sec:vpc:exp}.
\protect\subref{fig:sce:wormsim}: Expectation of the filtering distribution over body shape shown in blue with true body shape in black. 
\protect\subref{fig:sce:reconstruction}: The true voltages for $15$ neurons as generated by our simulator are shown as black dashed lines, with the SMC filtering distribution shown in blue. The rows correspond, top to bottom, to unobserved neurons, observed neurons, unobserved motor neurons, and observed motor neurons.
}
\label{fig:full_page}
\end{figure}

Techniques for investigating neural function at the cellular level traditionally require direct, invasive measurement and manipulation of physiological variables such as membrane potentials in a very small number of neurons, using tools including multi-electrode arrays~\cite{spira2013multi} and patch clamps~\cite{margrie2002vivo}.
Brain-wide {\em in vivo} application of such measurement techniques, at the fidelity of individual neurons, is logistically infeasible.
However covariates of a subset of an organism's neurons can be measured non-invasively {\em in vivo} using calcium fluorescence imaging~\cite{stosiek2003vivo, chung2013calcium, kato2015global, nguyen2016whole}.
Such techniques allow the calcium ion concentration, a time-smoothed covariate of the actual neural potential~\cite{rahmati2016inferring}, to be   non-invasively estimated {\em in vivo} for a spatially co-located subset of a specimen's neurons, with minimal effect on behaviour.  
The use of such measurements to infer unobserved states and test hypotheses about the mechanisms governing brain-wide state evolution {\em requires} the use of a model.
Connecting such a model to observational data so that hypotheses can be tested and tuned forms the bulk of our contribution in this paper.     
We refer to using an anatomically correct model to infer latent states and parameters, conditioned on partial data, as a ``virtual patch clamp'' (VPC).
The VPC also facilitates \emph{in silico} experimentation on ``digital'' {\ce} specimens, by programmatically modifying the simulator and observing the resulting simulations; enabling rapid, wide-reaching, fully observable and perfectly repeatable exploration and screening of experimental hypothesis.

\section{Simulating {\ce}}
\label{sec:simulator}

Due to the simplicity and regularity of its anatomy, and its predictable yet sophisticated behavioural repertoire~\cite{ardiel2008behavioral},  {\ce}, shown in Figure \ref{fig:full_page}\protect\subref{fig:ce_img}, is used as a ``model organism'' in biology and neuroscience research~\cite{wicks1996dynamic, ardiel2010elegant, varshney2011celegans}.
Notably, its connectome is regular across wild-type specimens~\cite{varshney2011celegans} and has been mapped at synapse and gap-junction fidelity using electron microscopy~\cite{white1986connectome, varshney2011celegans}.
Because of this fixed architecture, neural circuit simulators, imbued with anatomically correct structure, have been developed to produce feasible whole {\ce} connectome simulators~\cite{sce, gleeson2018c302, hasani2017sim} by leveraging highly accurate neural dynamics models~\cite{hines2006neuron, gewaltig2007nest, kunert2014low, wicks1996dynamic, kuramochi2010quantitative}.
Likewise, its simple anatomy has allowed body and locomotion simulators to be developed~\cite{boyle2012gait, palyanov2015sibernetic}.

The first contribution of this paper is a new {\ce} simulator that integrates existing simulators and models~\cite{sce, boyle2012gait,rahmati2016inferring} developed by the {\ce} community. 
The selection of models to integrate was influenced by the desire to produce a simulator capable of modelling {\ce} at single-neuron fidelity that is computationally tractable at desktop-scale for rapid exploration of the model; while also scaling to high performance compute clusters for parameter estimation and hypothesis testing. 
While higher fidelity models exist~\cite{gleeson2018c302, hines2006neuron, palyanov2015sibernetic}, they are significantly more computationally expensive.

At a high level, our simulator is comprised of three components: a simulator for the time-evolution of the membrane potential~\cite{sce} and intracellular calcium ion concentration~\cite{rahmati2016inferring} in all $302$ {\ce} neurons, a simulator for the physical form of the worm and the associated neural stimuli and proprioceptive feedback~\cite{boyle2012gait}, and a model relating the intracellular calcium to the observed florescence data~\cite{rahmati2016inferring, kato2015global}.
Part of our contribution is how we specifically integrate the different simulators.
In particular, our simulator introduces a physiologically motivated pathway~\cite{wen2012proprioceptive} for passing proprioceptive feedback from the body simulation to the correctly anatomically structured neural model.  


In the following we denote the number of neurons $N=302$, with $n\in \left\lbrace 1,\ldots,N \right\rbrace$ indexing an individual neuron.  The number of observed calcium traces is $M<N$ (for example $M=49$ in \citet{kato2015global}), and $t \in \left\lbrace 0,\ldots,T\right\rbrace$ indexes discrete time steps, where the time discretization used is $\delta t = 0.01$.  The graphical model for our simulator is shown in Figure \ref{fig:full_page}\protect\subref{fig:hmm_full}.

\paragraph{Neural Simulation}
\label{sec:prior:neural}
The first component of our model is a simulator of connectome-scale, single-neuron fidelity neural dynamics.
We selected and modified the simulator presented by \citet{sce}, which builds on developments presented by \citet{kunert2014low} and \citet{wicks1996dynamic}, called `simple {\ce}' (SCE).
SCE is designed to be an easily interpretable simulator of {\ce} neural membrane potential dynamics, individually denoted as $v_{t,n}$, via single-compartment neuron models~\cite{kunert2014low} connected by chemical synapses and electrical gap junctions.
SCE, unmodified,  uses an ordinary differential equation  (ODE) solver to iterate the differential equation governing membrane potential evolution.
By using a fine time discretization ($\delta t = 0.01$) we found that we could achieve a substantial computational speedup with negligible integration error by modifying SCE to use forward differencing to project neural voltages forward. 
We also add a small amount of independent, neuron-specific Gaussian noise to the membrane potential at each time step.

We also added to SCE an ODE model relating intracellular calcium ion concentration in each neuron, denoted $c_{t,n}$, to the membrane potential, as described by \citet{rahmati2016inferring}.

These simulators implicitly define the time-evolution of the neural state of the worm, denoted $p(\mathbf{v}_{t}, \mathbf{c}_{t} | \mathbf{v}_{t-1}, \mathbf{c}_{t-1}, \mathbf{r}_{t})$, where, $\mathbf{r}_t$ represents the yet to be discussed, proprioceptive feedback conditioned on the body shape, $\mathbf{b}_t$.
Exemplar voltage traces generated by our simulator are shown as black dashed lines in Figure \ref{fig:full_page}\protect\subref{fig:sce:reconstruction}.

\paragraph{Body Simulation}
The next component we incorporate is a simulator for the body shape of the worm.
For this we use WormSim~\cite{boyle2012gait}, which complements the experimental findings presented by \citet{wen2012proprioceptive}.
WormSim models the body shape and locomotion in two dimensions as a series of rigid rods, contractile units and springs driven by impulses generated by a simplified neural network.
These elements are defined by $49$ control points, comprised of the $x$, $y$ position and angle of each of the rods, and the first derivative of these terms.
A further $96$ values represent the instantaneous ``voltage'' in each of the $48$ dorsal and ventral contractile units.
The total body state is denoted $\mathbf{b}_t\in\mathcal{B}=\mathbb{R}^{49\times 3\times 2}\times \mathbb{R}^{48\times 2}$.
The model of evolution of body state, denoted $p(\mathbf{b}_t | \mathbf{b}_{t-1}, \mathbf{v}_{t-1})$, is dependent on both the previous state of the body $\mathbf{b}_{t-1}$ and the neural state at the previous timestep, $\mathbf{v}_{t-1}$, acting as driving neural input.
The body simulator then returns proprioceptive feedback, denoted $\mathbf{r}_t$, back to the neural simulator.
Our contribution here is specifically the interface for driving WormSim using the anatomically correct SCE model in place of the simplified network used in the original work, in addition to the specific formulation of how proprioceptive feedback flows back to SCE as current injected into neurons that are known to receive proprioceptive feedback~\cite{wen2012proprioceptive}.
A typical evolution of body state is shown in Figure \ref{fig:full_page}\subref{fig:sce:wormsim}. 

\paragraph{Observation Model}
The fluorescence signals observed in calcium imaging, examples of which are shown in Figure \ref{fig:full_page}\subref{fig:kato_calcium}, denoted $\mathbf{y}_t\in\mathcal{R}_+^{M}$, are a stochastic quantity dependent on the intracellular calcium concentration.
This dependence is described by a saturating Hill-type function~\cite{hill1938heat, rahmati2016inferring, grienberger2012imaging, yasuda2004imaging}, where details are presented in the supplementary materials.
Exemplar fluorescence traces, as recorded by \citet{kato2015global}, are shown in Figure \ref{fig:full_page}\protect\subref{fig:kato_calcium}.
Annotated are the neuron identities for which the source neuron could be determined by domain experts\cite{kato2015global}.

To summarize our model, the neuron states $\mathbf{v}_t$ and $\mathbf{c}_t$, body state $\mathbf{b}_t$,  proprioceptive feedback $\mathbf{r}_t$, and any sensory input $\mathbf{q}_t$ (not explicitly considered here as it can realistically be assumed to be constant over our simulation durations), define the latent ``brain'' and ``body'' state of the worm, collectively denoted at time $t$ as $\mathbf{x}_t\in\mathbb{R}^{994}$, indicated by the dashed box in Figure \ref{fig:full_page}\protect\subref{fig:hmm_full}.
The observed data, $\mathbf{y}$, is the calcium imaging signal, which, contrary to what is shown in the graphical model, is not actually observed at every timestep, a notational complication we intentionally avoid but correctly implement.

Note that our simulator defines a hidden Markov model, albeit one with complex, high-dimensional non-linear latent state transition dynamics, $p(\mathbf{x}_t | \mathbf{x}_{t-1})$, as well as a complex non-linear high-dimensional observation model $p(\mathbf{y}_t | \mathbf{x}_t).$  Our second contribution is showing how the tools of Bayesian inference can be employed to condition on partial observations, make predictions conditionally or unconditionally, and perform marginal maximum a posteriori parameter estimation.


\section{The Virtual Patch Clamp}
\label{sec:vpc}

The second contribution of this paper is the adoption and scaling of a method to impute the entire latent state, $\mathbf{x}_t$, conditioned on calcium imaging florescences emitted from neurons that have been successfully identified in existing calcium imaging literature~\cite{kato2015global}.
To be more specific, armed with our simulator and inference methods, we estimate the distribution at each timestep of the $994$ interpretable neural and physical simulator latent states, conditioned on $49$ florescence traces.
We condition on the same $49$ neurons that \citet{kato2015global} were able to identify in their experimental results, listed in the supplementary materials.
We describe this as a ``virtual patch clamp,'' as it permits the imputation of quantities such as membrane potential and ion currents, measured as part of the patch clamping procedure.
These variables are directly addressable in the simulator, and so their value can subsequently be programmatically ``clamped.'' 
The simulator is then initialized from the inferred latent distribution and iterated to simulate the effect of the clamping \emph{in silico}, as posterior predictive inference.

In the previous section we outlined our simulation model of {\ce}, implicitly defining the joint distribution over worm state and observed data, denoted $p(\mathbf{x}_{0:T}, \mathbf{y}_{1:T})$.
We wish to quantify the distribution over the latent states conditioned on the observed data, referred to as the posterior distribution $p(\mathbf{x}_{0:T} | \mathbf{y}_{1:T})$.
Direct sampling is intractable as the model is specified as a non-linear, non-differentiable, and non-invertible simulator.  
Therefore, approximation methods must be employed.  
Under the constraints imposed by our model, the available data, and our objective, we use sequential Monte Carlo (SMC) for estimating $p(\mathbf{x}_{0:T} | \mathbf{y}_{1:T})$.

\subsection{Sequential Monte Carlo}
\label{sec:vpc:smc}
Sequential Monte Carlo (SMC), similar to particle filtering in state-space models, produces a weighted discrete measure approximating the distribution $p(\mathbf{x}_{0:T} | \mathbf{y}_{1:T})$.
The variant of SMC we use samples from the prior and weights by the likelihood (see the tutorial by \citet{Doucet11atutorial} for details).
This approach iterates the particles, individually notated as $\mathbf{x}_t^{(n)},\ n \in 1:\mathcal{N}_p$, through the simulator, and then weights these particles by their probability under the observation density, $w_t^{(n)} = p(\mathbf{y}_t | \mathbf{x}_t^{(n)})$.  
Those particles that ``explain'' the observation well receive a high weight and are retained and continued, while those with low weight are not.  
SMC also provides, for no additional computation, an estimate of model evidence, calculated as $p( \mathbf{y}_{1:T} ) \approx \prod_{t \in 1:T} \frac{1}{\mathcal{N}_p} \Sigma_{n \in 1:\mathcal{N}_p} w_{t}^{(n)}$~\cite{Doucet11atutorial}.

To relate this process to our outlined objectives, the particles themselves represent the inferred distribution over all latent neural and physiological states, $\mathbf{x}_t$, providing the imputation element of the VPC.
Forward simulation of the particles provides posterior predictive inference over state evolution, providing the \emph{in silico} experimental facility.
Finally the evidence approximation allows us to objectively compare models and hypotheses, which will be used later for parameter estimation.

\paragraph{Initialization of Particles}
For the experiments we present, we assume the initial body pose, $\mathbf{b}_0$, of the worm is known.
Calcium imaging recordings are nearly always augmented with video recordings from which the pose can be determined, however this channel of \citet{kato2015global}'s data was not available and so we confine ourselves here to operating on synthetic data.
We initialize the muscular voltage to zero, noting it develops quickly from neural activity and body pose. 
The calcium concentration is initialized from a prior, where we perform a single importance sampling step for each observed neuron to further refine the initialization.
We found that directly sampling $302$ membrane potentials from the prior distribution led to gross particle degeneracy, due to finite particle sets and the high dimensional latent space, and hence poor particle filter sweeps.
We therefore refine the distribution from which membrane potentials are initialized using the model, where details are presented in the supplementary materials.
This refinement was observed to yield ``better'' initial particles, lower degeneracy, and better performance.

\subsection{Experiments}
\label{sec:vpc:exp}
In our first experiment we first generate a synthetic state trajectory by sampling from the model, then demonstrate that we can use SMC to condition our model on the simulated calcium data, by comparing the resulting reconstruction of $\mathbf{x}$ to the known ground truth.
Specifically we condition on the same $49$ neurons identified in the calcium imaging data released by \citet{kato2015global}, where fluorescence signals are simulated every $5$ timesteps.
Results for this are shown in Figure \ref{fig:full_page}\protect\subref{fig:sce:wormsim} and \ref{fig:full_page}\protect\subref{fig:sce:reconstruction}.
The particle distribution of a subset of $\mathbf{v}_t$ is shown in Figure \ref{fig:full_page}\protect\subref{fig:sce:reconstruction}.
The true state is shown in black, while the SMC filtering distribution is shown in blue.
The number of particles, $\mathcal{N}_p$, used in the SMC sweep was $1000$, with $5000$ particles used in the initialization procedure.
The wall-clock time was $800$ seconds to complete a single SMC sweep when run on a single node equipped with $48$ Intel Xeon Platinum 2.10GHz 8160F CPUs.

The blue reconstructions are congruent with the black trace, indicating that the latent behaviour of the complete system is being well-reconstructed despite partial observability.
Critically, neurons not directly connected to observed neurons (for instance VD$10$) are correctly reconstructed, indicating that the regularizing capacity of the model is sufficient to constrain these variables.
Further confirmation of the power of this method can be seen in Figure \ref{fig:full_page}\protect\subref{fig:sce:wormsim}, showing the predicted body shape closely matches the true state.

We observe a small drift in the position and orientation of the worm at early timesteps due to the particle approximation of the distribution over membrane potentials when the SMC sweep is initialized.
This initial drift cannot be corrected as the absolute position and rotation of the worm does not influence the neural activity.
Therefore, for ease of visual comparison of the body \emph{shape} reconstructions, we rigidly transform them to center them on the true body.
We perform the same centering post-processing in Figure \ref{fig:learning}\protect\subref{fig:learning:wormsim}.
The raw reconstructions are included in the supplementary materials.
Explicitly conditioning on body shape from video data, as suggested in the discussion section, would alleviate this issue.

This experiment shows that the VPC is tractable and is capable of yielding high-fidelity reconstructions of pertinent latent states given partial calcium imaging observations via the application of Bayesian inference to time series models of {\ce}.


\section{Parameter Estimation}
\label{sec:estimation}


The posterior inference and evidence approximation presented in the previous section is useful for imputing values and performing \emph{in silico} experimentation.
Thus far we have not discussed the parameters of our simulator-based model, as it was not necessary in order to demonstrate the effectiveness of SMC for posterior inference.  
These parameters, collectively denoted $\boldsymbol\theta$, include the non-directly observable electrical and chemical characteristics of individual synapses in the {\ce} connectome, as well as parameters of the body model, the calcium fluorescence model, etc.  
We conclude this paper by taking concrete steps towards performing such parameter estimation, as defined by the simulator-structured hypothesis class defined by the chosen model.

Our goal is to estimate the best simulator parameters $\boldsymbol\theta^*$ given observed data, i.e.~$\boldsymbol\theta^* = \argmax_{\theta} p(\boldsymbol\theta | \mathbf{y}) = \argmax_{\boldsymbol\theta} p(\mathbf{y} | \boldsymbol\theta) p(\boldsymbol\theta)$.
The method we employ for performing parameter estimation is a novel combination of variational optimization (VO) \cite{staines2012variational} and SMC evidence approximation.  This results in a stochastic gradient for parameter estimation that does not require a differentiable simulator and can deal with a large number of latent variables.


Variational optimization starts with the following inequality~\cite{staines2012variational}
\begin{equation*}
\min_{\boldsymbol\theta} f(\boldsymbol\theta) \leq \mathbb{E}_{\boldsymbol\theta \sim q(\boldsymbol\theta | \boldsymbol\phi)} \left[ f(\boldsymbol\theta) \right] = U(\boldsymbol\phi).
\end{equation*}
Intuitively, $U(\boldsymbol\phi)$ upper bounds the value of $\min f(\boldsymbol\theta)$, and so minimizing $U(\boldsymbol\phi)$ with respect to $\boldsymbol\phi$ minimizes the bound on $f(\boldsymbol\theta)$; in turn exactly minimizing $f(\boldsymbol\theta)$ if the variance of $q(\boldsymbol\theta|\boldsymbol\phi)$ is allowed to go to zero.
The gradient of $U(\boldsymbol\phi)$ with respect to $\boldsymbol\phi$ can then be computed as
\begin{align}
\nabla_{\boldsymbol\phi} U(\boldsymbol\phi) & = \nabla_{\boldsymbol\phi} \mathbb{E}_{\boldsymbol\theta \sim q(\boldsymbol\theta | \boldsymbol\phi)} \left[ f(\boldsymbol\theta) \right],\label{equ:vo_analytic}\\
& \approx  \frac{1}{\mathcal{N}_r} \Sigma_{n \in \left\lbrace 1, \dots, \mathcal{N}_r \right\rbrace}  f(\boldsymbol\theta^{(n)}) \nabla_{\boldsymbol\phi} \log q(\boldsymbol\theta^{(n)} ; \boldsymbol\phi) \qquad \boldsymbol\theta^{(n)} \sim q(\boldsymbol\theta | \boldsymbol\phi), \label{equ:vo_mc_grad}
\end{align}
where we have expanded the expectation and used the derivative of a logarithm to go from \eqref{equ:vo_analytic} to \eqref{equ:vo_mc_grad}, similar to the REINFORCE method~\cite{williams1992simple}.
Evaluation of this expectation is not analytically tractable so, also in going from \eqref{equ:vo_analytic} to \eqref{equ:vo_mc_grad}, we apply Monte Carlo integration, drawing $\mathcal{N}_r$ samples from the proposal distribution $q(\boldsymbol\theta | \boldsymbol\phi)$.
We then use ADAM~\cite{kingma2014adam} to minimize $U(\boldsymbol\phi)$ using this approximate gradient.
We investigated reducing the variance of the gradient operator using the optimal reward baseline~\cite{weaver2001optimal} as a control variate, but found it did not improve overall performance.

Here,  the objective function is the joint density $f(\boldsymbol\theta) = p(\mathbf{y} , \boldsymbol\theta) = p(\mathbf{y} | \boldsymbol\theta) p(\boldsymbol\theta)$, where the likelihood term is approximated via SMC as defined in Section \ref{sec:vpc:smc}.
To our knowledge, this is the first time that pseudo-marginal methods have been paired with variational optimization methods. 
We refer to this procedure as particle marginal variational optimization (PMVO).

\subsection{Autoregressive Model}
\label{sec:estimation:sub:ar}
To investigate the applicability of PMVO, we conducted experiments learning parameters in a simplified model family.
We investigate parameter estimation alternatives using a first order autoregressive generative model (AR) in-place of the neural simulator and data; where we use a sparse transition kernel, and prior distributions and observation model based on the {\ce} scenario.

Instead of a point-estimate, one might wish to have the full posterior distribution over model parameters. 
For this reason, we consider our PMVO method alongside particle marginal Metropolis Hastings (PMMH) and parallel tempering (PT), where ``optimal'' parameters are approximated by MAP sample values.  
Specific implementation details are provided in the supplementary materials.

We demonstrate our method on a $30$-dimensional AR process, where the transition kernel consists of $44$ parameters, $\boldsymbol\theta\in \mathbb{R}_{\geq 0}^{44}$.
Figure \ref{fig:ar}\protect\subref{fig:ar:reconstruction} shows, in black, the latent state of the AR process.
The SMC filtering distribution conditioned on the true parameters, randomly initialized parameters, and the learned parameters are shown in blue, red and green respectively.
Well-optimized parameters lead to better reconstructions.
Figure \ref{fig:ar}\protect\subref{fig:ar:parameters} shows the convergence of the parameter values for each of PMMT, PT, and PMVO, across $8$ random restarts.  
We see that PMVO recovers the true parameter values more quickly and reliably than PMMH and PT and in-turn produces better reconstructions.
This improved performance is reflected in Figure \ref{fig:ar}\protect\subref{fig:ar:posterior} by the error between the joint density of the true parameters and the MAP parameters reducing more quickly when using PMVO.


\subsection{ {\ce} Simulator}
\label{sec:estimation:sub:sce}
Our final experiment establishes the utility of our PMVO technique by demonstrating we can recover simulator parameters.
To show this we generate synthetic data using known parameters and demonstrate than we can recover suitable parameter values and feasible reconstructions.
For this work, we optimize the two parameters we introduced by integrating SCE and WormSim, namely the strength of motor stimulation, $w_{\text{m}}$, and proprioceptive feedback, $w_{\text{s}}$.  
The values of these parameters cannot be measured and so must be learned from data. 
The results of this experiment are shown in Figure \ref{fig:learning}. 
Figures \ref{fig:learning}\protect\subref{fig:learning:reconstruction} and \ref{fig:learning}\protect\subref{fig:learning:wormsim} show the imputed voltage traces and body poses when using the true parameters (blue), initial parameters (red) and optimized parameters (green).
As before, recovery of ``good'' parameter values facilitates good imputation and reconstruction of latent states.
Figure \ref{fig:learning}\protect\subref{fig:learning:parameters} shows the distribution of convergence paths of two parameters being addressed for $20$ initializations.
This experiment shows that parameter inference in {\ce} models using PMVO is viable.

For this optimization we take $500$ gradient steps, where $8$ parameters are sampled from the proposal at each approximate gradient calculation ($\mathcal{N}_r = 8$).
The SMC sweep uses $500$ particles in the initial step, sub-sampling to $90$ particles after the first observation.
The traces used were $500$ timesteps in length, corresponding to $5$ seconds of real time, with an observation every $5$ time steps.
Each individual SMC sweep takes approximately $90$ seconds to complete.
We implement and distribute a framework for embarrassingly parallel evaluation of multiple SMC sweeps on large, distributed high performance compute clusters, where each SMC sweep is executed on a single node, eliminating network overheads.
The wall-clock time was no more than $17$ hours for a single experimental repeat, when distributed across $10$ nodes equipped with $48$ Intel Xeon Platinum 2.10GHz 8160F CPUs.


\begin{figure}[t]			
		
\setcounter{figure}{2}    					
				
\begin{minipage}[b!]{0.3\textwidth}
	\subfloat[][]{
		\centering
		\includegraphics[width=\textwidth]{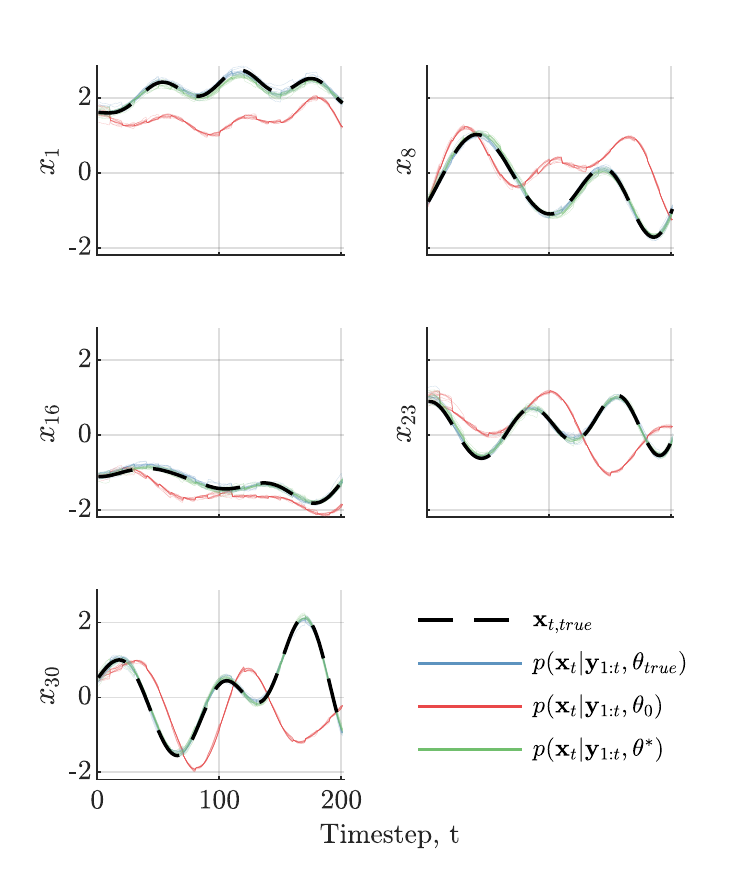}  
		\label{fig:ar:reconstruction}
	}
\end{minipage}
\hfill%
\begin{minipage}[b!]{0.3\textwidth}
	\subfloat[][]{
		\centering
		\includegraphics[width=\textwidth]{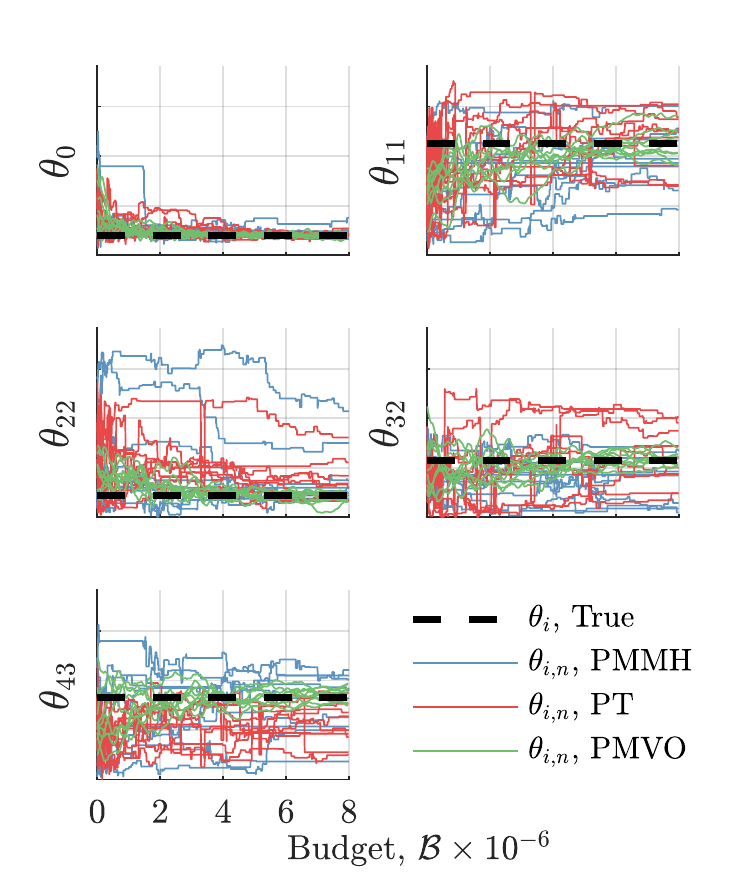}  
		\label{fig:ar:parameters}
	}
\end{minipage}
\hfill%
\begin{minipage}[b!]{0.3\textwidth}
	\subfloat[][]{
		\centering
		\includegraphics[width=\textwidth]{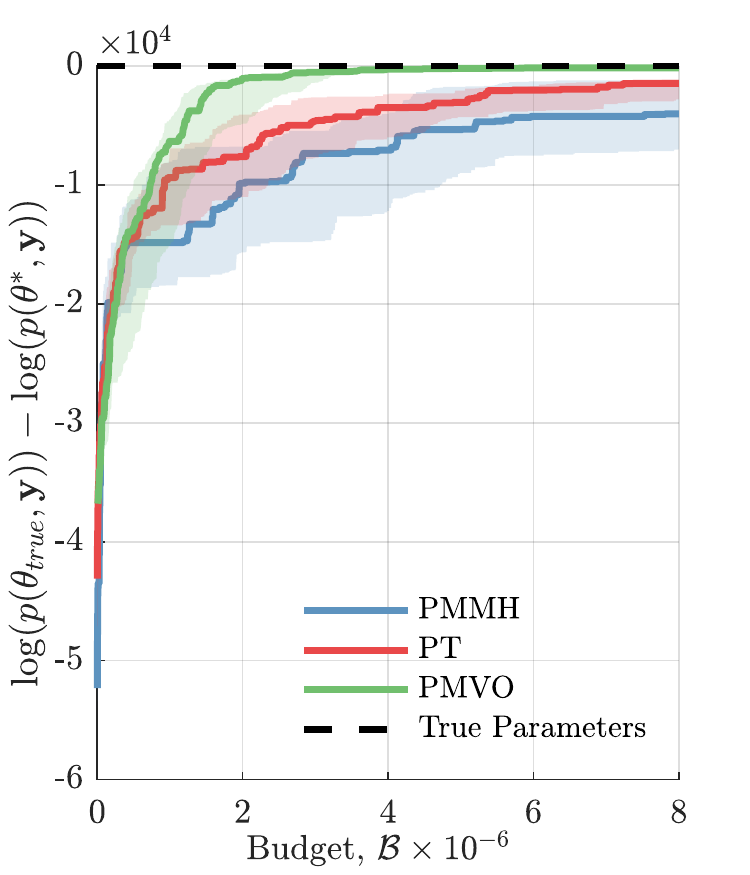}  
		\label{fig:ar:posterior}
	}
\end{minipage}
\setcounter{figure}{1}    					
\vspace*{-0.2cm}
\caption{
Comparison of PMVO, PMMH and PT methods for parameter estimation in the autoregressive example introduced in Section \ref{sec:estimation:sub:ar}. $8$ experimental repeats are shown. 
\protect\subref{fig:ar:reconstruction}: the true latent state sequence is shown as a black dashed line, and the filtering distribution from SMC using the true parameters (blue), initial parameters (red) and optimized parameters (green), for $5$ of the $30$ states.
\protect\subref{fig:ar:parameters} shows convergence of $5$ of the $44$ parameters to the true value, shown in black.
\protect\subref{fig:ar:posterior} shows the difference in the log-joint density between the true parameters and the learned parameters for three alternative optimization methods. 
We normalize all methods by computational budget.
}
\label{fig:ar}
\end{figure}


\begin{figure}[t]					

\setcounter{figure}{3}    																								  %

\begin{minipage}[b!]{0.48\textwidth}
	\subfloat[][]{
		\centering
		\includegraphics[width=\textwidth]{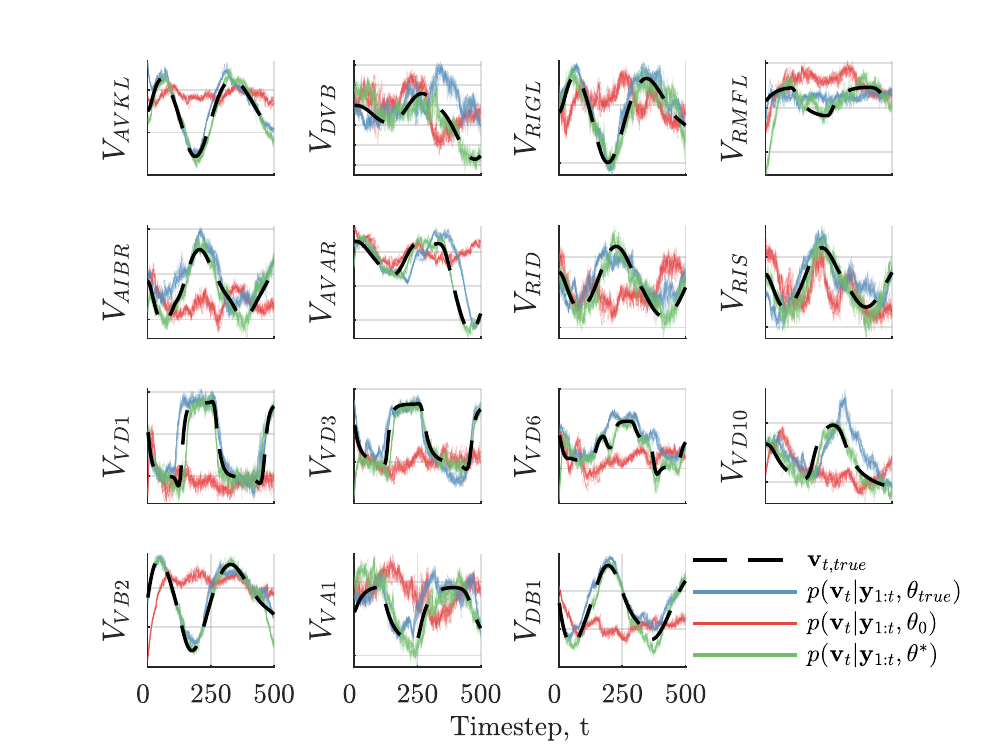}  
		\label{fig:learning:reconstruction}
	}
\end{minipage}
\hfill%
\begin{minipage}[b!]{0.48\textwidth}
	\subfloat[][]{
		\centering
		\includegraphics[width=\textwidth]{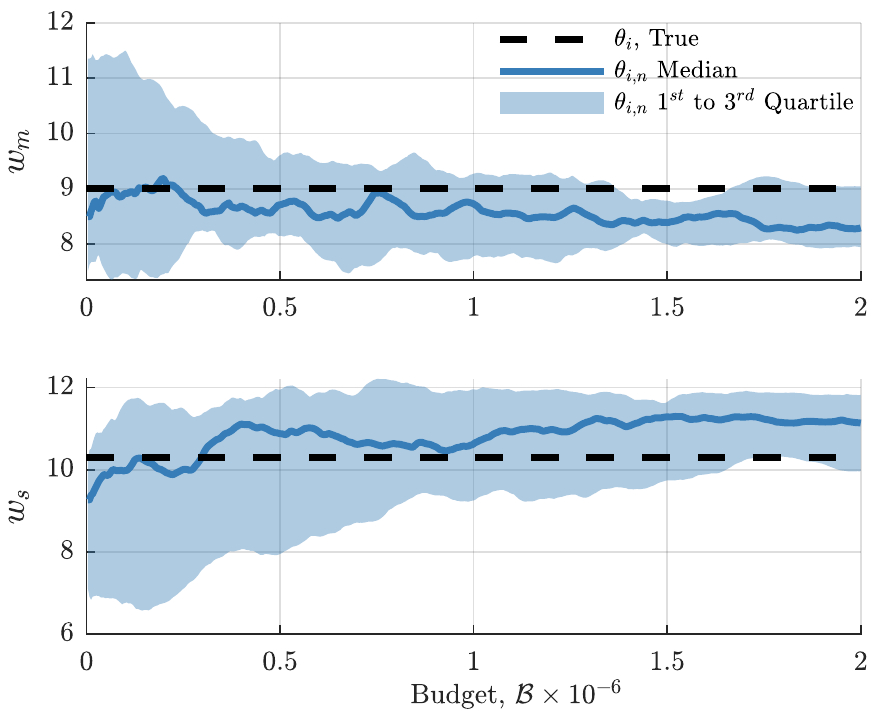}  
		\label{fig:learning:parameters}
	}
\end{minipage}

\vspace*{-0.3cm}
\subfloat[][]{
	\begin{minipage}[b!]{\textwidth}
		\centering
		\begin{minipage}[b!]{0.16\textwidth}
		\includegraphics[width=\textwidth]{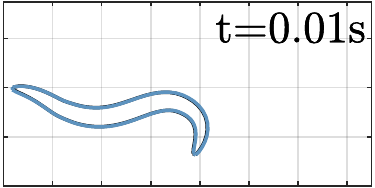} 
		\end{minipage}
		\begin{minipage}[b!]{0.16\textwidth}
		\includegraphics[width=\textwidth]{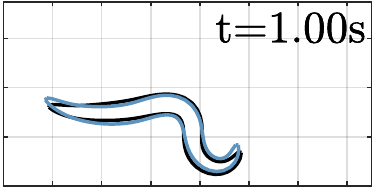} 
		\end{minipage}
		\begin{minipage}[b!]{0.16\textwidth}
		\includegraphics[width=\textwidth]{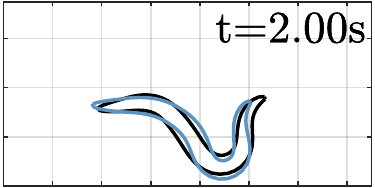} 
		\end{minipage}
		\begin{minipage}[b!]{0.16\textwidth}
		\includegraphics[width=\textwidth]{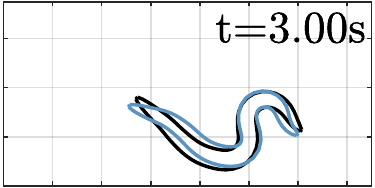} 
		\end{minipage}
		\begin{minipage}[b!]{0.16\textwidth}
		\includegraphics[width=\textwidth]{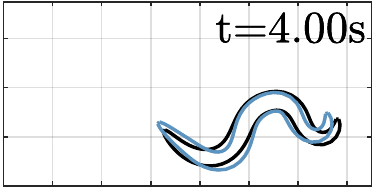} 
		\end{minipage}
		\begin{minipage}[b!]{0.16\textwidth}
		\includegraphics[width=\textwidth]{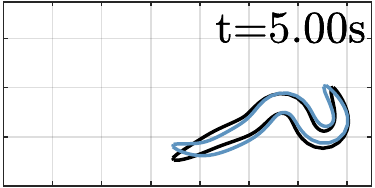} 
		\end{minipage}
		
		\begin{minipage}[b!]{0.16\textwidth}
		\includegraphics[width=\textwidth]{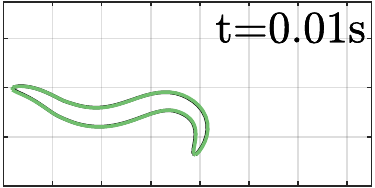} 
		\end{minipage}
		\begin{minipage}[b!]{0.16\textwidth}
		\includegraphics[width=\textwidth]{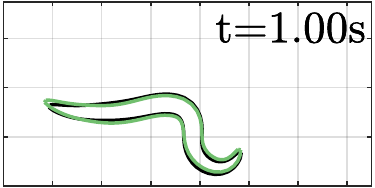} 
		\end{minipage}
		\begin{minipage}[b!]{0.16\textwidth}
		\includegraphics[width=\textwidth]{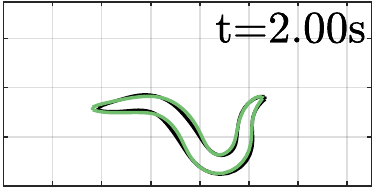} 
		\end{minipage}
		\begin{minipage}[b!]{0.16\textwidth}
		\includegraphics[width=\textwidth]{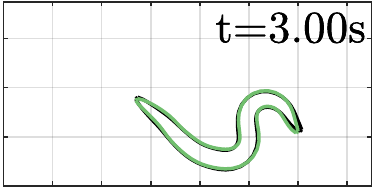} 
		\end{minipage}
		\begin{minipage}[b!]{0.16\textwidth}
		\includegraphics[width=\textwidth]{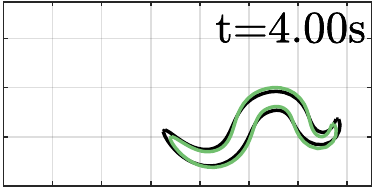} 
		\end{minipage}
		\begin{minipage}[b!]{0.16\textwidth}
		\includegraphics[width=\textwidth]{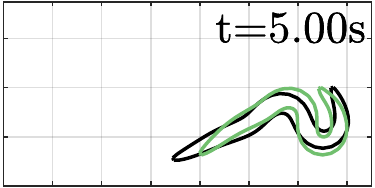} 
		\end{minipage}
		
		\begin{minipage}[b!]{0.16\textwidth}
		\includegraphics[width=\textwidth]{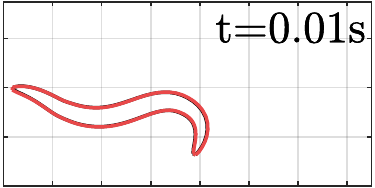} 
		\end{minipage}
		\begin{minipage}[b!]{0.16\textwidth}
		\includegraphics[width=\textwidth]{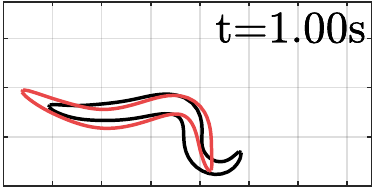} 
		\end{minipage}
		\begin{minipage}[b!]{0.16\textwidth}
		\includegraphics[width=\textwidth]{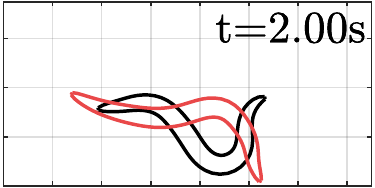} 
		\end{minipage}
		\begin{minipage}[b!]{0.16\textwidth}
		\includegraphics[width=\textwidth]{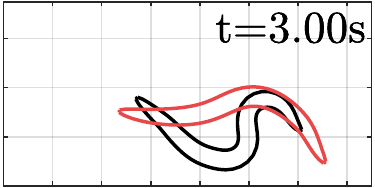} 
		\end{minipage}
		\begin{minipage}[b!]{0.16\textwidth}
		\includegraphics[width=\textwidth]{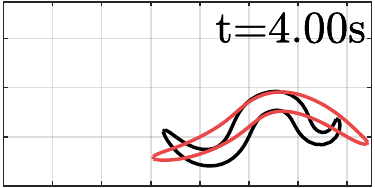} 
		\end{minipage}
		\begin{minipage}[b!]{0.16\textwidth}
		\includegraphics[width=\textwidth]{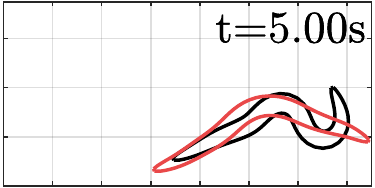} 
		\end{minipage}
		
	\end{minipage}
	\label{fig:learning:wormsim}
}
\setcounter{figure}{2}
\vspace*{-0.2cm}    					
\caption{Estimating {\ce} simulator parameters as described in Section \ref{sec:estimation:sub:sce}.
\protect\subref{fig:learning:reconstruction} and \protect\subref{fig:learning:wormsim} show the filtering distributions of SMC reconstructions of the membrane potentials of 15 cells given the true generative parameter (blue), optimization algorithm initial parameters (red), and optimized parameters (green). 
\protect\subref{fig:learning:parameters} shows the parameter optimization using PMVO, plotted as the median, upper and lower quartile across $20$ random restarts.}
\label{fig:learning}
\end{figure}


\section{Discussion}
In this work we have explored performing Bayesian inference in whole-connectome neural and whole-body {\ce} simulations.  
We describe the model-based Bayesian inference aspect of this as a ``virtual patch clamp,'' whereby unobserved latent membrane potentials can be inferred from partial observations gathered non-invasively.
Our choice of inference method facilitates estimation of the model evidence, a measure of how well the model explains the observed data.  
We presented a method for maximizing this evidence without requiring differentiable simulation components.

Previous work has investigated performing imputation of neural spikes, membrane potentials, calcium dynamics, model parameters and connectivities~\cite{vogelstein2009spike, vogelstein2010fast, friedrich2017calcium, aitchison2017model, speiser2017fast, gerkin2018towards}.
However these studies do not operate under a biologically accurate model of the dynamics of a \emph{whole} connectome, instead investigating individual or small networks of fully observed synthetic neurons.
We propose imputation of the state of neurons only distantly connected to observed neurons and performing parameter inference, regularized by connectome-scale dynamics.
Not only is this the first instance of whole-connectome neural simulators being conditioned on data, we also believe this to be one of the largest non-differentiable, non-linear state-space models in which inference has been performed.

Our open-source implementation of simulator and PMVO technique is richly extensible.  
Better models for elements of {\ce} behaviours, such as body simulators~\cite{palyanov2015sibernetic}, neural simulators~\cite{gleeson2018c302}, multi-compartment ion dynamics~\cite{Kuramochi2017multicompartment} and sensory stimuli~\cite{izquierdo2013} exist; although these models incur significantly more computational cost.
Developments even to these models are still required to explain specific behaviours, for instance, the kinds of habituation that {\ce} exhibits~\cite{ardiel2010elegant}, and newly discovered action potential generating {\ce} neurons~\cite{liu2018spiking} (contrary to long-held belief~\cite{goodman1998active}).
When additional models of these dynamics are developed, by design, they can be straightforwardly integrated into our software toolchain.

Additional data is also becoming available.
As \emph{in vivo} calcium imaging techniques improve, more neurons can be simultaneously observed, allowing the SMC sweep to be conditioned on more data.
An experiment presented in the supplementary material where florescence of all neurons is observed demonstrates improved reconstructions and recovery of parameter values than when conditioned on just $49$ neurons.
We also suggest that the simulator can be conditioned on easily recorded worm body pose data.
The simulator includes a body pose, and so a promising research direction is to develop the likelihoods terms that allow for conditioning on an observed pose, in addition to fluorescence. 

To conclude we note that in the past year several articles discussing open research issues pertaining to {\ce} simulation have been produced by the {\ce} community~\cite{Stiefel2019, larson2018connectome, sarma2018openworm, gerkin2018towards}.
Figure \ref{fig:full_page}\protect\subref{fig:openworm} outlines the community planned development pipeline for {\ce} simulation.
Our work addresses the implementation of the box simply labelled ``optimization.'' 
We propose performing this optimization by combining state-space inference techniques and variational optimization, and show on representative synthetic data that our method is capable of performing the desired optimization.

\section{Acknowledgements}
Andrew Warrington is funded by the Shilston Scholarship, Keble College, Oxford. Arthur Spencer is supported by the Wellcome Trust. We also acknowledge the support of the Natural Sciences and Engineering Research Council of Canada (NSERC), the Canada CIFAR AI Chairs Program, Intel, and DARPA under its D3M program.

\bibliographystyle{abbrv}
{\footnotesize
\bibliography{arxiv}}

\begin{thebibliography}{84}
\providecommand{\natexlab}[1]{#1}
\providecommand{\url}[1]{\texttt{#1}}
\expandafter\ifx\csname urlstyle\endcsname\relax
  \providecommand{\doi}[1]{doi: #1}\else
  \providecommand{\doi}{doi: \begingroup \urlstyle{rm}\Url}\fi

\bibitem[Seung(2011)]{seung2011neuroscience}
H. Seung.
\newblock Neuroscience: towards functional connectomics.
\newblock \emph{Nature}, 471\penalty0 (7337):\penalty0 170--172, 2011.

\bibitem[Kristan and Katz(2006)]{kristan2006form}
W.~B. Kristan and P. Katz.
\newblock Form and function in systems neuroscience.
\newblock \emph{Current biology}, 16\penalty0 (19):\penalty0 R828--R831, 2006.

\bibitem[Doty(1975)]{doty1975consciousness}
R.~W. Doty.
\newblock Consciousness from neurons.
\newblock \emph{Acta neurobiologiae experimentalis}, 35\penalty0
  (5-6):\penalty0 791--804, 1975.

\bibitem[Sarma et~al.(2018)Sarma, Lee, Portegys, Ghayoomie, Jacobs, Alicea,
  Cantarelli, Currie, Gerkin, Gingell, et~al.]{sarma2018openworm}
G.~P. Sarma, C.~W. Lee, T. Portegys, V. Ghayoomie, T. Jacobs, B. Alicea, M.
  Cantarelli, M. Currie, R.~C. Gerkin, S. Gingell, et~al.
\newblock Openworm: overview and recent advances in integrative biological
  simulation of caenorhabditis elegans.
\newblock \emph{Philosophical Transactions of the Royal Society B: Biological
  Sciences}, 373\penalty0 (1758):\penalty0 20170382, 2018.

\bibitem[Palyanov et~al.(2016)Palyanov, Khayrulin, and
  Larson]{palyanov2016sibernetic}
A. Palyanov, S. Khayrulin, and S.~D. Larson.
\newblock {Application of smoothed particle hydrodynamics to modeling
  mechanisms of biological tissue}.
\newblock \emph{Advances in Engineering Software}, 98:\penalty0 1--11, 2016.
\newblock ISSN 18735339.
\newblock \doi{10.1016/j.advengsoft.2016.03.002}.

\bibitem[Boyle et~al.(2012)Boyle, Berri, and Cohen]{boyle2012gait}
J.~H. Boyle, S. Berri, and N. Cohen.
\newblock Gait modulation in c. elegans: an integrated neuromechanical model.
\newblock \emph{Frontiers in computational neuroscience}, 6:\penalty0 10, 2012.

\bibitem[Gleeson et~al.(2018)Gleeson, Lung, Grosu, Hasani, and
  Larson]{gleeson2018c302}
P. Gleeson, D. Lung, R. Grosu, R. Hasani, and S.~D. Larson.
\newblock c302: a multiscale framework for modelling the nervous system of
  caenorhabditis elegans.
\newblock \emph{Philosophical Transactions of the Royal Society B: Biological
  Sciences}, 373\penalty0 (1758):\penalty0 20170379, 2018.

\bibitem[Kunert et~al.(2014)Kunert, Shlizerman, and Kutz]{kunert2014low}
J. Kunert, E. Shlizerman, and J.~N. Kutz.
\newblock Low-dimensional functionality of complex network dynamics:
  Neurosensory integration in the caenorhabditis elegans connectome.
\newblock \emph{Physical Review E}, 89\penalty0 (5):\penalty0 052805, 2014.

\bibitem[Marblestone(2016)]{sce}
A. Marblestone.
\newblock Simple c. elegans.
\newblock \url{https://https://github.com/adammarblestone/simple-C-elegans},
  2016.

\bibitem[Rahmati et~al.(2016)Rahmati, Kirmse, Markovi{\'c}, Holthoff, and
  Kiebel]{rahmati2016inferring}
V. Rahmati, K. Kirmse, D. Markovi{\'c}, K. Holthoff, and S.~J. Kiebel.
\newblock Inferring neuronal dynamics from calcium imaging data using
  biophysical models and bayesian inference.
\newblock \emph{PLoS computational biology}, 12\penalty0 (2):\penalty0
  e1004736, 2016.

\bibitem[Kato et~al.(2015)Kato, Kaplan, Schr{\"o}del, Skora, Lindsay, Yemini,
  Lockery, and Zimmer]{kato2015global}
S. Kato, H.~S. Kaplan, T. Schr{\"o}del, S. Skora, T.~H. Lindsay, E. Yemini, S.
  Lockery, and M. Zimmer.
\newblock Global brain dynamics embed the motor command sequence of
  caenorhabditis elegans.
\newblock \emph{Cell}, 163\penalty0 (3):\penalty0 656--669, 2015.

\bibitem[Nguyen et~al.(2016)Nguyen, Shipley, Linder, Plummer, Liu, Setru,
  Shaevitz, and Leifer]{nguyen2016whole}
J.~P. Nguyen, F.~B. Shipley, A.~N. Linder, G.~S. Plummer, M. Liu, S.~U. Setru,
  J.~W. Shaevitz, and A.~M. Leifer.
\newblock Whole-brain calcium imaging with cellular resolution in freely
  behaving caenorhabditis elegans.
\newblock \emph{Proceedings of the National Academy of Sciences}, 113\penalty0
  (8):\penalty0 E1074--E1081, 2016.

\bibitem[Szigeti et~al.(2014)Szigeti, Gleeson, Vella, Khayrulin, Palyanov,
  Hokanson, Currie, Cantarelli, Idili, and Larson]{szigeti2014openworm}
B. Szigeti, P. Gleeson, M. Vella, S. Khayrulin, A. Palyanov, J. Hokanson, M.
  Currie, M. Cantarelli, G. Idili, and S. Larson.
\newblock {OpenWorm: an open-science approach to modeling Caenorhabditis
  elegans}.
\newblock \emph{Frontiers in Computational Neuroscience}, 8\penalty0
  (November):\penalty0 1--7, 2014.
\newblock ISSN 1662-5188.
\newblock \doi{10.3389/fncom.2014.00137}.

\bibitem[Spira and Hai(2013)]{spira2013multi}
M.~E. Spira and A. Hai.
\newblock Multi-electrode array technologies for neuroscience and cardiology.
\newblock \emph{Nature nanotechnology}, 8\penalty0 (2):\penalty0 83, 2013.

\bibitem[Margrie et~al.(2002)Margrie, Brecht, and Sakmann]{margrie2002vivo}
T.~W. Margrie, M. Brecht, and B. Sakmann.
\newblock In vivo, low-resistance, whole-cell recordings from neurons in the
  anaesthetized and awake mammalian brain.
\newblock \emph{Pfl{\"u}gers Archiv}, 444\penalty0 (4):\penalty0 491--498,
  2002.

\bibitem[Stosiek et~al.(2003)Stosiek, Garaschuk, Holthoff, and
  Konnerth]{stosiek2003vivo}
C. Stosiek, O. Garaschuk, K. Holthoff, and A. Konnerth.
\newblock In vivo two-photon calcium imaging of neuronal networks.
\newblock \emph{Proceedings of the National Academy of Sciences}, 100\penalty0
  (12):\penalty0 7319--7324, 2003.

\bibitem[Chung et~al.(2013)Chung, Sun, and Gabel]{chung2013calcium}
S.~H. Chung, L. Sun, and C.~V. Gabel.
\newblock {{I}n vivo neuronal calcium imaging in {C}. elegans}.
\newblock \emph{J Vis Exp}, \penalty0 (74), Apr 2013.

\bibitem[Ardiel and Rankin(2008)]{ardiel2008behavioral}
E.~L. Ardiel and C.~H. Rankin.
\newblock Behavioral plasticity in the c. elegans mechanosensory circuit.
\newblock \emph{Journal of neurogenetics}, 22\penalty0 (3):\penalty0 239--255,
  2008.

\bibitem[Wicks et~al.(1996)Wicks, Roehrig, and Rankin]{wicks1996dynamic}
S.~R. Wicks, C.~J. Roehrig, and C.~H. Rankin.
\newblock A dynamic network simulation of the nematode tap withdrawal circuit:
  predictions concerning synaptic function using behavioral criteria.
\newblock \emph{Journal of Neuroscience}, 16\penalty0 (12):\penalty0
  4017--4031, 1996.

\bibitem[Ardiel and Rankin(2010)]{ardiel2010elegant}
E.~L. Ardiel and C.~H. Rankin.
\newblock An elegant mind: learning and memory in caenorhabditis elegans.
\newblock \emph{Learning \& memory}, 17\penalty0 (4):\penalty0 191--201, 2010.

\bibitem[Varshney et~al.(2011)Varshney, Chen, Paniagua, Hall, and
  Chklovskii]{varshney2011celegans}
L.~R. Varshney, B.~L. Chen, E. Paniagua, D.~H. Hall, and D.~B. Chklovskii.
\newblock Structural properties of the caenorhabditis elegans neuronal network.
\newblock \emph{PLOS Computational Biology}, 7\penalty0 (2):\penalty0 1--21, 02
  2011.
\newblock \doi{10.1371/journal.pcbi.1001066}.

\bibitem[White et~al.(1986)White, Southgate, Thomson, and
  Brenner]{white1986connectome}
J.~G. White, E. Southgate, J.~N. Thomson, and S. Brenner.
\newblock {{T}he structure of the nervous system of the nematode
  {C}aenorhabditis elegans}.
\newblock \emph{Philos. Trans. R. Soc. Lond., B, Biol. Sci.}, 314\penalty0
  (1165):\penalty0 1--340, Nov 1986.

\bibitem[Hasani et~al.(2017)Hasani, Beneder, Fuchs, Lung, and
  Grosu]{hasani2017sim}
R.~M. Hasani, V. Beneder, M. Fuchs, D. Lung, and R. Grosu.
\newblock Sim-ce: An advanced simulink platform for studying the brain of
  caenorhabditis elegans.
\newblock \emph{arXiv preprint arXiv:1703.06270}, 2017.

\bibitem[Hines and Carnevale(2006)]{hines2006neuron}
M. Hines and N. Carnevale.
\newblock The neuron simulation environment.
\newblock \emph{NEURON}, 9\penalty0 (6), 2006.

\bibitem[Gewaltig and Diesmann(2007)]{gewaltig2007nest}
M.-O. Gewaltig and M. Diesmann.
\newblock Nest (neural simulation tool).
\newblock \emph{Scholarpedia}, 2\penalty0 (4):\penalty0 1430, 2007.

\bibitem[Kuramochi and Iwasaki(2010)]{kuramochi2010quantitative}
M. Kuramochi and Y. Iwasaki.
\newblock Quantitative modeling of neuronal dynamics in c. elegans.
\newblock In \emph{International Conference on Neural Information Processing},
  pages 17--24. Springer, 2010.

\bibitem[Palyanov and Khayrulin(2015)]{palyanov2015sibernetic}
A.~Y. Palyanov and S.~S. Khayrulin.
\newblock Sibernetic: A software complex based on the pci sph algorithm aimed
  at simulation problems in biomechanics.
\newblock \emph{Russian Journal of Genetics: Applied Research}, 5\penalty0
  (6):\penalty0 635--641, Nov 2015.
\newblock \doi{10.1134/S2079059715060052}.

\bibitem[Wen et~al.(2012)Wen, Po, Hulme, Chen, Liu, Kwok, Gershow, Leifer,
  Butler, Fang-Yen, et~al.]{wen2012proprioceptive}
Q. Wen, M.~D. Po, E. Hulme, S. Chen, X. Liu, S.~W. Kwok, M. Gershow, A.~M.
  Leifer, V. Butler, C. Fang-Yen, et~al.
\newblock Proprioceptive coupling within motor neurons drives c. elegans
  forward locomotion.
\newblock \emph{Neuron}, 76\penalty0 (4):\penalty0 750--761, 2012.

\bibitem[Hill(1938)]{hill1938heat}
A.~V. Hill.
\newblock The heat of shortening and the dynamic constants of muscle.
\newblock \emph{Proc. R. Soc. Lond. B}, 126\penalty0 (843):\penalty0 136--195,
  1938.

\bibitem[Grienberger and Konnerth(2012)]{grienberger2012imaging}
C. Grienberger and A. Konnerth.
\newblock Imaging calcium in neurons.
\newblock \emph{Neuron}, 73\penalty0 (5):\penalty0 862--885, 2012.

\bibitem[Yasuda et~al.(2004)Yasuda, Nimchinsky, Scheuss, Pologruto, Oertner,
  Sabatini, and Svoboda]{yasuda2004imaging}
R. Yasuda, E.~A. Nimchinsky, V. Scheuss, T.~A. Pologruto, T.~G. Oertner, B.~L.
  Sabatini, and K. Svoboda.
\newblock Imaging calcium concentration dynamics in small neuronal
  compartments.
\newblock \emph{Sci. STKE}, 2004\penalty0 (219):\penalty0 pl5--pl5, 2004.

\bibitem[Doucet and Johansen(2010)]{Doucet11atutorial}
A. Doucet and A. Johansen.
\newblock A tutorial on particle filtering and smoothing: fifteen years later,
  2010.

\bibitem[Staines and Barber(2012)]{staines2012variational}
J. Staines and D. Barber.
\newblock Variational optimization.
\newblock \emph{arXiv preprint arXiv:1212.4507}, 2012.

\bibitem[Williams(1992)]{williams1992simple}
R.~J. Williams.
\newblock Simple statistical gradient-following algorithms for connectionist
  reinforcement learning.
\newblock \emph{Machine learning}, 8\penalty0 (3-4):\penalty0 229--256, 1992.

\bibitem[Kingma and Ba(2014)]{kingma2014adam}
D.~P. Kingma and J. Ba.
\newblock Adam: A method for stochastic optimization.
\newblock \emph{arXiv preprint arXiv:1412.6980}, 2014.

\bibitem[Weaver and Tao(2001)]{weaver2001optimal}
L. Weaver and N. Tao.
\newblock The optimal reward baseline for gradient-based reinforcement
  learning.
\newblock In \emph{Proceedings of the Seventeenth conference on Uncertainty in
  artificial intelligence}, pages 538--545. Morgan Kaufmann Publishers Inc.,
  2001.

\bibitem[Vogelstein et~al.(2009)Vogelstein, Watson, Packer, Yuste, Jedynak, and
  Paninski]{vogelstein2009spike}
J.~T. Vogelstein, B.~O. Watson, A.~M. Packer, R. Yuste, B. Jedynak, and L.
  Paninski.
\newblock Spike inference from calcium imaging using sequential monte carlo
  methods.
\newblock \emph{Biophysical journal}, 97\penalty0 (2):\penalty0 636--655, 2009.

\bibitem[Vogelstein et~al.(2010)Vogelstein, Packer, Machado, Sippy, Babadi,
  Yuste, and Paninski]{vogelstein2010fast}
J.~T. Vogelstein, A.~M. Packer, T.~A. Machado, T. Sippy, B. Babadi, R. Yuste,
  and L. Paninski.
\newblock Fast nonnegative deconvolution for spike train inference from
  population calcium imaging.
\newblock \emph{Journal of neurophysiology}, 104\penalty0 (6):\penalty0
  3691--3704, 2010.

\bibitem[Friedrich et~al.(2017)Friedrich, Zhou, and
  Paninski]{friedrich2017calcium}
J. Friedrich, P. Zhou, and L. Paninski.
\newblock Fast online deconvolution of calcium imaging data.
\newblock \emph{PLOS Computational Biology}, 13\penalty0 (3):\penalty0 1--26,
  03 2017.
\newblock \doi{10.1371/journal.pcbi.1005423}.

\bibitem[Aitchison et~al.(2017)Aitchison, Russell, Packer, Yan, Castonguay,
  Hausser, and Turaga]{aitchison2017model}
L. Aitchison, L. Russell, A.~M. Packer, J. Yan, P. Castonguay, M. Hausser, and
  S.~C. Turaga.
\newblock Model-based bayesian inference of neural activity and connectivity
  from all-optical interrogation of a neural circuit.
\newblock In \emph{Advances in Neural Information Processing Systems}, pages
  3486--3495, 2017.

\bibitem[Speiser et~al.(2017)Speiser, Yan, Archer, Buesing, Turaga, and
  Macke]{speiser2017fast}
A. Speiser, J. Yan, E.~W. Archer, L. Buesing, S.~C. Turaga, and J.~H. Macke.
\newblock Fast amortized inference of neural activity from calcium imaging data
  with variational autoencoders.
\newblock In \emph{Advances in Neural Information Processing Systems}, pages
  4024--4034, 2017.

\bibitem[Gerkin et~al.(2018)Gerkin, Jarvis, and Crook]{gerkin2018towards}
R.~C. Gerkin, R.~J. Jarvis, and S.~M. Crook.
\newblock Towards systematic, data-driven validation of a collaborative,
  multi-scale model of caenorhabditis elegans.
\newblock \emph{Philosophical Transactions of the Royal Society B: Biological
  Sciences}, 373\penalty0 (1758):\penalty0 20170381, 2018.

\bibitem[Kuramochi and Doi(2017)]{Kuramochi2017multicompartment}
M. Kuramochi and M. Doi.
\newblock A computational model based on multi-regional calcium imaging
  represents the spatio-temporal dynamics in a caenorhabditis elegans sensory
  neuron.
\newblock \emph{PLOS ONE}, 12\penalty0 (1):\penalty0 1--19, 01 2017.
\newblock \doi{10.1371/journal.pone.0168415}.

\bibitem[Izquierdo and Beer(2013{\natexlab{a}})]{izquierdo2013}
E.~J. Izquierdo and R.~D. Beer.
\newblock {Connecting a Connectome to Behavior: An Ensemble of Neuroanatomical
  Models of C. elegans Klinotaxis}.
\newblock \emph{PLoS Computational Biology}, 9\penalty0 (2),
  2013{\natexlab{a}}.
\newblock ISSN 1553734X.
\newblock \doi{10.1371/journal.pcbi.1002890}.

\bibitem[Liu et~al.(2018)Liu, Kidd, Dobosiewicz, and Bargmann]{liu2018spiking}
Q. Liu, P.~B. Kidd, M. Dobosiewicz, and C.~I. Bargmann.
\newblock C. elegans awa olfactory neurons fire calcium-mediated all-or-none
  action potentials.
\newblock \emph{Cell}, 175\penalty0 (1):\penalty0 57 -- 70.e17, 2018.
\newblock ISSN 0092-8674.
\newblock \doi{https://doi.org/10.1016/j.cell.2018.08.018}.

\bibitem[Goodman et~al.(1998)Goodman, Hall, Avery, and
  Lockery]{goodman1998active}
M.~B. Goodman, D.~H. Hall, L. Avery, and S.~R. Lockery.
\newblock Active currents regulate sensitivity and dynamic range in c. elegans
  neurons.
\newblock \emph{Neuron}, 20\penalty0 (4):\penalty0 763--772, 1998.

\bibitem[Stiefel and Brooks(2019)]{Stiefel2019}
K.~M. Stiefel and D.~S. Brooks.
\newblock Why is there no successful whole brain simulation (yet)?
\newblock \emph{Biological Theory}, Mar 2019.
\newblock ISSN 1555-5550.
\newblock \doi{10.1007/s13752-019-00319-5}.

\bibitem[Larson et~al.(2018)Larson, Gleeson, and Brown]{larson2018connectome}
S.~D. Larson, P. Gleeson, and A.~E. Brown.
\newblock Connectome to behaviour: modelling caenorhabditis elegans at cellular
  resolution, 2018.

\bibitem[Kunert et~al.(2017)Kunert, Proctor, Brunton, and
  Kutz]{kunert2017spatiotemporal}
J.~M. Kunert, J.~L. Proctor, S.~L. Brunton, and J.~N. Kutz.
\newblock Spatiotemporal feedback and network structure drive and encode
  caenorhabditis elegans locomotion.
\newblock \emph{PLoS computational biology}, 13\penalty0 (1):\penalty0
  e1005303, 2017.

\bibitem[Cohen and Sanders(2014)]{cohen2014nematode}
N. Cohen and T. Sanders.
\newblock Nematode locomotion: dissecting the neuronal--environmental loop.
\newblock \emph{Current opinion in neurobiology}, 25:\penalty0 99--106, 2014.

\bibitem[Yemini et~al.(2013)Yemini, Jucikas, Grundy, Brown, and
  Schafer]{yemini2013elegansvideo}
E. Yemini, T. Jucikas, L. Grundy, A. Brown, and W. Schafer.
\newblock A database of c. elegans behavioral phenotypes.
\newblock \emph{Nature Methods}, 10\penalty0 (9):\penalty0 877--879, 2013.
\newblock \doi{10.1038/nmeth.2560.A}.

\bibitem[Mujika et~al.(2017)Mujika, Le{\v{s}}kovsk{\`y}, {\'A}lvarez, Otaduy,
  and Epelde]{mujika2017modeling}
A. Mujika, P. Le{\v{s}}kovsk{\`y}, R. {\'A}lvarez, M.~A. Otaduy, and G. Epelde.
\newblock Modeling behavioral experiment interaction and environmental stimuli
  for a synthetic c. elegans.
\newblock \emph{Frontiers in neuroinformatics}, 11:\penalty0 71, 2017.

\bibitem[Abbott and Kepler(1990)]{abbott1990model}
L. Abbott and T.~B. Kepler.
\newblock Model neurons: from hodgkin-huxley to hopfield.
\newblock In \emph{Statistical mechanics of neural networks}, pages 5--18.
  Springer, 1990.

\bibitem[Abbott(1999)]{abbott1999lapicque}
L.~F. Abbott.
\newblock Lapicque’s introduction of the integrate-and-fire model neuron
  (1907).
\newblock \emph{Brain research bulletin}, 50\penalty0 (5-6):\penalty0 303--304,
  1999.

\bibitem[Hodgkin and Huxley(1952)]{hodgkin1952quantitative}
A.~L. Hodgkin and A.~F. Huxley.
\newblock A quantitative description of membrane current and its application to
  conduction and excitation in nerve.
\newblock \emph{The Journal of physiology}, 117\penalty0 (4):\penalty0
  500--544, 1952.

\bibitem[Hertz(2018)]{hertz2018introduction}
J.~A. Hertz.
\newblock \emph{Introduction to the theory of neural computation}.
\newblock CRC Press, 2018.

\bibitem[Saarinen et~al.(2006)Saarinen, Linne, and Yli-Harja]{SAARINEN20061091}
A. Saarinen, M.-L. Linne, and O. Yli-Harja.
\newblock Modeling single neuron behavior using stochastic differential
  equations.
\newblock \emph{Neurocomputing}, 69\penalty0 (10):\penalty0 1091 -- 1096, 2006.
\newblock ISSN 0925-2312.
\newblock \doi{https://doi.org/10.1016/j.neucom.2005.12.052}.
\newblock Computational Neuroscience: Trends in Research 2006.

\bibitem[Vidal-Gadea et~al.(2011)Vidal-Gadea, Topper, Young, Crisp, Kressin,
  Elbel, Maples, Brauner, Erbguth, Axelrod, et~al.]{vidal2011caenorhabditis}
A. Vidal-Gadea, S. Topper, L. Young, A. Crisp, L. Kressin, E. Elbel, T. Maples,
  M. Brauner, K. Erbguth, A. Axelrod, et~al.
\newblock Caenorhabditis elegans selects distinct crawling and swimming gaits
  via dopamine and serotonin.
\newblock \emph{Proceedings of the National Academy of Sciences}, 108\penalty0
  (42):\penalty0 17504--17509, 2011.

\bibitem[Lebois et~al.(2012)Lebois, Sauvage, Py, Cardoso, Ladoux, Hersen, and
  Di~Meglio]{lebois2012locomotion}
F. Lebois, P. Sauvage, C. Py, O. Cardoso, B. Ladoux, P. Hersen, and J.-M.
  Di~Meglio.
\newblock Locomotion control of caenorhabditis elegans through confinement.
\newblock \emph{Biophysical journal}, 102\penalty0 (12):\penalty0 2791--2798,
  2012.

\bibitem[WormAtlas()]{wormatlas}
WormAtlas.
\newblock \url{http://www.wormatlas.org}.
\newblock Accessed: 2018-05-15.

\bibitem[Linderman et~al.(2017)Linderman, Mena, Cooper, Paninski, and
  Cunningham]{linderman2017reparameterizing}
S.~W. Linderman, G.~E. Mena, H. Cooper, L. Paninski, and J.~P. Cunningham.
\newblock Reparameterizing the birkhoff polytope for variational permutation
  inference.
\newblock \emph{arXiv preprint arXiv:1710.09508}, 2017.

\bibitem[Mena et~al.(2018)Mena, Belanger, Linderman, and
  Snoek]{mena2018learning}
G. Mena, D. Belanger, S. Linderman, and J. Snoek.
\newblock Learning latent permutations with gumbel-sinkhorn networks.
\newblock \emph{arXiv preprint arXiv:1802.08665}, 2018.

\bibitem[Bretscher et~al.(2011)Bretscher, Kodama-Namba, Busch, Murphy, Soltesz,
  Laurent, and de~Bono]{bretscher2011temperature}
A.~J. Bretscher, E. Kodama-Namba, K.~E. Busch, R.~J. Murphy, Z. Soltesz, P.
  Laurent, and M. Bono.
\newblock Temperature, oxygen, and salt-sensing neurons in c. elegans are
  carbon dioxide sensors that control avoidance behavior.
\newblock \emph{Neuron}, 69\penalty0 (6):\penalty0 1099--1113, 2011.

\bibitem[Metaxakis et~al.(2018)Metaxakis, Petratou, and
  Tavernarakis]{metaxakis2018multimodal}
A. Metaxakis, D. Petratou, and N. Tavernarakis.
\newblock Multimodal sensory processing in caenorhabditis elegans.
\newblock \emph{Open biology}, 8\penalty0 (6):\penalty0 180049, 2018.

\bibitem[Cohen and Denham(2018)]{cohen2018whole}
N. Cohen and J.~E. Denham.
\newblock Whole animal modeling: piecing together nematode locomotion.
\newblock \emph{Current Opinion in Systems Biology}, 2018.

\bibitem[Izquierdo and Beer(2013{\natexlab{b}})]{izquierdo2013connecting}
E.~J. Izquierdo and R.~D. Beer.
\newblock Connecting a connectome to behavior: an ensemble of neuroanatomical
  models of c. elegans klinotaxis.
\newblock \emph{PLoS computational biology}, 9\penalty0 (2):\penalty0 e1002890,
  2013{\natexlab{b}}.

\bibitem[Olivares et~al.(2018)Olivares, Izquierdo, and
  Beer]{olivares2018potential}
E.~O. Olivares, E.~J. Izquierdo, and R.~D. Beer.
\newblock Potential role of a ventral nerve cord central pattern generator in
  forward and backward locomotion in caenorhabditis elegans.
\newblock \emph{Network Neuroscience}, 2\penalty0 (3):\penalty0 323--343, 2018.

\bibitem[Fouad et~al.(2018)Fouad, Teng, Mark, Liu, Alvarez-Illera, Ji, Du,
  Bhirgoo, Cornblath, Guan, et~al.]{fouad2018distributed}
A.~D. Fouad, S. Teng, J.~R. Mark, A. Liu, P. Alvarez-Illera, H. Ji, A. Du,
  P.~D. Bhirgoo, E. Cornblath, S.~A. Guan, et~al.
\newblock Distributed rhythm generators underlie caenorhabditis elegans forward
  locomotion.
\newblock \emph{Elife}, 7:\penalty0 e29913, 2018.

\bibitem[Gao et~al.(2017)Gao, Guan, Fouad, Meng, Huang, Li, Alcaire, Hung,
  Kawano, Lu, et~al.]{gao2017excitatory}
S. Gao, S.~A. Guan, A.~D. Fouad, J. Meng, Y.-C. Huang, Y. Li, S. Alcaire, W.
  Hung, T. Kawano, Y. Lu, et~al.
\newblock Excitatory motor neurons are local central pattern generators in an
  anatomically compressed motor circuit for reverse locomotion.
\newblock \emph{bioRxiv}, page 135418, 2017.

\bibitem[Teng and Wood(2018)]{teng2018bayesian}
M. Teng and F. Wood.
\newblock Bayesian distributed stochastic gradient descent.
\newblock In \emph{Advances in Neural Information Processing Systems}, pages
  6378--6388, 2018.

\bibitem[Chen et~al.(2016)Chen, Pan, Monga, Bengio, and
  Jozefowicz]{chen2016revisiting}
J. Chen, X. Pan, R. Monga, S. Bengio, and R. Jozefowicz.
\newblock Revisiting distributed synchronous sgd.
\newblock \emph{arXiv preprint arXiv:1604.00981}, 2016.

\bibitem[Hastings(1970)]{hastings1970monte}
W.~K. Hastings.
\newblock Monte carlo sampling methods using markov chains and their
  applications.
\newblock \emph{Biometrika}, 57\penalty0 (1):\penalty0 97--109, 1970.

\bibitem[Chib and Greenberg(1995)]{chib1995mh}
S. Chib and E. Greenberg.
\newblock Understanding the metropolis-hastings algorithm.
\newblock \emph{The american statistician}, 49\penalty0 (4):\penalty0 327--335,
  1995.

\bibitem[Andrieu et~al.(2010)Andrieu, Doucet, and
  Holenstein]{andrieu2010particle}
C. Andrieu, A. Doucet, and R. Holenstein.
\newblock Particle markov chain monte carlo methods.
\newblock \emph{Journal of the Royal Statistical Society: Series B (Statistical
  Methodology)}, 72\penalty0 (3):\penalty0 269--342, 2010.

\bibitem[Kantas et~al.(2015)Kantas, Doucet, Singh, Maciejowski, Chopin,
  et~al.]{kantas2015particle}
N. Kantas, A. Doucet, S.~S. Singh, J. Maciejowski, N. Chopin, et~al.
\newblock On particle methods for parameter estimation in state-space models.
\newblock \emph{Statistical science}, 30\penalty0 (3):\penalty0 328--351, 2015.

\bibitem[Haario et~al.(2005)Haario, Saksman, and
  Tamminen]{haario2005componentwise}
H. Haario, E. Saksman, and J. Tamminen.
\newblock Componentwise adaptation for high dimensional mcmc.
\newblock \emph{Computational Statistics}, 20\penalty0 (2):\penalty0 265--273,
  2005.

\bibitem[van~de Meent et~al.(2014)van~de Meent, Paige, and
  Wood]{van2014tempering}
J.-W. Meent, B. Paige, and F. Wood.
\newblock Tempering by subsampling.
\newblock \emph{arXiv preprint arXiv:1401.7145}, 2014.

\bibitem[Maclaurin and Adams(2015)]{maclaurin2015firefly}
D. Maclaurin and R.~P. Adams.
\newblock Firefly monte carlo: Exact mcmc with subsets of data.
\newblock In \emph{Twenty-Fourth International Joint Conference on Artificial
  Intelligence}, 2015.

\bibitem[Angelopoulos and Cussens(2008)]{angelopoulos2008bayesian}
N. Angelopoulos and J. Cussens.
\newblock Bayesian learning of bayesian networks with informative priors.
\newblock \emph{Annals of Mathematics and Artificial Intelligence}, 54\penalty0
  (1-3):\penalty0 53--98, 2008.

\bibitem[Gupta et~al.(2018)Gupta, Hainsworth, Hogg, Lee, and
  Faeder]{gupta2018evaluation}
S. Gupta, L. Hainsworth, J. Hogg, R. Lee, and J. Faeder.
\newblock Evaluation of parallel tempering to accelerate bayesian parameter
  estimation in systems biology.
\newblock In \emph{2018 26th Euromicro International Conference on Parallel,
  Distributed and Network-based Processing (PDP)}, pages 690--697. IEEE, 2018.

\bibitem[Swendsen and Wang(1986)]{swendsen1986replica}
R.~H. Swendsen and J.-S. Wang.
\newblock Replica monte carlo simulation of spin-glasses.
\newblock \emph{Physical review letters}, 57\penalty0 (21):\penalty0 2607,
  1986.

\bibitem[Altekar et~al.(2004)Altekar, Dwarkadas, Huelsenbeck, and
  Ronquist]{Altekar2003PT}
G. Altekar, S. Dwarkadas, J.~P. Huelsenbeck, and F. Ronquist.
\newblock {Parallel Metropolis coupled Markov chain Monte Carlo for Bayesian
  phylogenetic inference}.
\newblock \emph{Bioinformatics}, 20\penalty0 (3):\penalty0 407--415, 01 2004.
\newblock ISSN 1367-4803.
\newblock \doi{10.1093/bioinformatics/btg427}.

\bibitem[Miasojedow et~al.(2013)Miasojedow, Moulines, and
  Vihola]{miasojedow2013adaptive}
B. Miasojedow, E. Moulines, and M. Vihola.
\newblock An adaptive parallel tempering algorithm.
\newblock \emph{Journal of Computational and Graphical Statistics}, 22\penalty0
  (3):\penalty0 649--664, 2013.

\bibitem[{\L}{\k{a}}cki and Miasojedow(2016)]{lkacki2016state}
M.~K. {\L}{\k{a}}cki and B. Miasojedow.
\newblock State-dependent swap strategies and automatic reduction of number of
  temperatures in adaptive parallel tempering algorithm.
\newblock \emph{Statistics and Computing}, 26\penalty0 (5):\penalty0 951--964,
  2016.

\end{thebibliography}
\newpage


\appendix
\newpage

\section{Supplementary Materials}
In this supplement, we offer additional proofs, experimental details and configurations, and intuitions about the methods presented in the main text.

We first present additional information relating to the simulation platform and software implementations we distribute.
We then present extended experimental details and configurations for the experiments presented, along with additional related results not included in the main text.
The then present the mathematical machinery required to define and understand the sequential Monte Carlo method used throughout, and the particle marginal variational optimization algorithm we present.
We conclude by including, for completeness as opposed to being a ``tutorial,'' explanation of Metropolis Hastings, particle marginal Metropolis Hastings and parallel tempering methods compared to in Section 4 of the main text, as well as some qualitative evaluation of the merits of optimization compared to inference.
No additional experimental details or methodological innovations are presented in this final section, and so this section is included for reference only. 

Source code for the simulator, inference and optimization algorithms, and for reproducing results figures are available on request.

\section{Simulating \cef}
We now give more detail on the simulator we assemble and perform inference in.
The graphical model of the simulator is shown in Figure 1(d) of the main text.
The guiding principal for this simulator is computational tractability and modularity.
The most accurate simulators of {\ce} neural dynamics~\cite{gleeson2018c302} and body~\cite{palyanov2015sibernetic, palyanov2016sibernetic} are prohibitively computationally expensive to run the number of particles and iterations required for inference or optimization, and for rapid experimentation and exploration.
Therefore, we design a simulator prioritising throughput, making the inference task is at least computationally tractable. 
Our ambition is that the simulator will have the fidelity and run-time cost that rapid \emph{in silico} experimentation can be performed by practitioners on standard desktop machinery.
This \emph{in silico} virtual patch clamp experiment can be as simple as pinning the value of a voltage or current (a variable in the model) to a particular value (or series of values) and inspecting the resulting simulations.
The computational cost of our model means that tens of seconds of data can be simulated in a matter of minutes on widely available hardware, whereas more complex simulators would require hours of simulation time and/or super-computing power not readily available.
The modularity of our design allows better models, once developed, to be used, with suitable hardware, as all of the inference modules we describe are agnostic to the particulars of the models used and vice versa.

\subsection{Simulator Components}
The key component of any {\ce} simulator is a model of neural activity in all $302$ neurons, including the interactions at synapses and gap junctions. 
We also include a body simulator~\cite{boyle2012gait}, driven by the neural activity, which is capable of providing a proprioceptive feedback signal to the network.
Evidence suggests that this proprioception is an important element in capturing the neural dynamics of the worm~\cite{kunert2014low, kunert2017spatiotemporal}, providing the closed-loop feedback required to generate the oscillations which induce locomotion~\cite{cohen2014nematode, wen2012proprioceptive, boyle2012gait}.
There also exist large repositories of body pose data~\cite{yemini2013elegansvideo} which we wish to condition on in future iterations of the project -- further motivating the inclusion of the body simulator in the pipeline, although leveraging this data is deferred to the future work.
We have identified \emph{in vivo} calcium imaging as a source of data on which to condition our learning, and therefore include a model of the data generation process.
Finally, we also suggest that quantification of external stimuli is important for faithful simulation, as is studied in \cite{mujika2017modeling}, and so we provision for this `1direct stimulation,'' but defer detailed study to future work.
We now describe the specifics of each of our implementations of these elements in more details.

\paragraph{Neural Simulation}
\label{sec:prior:neural}
The behaviour of neurons is well characterised by sets of ordinary differential equations (ODEs)~\cite{abbott1990model, abbott1999lapicque, hodgkin1952quantitative, hertz2018introduction, SAARINEN20061091} which can be numerically integrated over time to simulate networks of neurons~\cite{hines2006neuron, gewaltig2007nest}.
We use the simulator presented by \citet{sce}, which builds on developments presented in \citet{kunert2014low} and \citet{wicks1996dynamic}, called ``simple {\ce}'' (SCE).
SCE represents each of the $302$ neurons,\footnote{The original release of SCE only simulated $299$ neurons, but adding the $3$ omitted neurons was trivial under the modelling assumptions.} which we denote $N$, as single compartments, and models voltage-dependent currents due to leakage, connections between neurons.
Current through synapses is approximated as a single current due to all ion mechanisms in order to reduce the complexity and make the simulation faster.
The complexity is further reduced by combining, with an additive effect, multiple connections of the same type between a pair of neurons, into a single connection.
We refer the reader to \citet{kunert2014low} for a detailed description of the underlying model used in SCE.
To this we added simulation of intracellular calcium, leveraging the relationships defined in \citet{rahmati2016inferring}.
This defines the calcium ion concentration as a low-pass filtered function of the membrane potential.

Together, these expressions describe the time-evolution of the neural circuit, denoted as $p(\mathbf{v}_{t}, \mathbf{c}_{t} | \mathbf{v}_{t-1}, \mathbf{c}_{t}, \boldsymbol\theta)$, where $\mathbf{v}_{t} \in \mathcal{V} = \mathbb{R}^{N\times 2}$, represents the voltage and calcium concentration for each of the $N$ neurons at time $t$.
Here $\boldsymbol\theta$ corresponds to the electrophysiological constants that govern the properties of neurons, as well as the relative strengths of the connections between neurons.
For the purposes of this work, we assume these physiological parameters are fixed and known, although defining a formal mechanism for refining these parameters is a driving ambition for this work.

We re-implemented much of the original SCE implementation to use vectorized NumPy calculations, resulting in orders of magnitude speedup.
We do not consider this a ``contribution'' however as the original implementation was designed to make the code highly interpretable, as opposed to computationally streamlined.
We also implemented a parallel processing system, allowing individual particles to be iterated embarrassingly parallelly, within a node.
We experimented with distributing across nodes, but found that communication overheads led to sub-linear scaling in the performance, and so exploit multi-node architectures in other ways, addressed later in this text.

A further efficiency saving we identified is no longer using an ODE solver, instead using finite difference.
We conducted experiments into the reduction in accuracy and the time saving.
The results of this experiment are shown below in Figure \ref{fig:ode_diff}.
Plotted are the voltage trajectories for the most active of the neurons.
The difference between red (ODE) and blue (finite difference) curves is minimal.
Importantly we found that the average time for iterating using the ODE solver was $16.69$ milliseconds, was reduced to just $0.7$ milliseconds by using forward difference  -- more than a $20\times$ speed-up.
For all simulations presented, we use an integration timestep of $\delta t = 0.01$. 

\begin{figure}[h!]
\centering
\includegraphics[width=\textwidth]{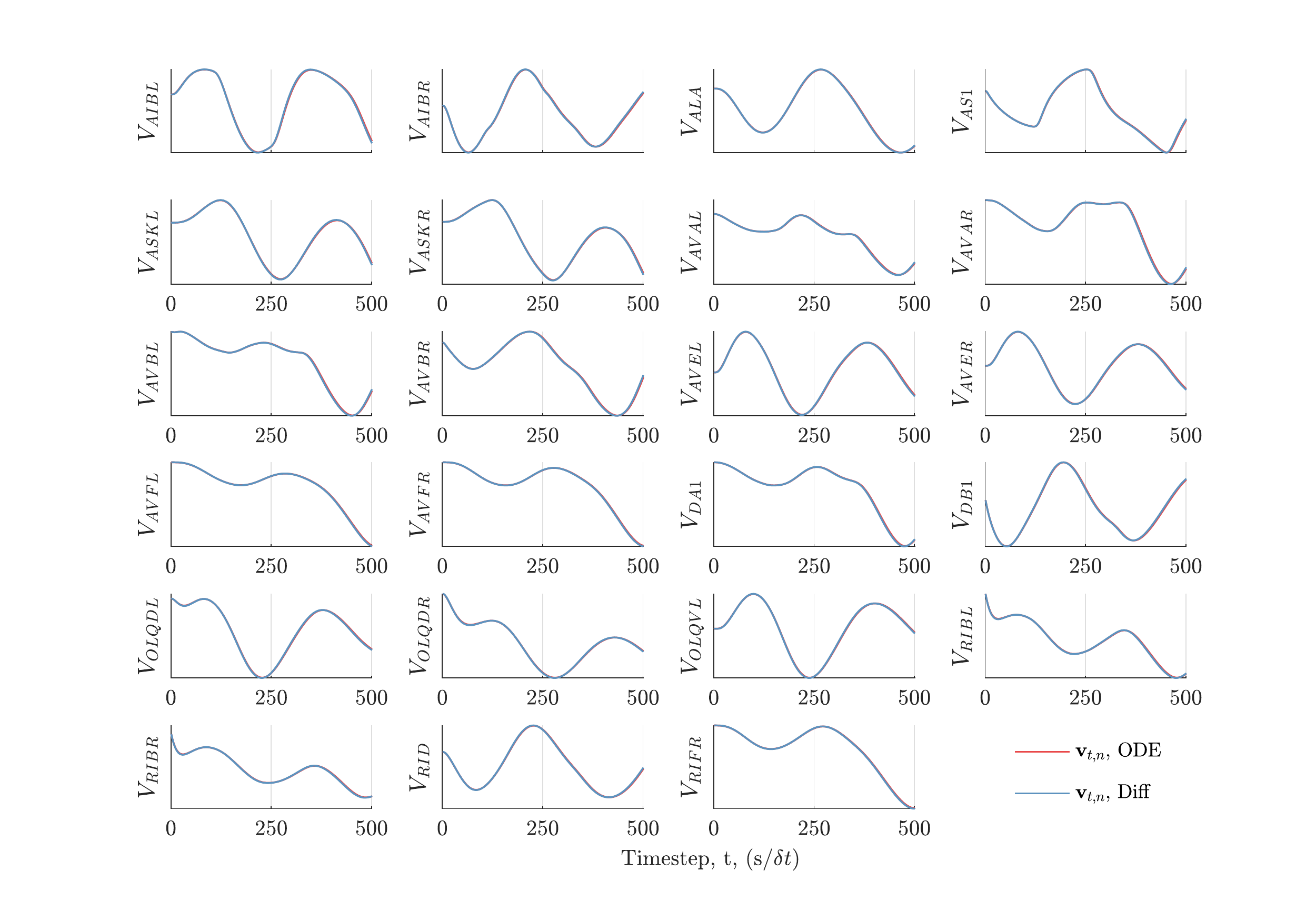}  
\caption{Experiment showing the accumulated error when using and ODE solver (\texttt{odeint} provided by SciPy) and simply using finite difference.}
\label{fig:ode_diff}
\end{figure}

While no longer using an ODE solver does introduce integration errors, as we justify in the next section, we add noise to the state at each timestep to improve the performance of SMC, where the magnitude of the noise we add is far larger than the inaccuracy introduced by the discretization.
The speedup is significant, and hence we can run more particles for the same computational cost, which will likely lead to a more accurate SMC sweep overall, compared to using the more accurate, but slower ODE integrator.
We note that we still use an ODE solver in WormSim, as this integrator ``fails'' when the worms' body position is not physioloigcally plausible, allowing us to remove that particle from the sweep.

We investigated using the more physiologically accurate simulation platform ``c302'' presented by~\citet{gleeson2018c302} building on the NEURON simulation environment~\cite{hines2006neuron}.
This simulator provides a ``wrapper'' for constructing a NEURON simulation environment with the structure of the {\ce} connectome. 
This environment is more accurate, simulating the neurons as multi-compartment differential equations, simulating multiple ion mechanisms, and using more sophisticated models of synaptic conductance.
However, we elected not to use c302 for several reasons.
The main reason was c302 is considerably more computationally expensive than SCE, as much as two orders of magnitude.
This computational burden would severely limit the number of samples that can be taken to the detriment of the inference result, and therefore we select the more computationally tractable SCE package. 
Although modelling assumptions in SCE fundamentally limit its absolute fidelity, we believe that it is sufficiently accurate to make progress on the inference challenge, where ``upgrading'' the simulator to c302, or including bespoke NEURON components, at a later date is possible.
Secondly, SCE is less parametrized than the c302 environment and has a lower dimensional state representation, making the connectome-scale operation easier to interpret and form and test hypothesis around.

\paragraph{Body Simulation}
The body simulator, WormSim, was developed by \citet{boyle2012gait} to demonstrate the need for proprioceptive feedback to drive locomotion in {\ce}~\cite{wen2012proprioceptive, vidal2011caenorhabditis, lebois2012locomotion, cohen2014nematode}.
This model represents the body of the worm in two dimensions as a series of rigid rods, tensile units and springs.
The springs define the elastic nature of the worm's body, while the rods serve to maintain the worm bodies overall form, achieved by cell tension and internal pressure in the real worm.

We incorporated this model into SCE by defining an interface for driving WormSim using the anatomically correct network instead of the simplified network used in the original work.
We mesh the WormSim simulator onto SCE by using neural activity from SCE to drive the body simulation, and integrate the proprioceptive feedback estimated by WormSim to SCE.
The nature and precise implementation of this meshing was determined by the structure of WormSim.
Since this model is a departure from the true physioloigy of the worm, modelling only the major contributors to locomotion, our meshing works within the framework defined by WormSim accordingly.

WormSim translates the body shape into proprioceptive feedback by calculating a current that is injected back into the controlling neural network.
WormSim uses $12$ identical neural units inplace of the biologically correct network.
Therefore, we linearly interpolate the signal received by each biological neuron based on its location as specified in WormAtlas~\cite{wormatlas} and the relative location of each neural unit in the WormSim model.
The neurons we allow proprioceptive feedback to flow into are $DB\{1,2,3,4,5,6,7\}$ on the dorsal side and $VB\{1,2,3,4,5,6,7,8,9,10,11\}$ on the ventral side.
While other neurons may receive feedback, these are the only neurons provisioned by WormSim, as suggested by \citet{wen2012proprioceptive}.
We then introduce a ``strength'' parameter, notated as $w_{\text{s}}$, for multiplying the WormSim calculated stretch receptive current into to the current injected into the aforementioned biological neurons.
If the representations simulators' were already compatible, this parameter would take unity value. 
However, we find that tuning this parameter is highly important to get locomotion in the body.

We use a similar approach for converting the neural stimuli from the biological representation used by SCE to the $12$ units used by WormSim.
We use the $DB$ and $DD$ for dorsal muscle excitation, and $VB$ and $VD$ for ventral excitation, again, inspired by \citet{wen2012proprioceptive} and the implementation of WormSim.
Specifically, the neurons used are $DB\{1,2,3,4,5,6,7\}$, $DD\{1,2,3,4,5,6\}$, $VB\{1,2,3,4,5,6,7,8,9,10,11\}$ and $VD\{1,2,3,4,5,6,7,8,9,10,11,12,13\}$
To convert between these neurons and the $12$ repeating units we linearly interpolate again, based on the location of the neurons as defined in WormAtlas~\cite{wormatlas} and the position of the target repeating unit in WormSim.
Again, this conversion introduces an associated conversion factor, denoted $w_{\text{m}}$.
WormSim then passes each of our ``interpolated'' neurons through a low-pass filter to smooth the excitation in the muscle, somewhat reminiscent of muscle excitation with spiking neurons.
Like $w_{\text{s}}$, we found the simulations were very sensitive to this parameter, both in terms of the stability of WormSim and the quality of the simulation.

If either of these parameters are too low, the worm does not exhibit movement and the neural circuit quickly returns to its quiescent point.
If either of these parameters are too high, the integrator used in WormSim (SundialsODE) fails to integrate the function, often typified by membrane potentials growing exponentially just before this ``crash.''
We leverage this crash to assert that that particle has zero probability, effectively removing it from the SMC sweeps used.
This implies, as well, that the parameter settings being used are not physioloigcally plausible, and hence define the likelihood of the parameter values to be zero if \emph{all} particles used in the inference sweep cannot be integrated at a particular timestep.

An example simulation, using coarsely hand-tuned parameters, is shown in Figure 1(c) of the main text.
The body state at time $t$, is denoted $\mathbf{b}_t \in \mathcal{B}$.
This body state is defined by the x-y coordinate of each of the $49$ control points, as well as the angle of the associated rods centered on the control points, and the first derivative of these quantities, resulting in $49\times 3\times 2$ states.
Also included is the ``muscle voltage'' in each of the contractile units.
There are $48$ contractile units on both dorsal and ventral sides, resulting in an additional $96$ states.
These muscle voltages are dependent on the motor stimulation described above.
Accordingly, the entire body state has dimensionality $\mathcal{B} = \mathbb{R}^{49\times 3 \times 4} \times \mathbb{R}^{48 \times 2} = \mathbb{R}^{390}$.
The body simulator, $p(\mathbf{b}_t | \mathbf{b}_{t-1}, \mathbf{v}_{t-1}, \boldsymbol\theta)$, is dependent on both the previous state of the body $\mathbf{b}_{t-1}$ and the neural state at the previous timestep, $\mathbf{v}_{t-1}$, acting as the driving neural input.
The proprioceptive feedback, denoted $\mathbf{r}_t$, conditioned on the body shape, is returned to the neural simulator as current inputs for the next time step, through the interpolation procedure described above.
Accordingly, we modify the definition of the neural simulator to also be conditioned on the proprioceptive feedback: $p(\mathbf{v}_{t} | \mathbf{v}_{t-1}, \mathbf{r}_{t}, \boldsymbol\theta)$.
The parameters we consider here are the two parameters we introduce as part of the ``meshing,'' and hence $\boldsymbol\theta = \left\lbrace w_{\text{m}}, w_{\text{sr}} \right\rbrace \in \mathbb{R}^2_{\geq 0}$.
Here we have given more details on the specific implementation details we used in meshing SCE and WormSim. 
We refer the reader to the original text by \citet{boyle2012gait}, and \citet{wen2012proprioceptive} for more information on the underlying model.

The WormSim code is implemented in C++, and so we use the interprocess communication package ZeroMQ (ZMQ) to communicate between the Python implementation of SCE and our inference packages.
We modified the WormSim implementation to run in a separate process.
It accepts, via ZMQ communication, an entire body state, and performs the forward iteration for one timestep. 
The iterated body state is then returned back to the calling process via ZMQ.
As such, a single WormSim process can be used to iterate multiple particles sequentially, or, in process many particles in parallel by running multiple instantiations.
We couple this with a parallelized implementation of SCE, such that a single process is passed an entire worm state and alternates between iterating the neural and the body state for the desired number of iterations, returning the iterated state, and the intermediate states as a side effect for later retrieval for super-resolution (in time) reconstruction of traces.

As with the neural dynamics simulators, we also investigated the use of the more accurate body simulation platform ``Sibernetic,'' presented by~\citet{palyanov2015sibernetic}.
Sibernetic is a highly accurate particle physics and solid body simulator representing the physiology of {\ce} on a very fine-grained scale.
However, much like c302, this marked increase in fidelity incurs an enormous computational burden, and hence we do not use Sibernetic, instead using the much computationally cheaper WormSim.

\paragraph{Observation Model}
The fluorescence signal provided by calcium imaging~\cite{kato2015global, nguyen2016whole}, denoted $\mathbf{y}\in\mathbb{R}_+^{M}, M\leq N$, is a stochastic quantity dependent on the intracellular calcium concentration, thus determined by $p(\mathbf{y}_t|\mathbf{v}_{t}, \boldsymbol\theta)$, where $M$ of the $N$ neurons are observed.
Here, we formally define ``observed'' to mean those neurons for which a fluorescence trace could be confidently attributed to a particular neuron by expert annotators.
In the dataset released to us by \citet{kato2015global}, the number of neurons identified varies between datasets, and so we take the dataset with the most observed neurons as our benchmark for observability, such that $M=49$. 
This corresponds to observing the florescence for a fixed and a priori known subset of neurons.
To use the remaining, unlabelled neurons introduces a challenging permutation problem.
This problem is further investigated by \citet{linderman2017reparameterizing, mena2018learning}.
We investigated the permutation inference challenge, but instead chose to focus here on performing inference in the state-space model and performing parameter estimation, and defer inferences over the use of unlabelled fluorescence traces to future work.

We utilize a saturating Hill-type function~\cite{hill1938heat}, a parametrized non-linear function, for this dependence as suggested by \citet{rahmati2016inferring, grienberger2012imaging, yasuda2004imaging} distorted with zero mean Gaussian noise per observed neuron:
\begin{equation}
y_{t,m} \sim p(y_{t,m} | c_{t,m}) = F \times \frac{c_{t,m}}{c_{t,m} + K_d} + D + \mathcal{N}(0, \sigma_m^2),
\end{equation}
where $F$, $K_d$ $D$ and $\sigma_m^2$ are constants that can be independently calibrated or estimated on-line from data~\cite{kato2015global, rahmati2016inferring}, and $c_{t,m}$ is the calcium concentration of the $m^{\text{th}}$ observed neuron.
When we generate synthetic data, we generate noise-free data by setting $\sigma_m^2 = 0$.
This model, importantly, provides us with the reweighting distribution that is required to score different realizations of state, as part of sequential Monte Carlo, and allow for objective comparison of the similarity between the observed data and different simulations.

\paragraph{Central Pattern Generation and Artificial Stimulation}
Finally, we include the ability to directly stimulate the network through sensory inputs~\cite{bretscher2011temperature, izquierdo2013}, denoted $\mathbf{q}\in\mathcal{Q}$, although we do not use this here and defer inclusion of external stimuli to the future work.
Importantly, once mechanisms of sensory stimuli are better understood and quantified~\cite{metaxakis2018multimodal, cohen2018whole,izquierdo2013connecting}, they can, at least in principal, simply be added as additional random variables in the HMM underpinning the simulation, shown in Figure 1(d) of the main text, like the proprioception and body simulation ``loop'' we use.

\begin{figure}[h!]
\centering
\includegraphics[width=\textwidth]{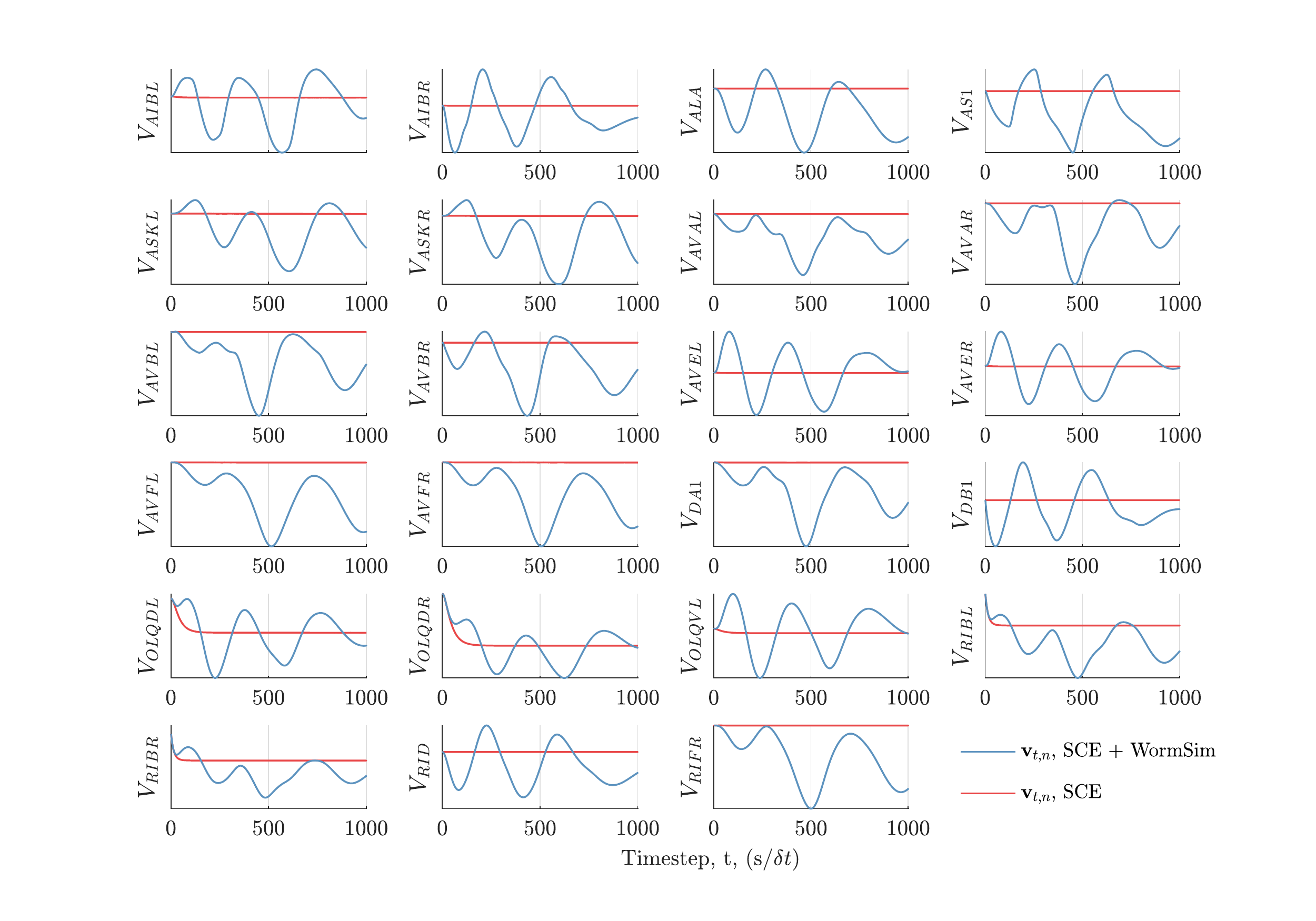}  
\caption{Experiment showing the success of stimulating the whole connectome with motor feedback provided by WormSim, compared to SCE alone. Both without the use of a central pattern generator or artificial stimuli.}
\label{fig:supp:cpg}
\end{figure}

There is discussion in the {{\ce}} community as to the existence of a ``central pattern generator'' (CPG) in the {{\ce}} connectome~\cite{olivares2018potential, fouad2018distributed,gao2017excitatory, boyle2012gait, cohen2014nematode}, particularly for facilitating locomotion.
In other organisms, a CPG is hypothesised to act as a central ``clock pulse'' for coordinating and activating the rest of the connectome.
No CPG has been found in the connectome of the {{\ce}}, and so to produce sustained oscillatory activity, computational {{\ce}}  simulators~\cite{gleeson2018c302, sce} are often driven by physiologically unrealistic and arbitrary stimuli~\cite{kunert2014low, kunert2017spatiotemporal}.
However, \citet{boyle2012gait} demonstrates in the WormSim simulator that the oscillating signals required to drive locomotion in {{\ce}} can be generated by a proprioceptive feedback circuit.
In this model, a stretch-receptive current, which is calculated according to the mechanical action of the muscle cells, is fed back into B-type motor neurons. 
This model is supported by experimental evidence from \citet{wen2012proprioceptive},  in which movement restriction studies demonstrate proprioception in {{\ce}}, and subsequent laser ablation studies identify the B-type motor neurons as the sole candidates for receiving such feedback.
The effect of this proprioceptive feedback mechanism are seen in Figure \ref{fig:supp:cpg}, which compares our simulator (a combination of SCE and WormSim, incorporating proprioception) to the ``vanilla'' SCE package (without proprioception). 
We find that vanilla SCE does not produce oscillatory behaviour without central pattern generation, while our simulation provides sustained simulation for approximately $10$ seconds.
Note that, even with proprioception, the oscillations eventually decay due to leakage currents in each neuron and muscle cell.
The rate of this decay is highly dependent on model parameters, thus the activity can be prolonged by tuning certain parameter values.
Dynamics would also be perpetuated if the model were expanded to encompass sensory stimuli, which could provide the ``boost'' in activity required to continue driving the oscillations.

To get sufficient excitement of the body and stretch-receptive feedback, we found that the membrane potentials of the most-excited neurons (some of the motor neurons) exist in unphysioloigcal ranges.
We believe that this is a deficiency in both the original simulators (SCE and WormSim) and our ``meshing'' of the simulators through employment of poor parameter values.
This motivates the use of the parameter estimation as described in Section 4 of the main text, coupled with refinement of parameter values after further consultation with neuroscientists. 
Although undesirable, we do not believe this flaw fundamentally limits the functionality of our simulator, and rather provides an opportunity for further refinement and demonstration of the proposed methodology on \emph{real data}.

\subsection{Summary}
Before we proceed, we pause to take stock of the components introduced above, and concretely reinforce the relationship between these components and the notation, aims and objectives introduced in the introduction. 
Collectively, the neural simulator, body simulator and any additional stimulation define the state of the worm. 
This state is represented as a $994$ dimensional vector, $604$ of which is attributed to the voltage and calcium concentrations of the $302$ neurons, and $243$ originating from the $x$, $y$ and $\alpha$ orientation of the $49$ rods, as well as the first derivative of these quantities,  and the $96$ contractile units used as the representation of the body in WormSim.
SCE, the calcium dynamics and WormSim define the conditional probabilities that define the evolution of this state, $p(\mathbf{x}_t | \mathbf{x}_{t-1}, \boldsymbol\theta)$.
To condition our model in real data we include a parameterized observation model, denoted $p(\mathbf{y}_t | \mathbf{x}_t , \boldsymbol\theta)$.
Together, these probabilities define time evolution of the simulator as their product:
\begin{equation}
p(\mathbf{x}_t , \mathbf{y}_t | \mathbf{x}_{t-1} , \boldsymbol\theta) = p(\mathbf{y}_t | \mathbf{x}_t , \boldsymbol\theta) p(\mathbf{x}_t | \mathbf{x}_{t-1}, \boldsymbol\theta).
\end{equation}

\subsection{Software Implementation} 
We now present details on the extensive software implementations we develop to execute the required inferences, making use of parallel architectures.
We distribute all our simulation code and data.

\subsubsection{State Space Estimation}
We have already described how we modified the SCE implementation to make use of vectorized NumPy calculations and opted to use forward difference in place of an ODE integrator.
These choices led to orders of magnitude speedups.
We also describe how we implement WormSim in a separate process, with interprocess communication provided by ZMQ.
We further develop the software to allow multiple particles to be iterated in parallel.
To do this, a pool of Python processes is opened, each one running a single WormSim executable. 
Each of these processes is then linked with a unique ZMQ socket.
A second pool of processes is opened, each running a single SCE instance.
Each SCE processes is paired with a single WormSim process, such that the state is iterated back and forth between them until the required number of iterations has been reached.
This design also means that SCE and WormSim, in a single SCE-WormSim process pair are never concurrently executing -- while SCE is iterating, WormSim is waiting for data, and vice versa.
This means it is easy to select the number of processes in each pool as the number of processors available on the machine.
By doing this, we find that, for moderate sized particle pools, we get a utilization of over $90\%$.

\subsubsection{Parameter Estimation}
We further developed our implementation for embarrassingly parallel evaluation of likelihoods as part of the parameter estimation task.
We achieve this in our software in two different ways, primarily separated by the computationally cheaper autoregressive experiments that can be run on desktop machines, or the more expensive {\ce} optimizations, which are more likely to be run on distributed, high performance compute (HPC) clusters.

For autoregressive models we simply leverage Pythons inbuilt multiprocessing library to parallelize function evaluations if NumPy is configured to only use part of the computational resources.

More interestingly however is the implementation for distributing {\ce}, where we interleave calls to the message passing interface MPI, for inter-node communication, and calls to Pythons multiprocessing library for intra-node communication.
The workflow is as follows: There exists a ``controller'' MPI process, and one ``worker'' MPI process per node available (for the experiments we present in the main text, there is one controller, and nine workers, equally distributed one per each of the ten nodes).
The controller communicates strictly with the worker MPI processes, while each worker communicates both with the controller and also opening a pool of node-internal worker processes using Pythons multiprocessing library.
The controller will then transmit the $\mathcal{N}_r$ parameter values for which the corresponding likelihood needs calculating to each of the worker process, where one worker may receive multiple parameters as required.
The likelihood is then calculated \emph{within} the node, where individual particles are iterated in parallel on multiple processes and synchronised when a resample statement is hit.
The worker thread then returns the likelihood via MPI back to the controller.
Once the controller has collected all required likelihoods, the gradient/MH step is taken and the new parameters are distributed to the workers.
No particles pass over MPI keeping overheads low.
Individual workers may also be instructed to write their particles to disk for later retrieval. 
This writing process can be done to storage local to the node for speed, as these intermediate results will be moved to more permanent memory as and when determined by the controller.

We have tested our code on two HPC clusters, Cori, administered by NERSC, and Cedar, administered by ComputeCanada, and have run the code on over $30$ nodes with minimal communication overhead.
A minor drawback is in nodes sitting idle as part of synchronization between workers.
In reality, we find this is minimal once ``good'' parameter values have been found, although we are investigating ``cut-off'' techniques~\cite{teng2018bayesian,chen2016revisiting} for minimizing this inefficiency.
We also distribute ``dummy'' code to demonstrate the functionality of our approach for parallelization, for debugging on new HPC clusters and, ultimately, for other developers to use as they see fit, allows for nearly embarrassingly parallelization of our code, allowing for large HPC clusters to be used.

\section{Experimental Details}
We now present some more details on our software implementations of our inference methods, and then the experimental configurations used.

\subsection{Virtual Patch Clamp Experiments}
We now give more fine-grained details on the configuration for the virtual patch clamp experiment introduced in the main text.
Before we proceed, we note that calcium imaging experiments are often paired with visual light recording devices from which the configuration of the worm can also be recovered.
The shape of the worm can be well-approximated by a cubic spline, through which the position and velocity components of the WormSim representation can fitted to ascertain the shape of the worm.
The velocity can then be estimated by calculating the Euler difference these representations between time steps.
Therefore, we assume that we can initialize the shape and velocity of the worm by fitting of the control points used in the WormSim representation, to the spline fit of the true shape and velocity.
The net result of this approximation is that the initial distribution over body shapes is a Dirac-$\delta$ function centred on, or very close to, the true body.
We clarify that do not subsequently use the body shape in inference, due to the intractability of the likelihood terms, and only use the true body shape data for initialization.

For this experiment, we simulate the neural and body dynamics for a total of $500$ time steps, corresponding to a real simulation time of $5$ seconds, as we take $\delta t = 0.01$.
We also generate calcium imaging florescence observations for the $49$ neurons identified in one of the datasets provided to us by the authors of \citet{kato2015global}.
The identities of the $49$ neurons, in alphabetical order, are as follows:
AIBL, AIBR, ALA, AS1, ASKL, ASKR, AVAL, AVAR, AVBL, AVBR, AVEL, AVER, AVFL, AVFR, DA1, DB1,
OLQDL, OLQDR, OLQVL, RIBL, RIBR, RID, RIFR, RIML, RIMR, RIS, RIVL, RIVR, RMED, RMEL, RMER,
RMEV, SABD, SABVL, SABVR, SIBVL, SMBDL, SMBDR, SMDVL, SMDVR, URADL, URADR, URYDL, URYDR,
URYVL, URYVR, VA1, VB1, VB2.
We also simulate observations every $0.05$ seconds, as opposed to every $0.343$ seconds as in the Kato dataset~\cite{kato2015global}.
The synthetic data we generate is noise-free.
At inference time, we use additive noise kernels in the plant model, where the variance of the term is scaled according to expected variance of that state.
The standard deviation of the noise is $5$mV for the neuron with the largest voltage range, while the smallest noise term used is $0.0005$mV. 
These variances are determined at the start of inference by simulating a corpus of datasets $48$ noise-free datasets from the model and calculating the per-neuron variance of the voltage.
From this corpus, we also estimate the ``learned'' prior distribution, by evaluating the per-neuron mean and variance, and using these values as the mean and variance of the learned prior.
The ``initial prior'' used for generating this corpus of data was $\mathcal{N}(-20\text{mV}, 0.033\text{mV}^2)$.
Similarly, we scale the variance of the reweighting distribution (likelihood) according to the expected variance of the observations generated in the aforementioned corpus.
For this, we add a ``stabilizer,'' with value $0.1$, to ensure that those neurons that do not vary do not get ``overfitted'' to.
Therefore, we scale the noise kernel according to:
\begin{equation}
\sigma^2_m = 0.02 \times \sqrt{\frac{1.1 \times \hat{\sigma}^2_m}{\max_{i\in 1:M} \hat{\sigma}^2_i + 0.1}},
\end{equation}
where $\hat{\sigma}^2_n$ is the variance of the $m^{\text{th}}$ observation in the synthetic corpus.
We found these scalings improved the SMC inference result, preventing those neurons whose dynamic range far outweighs those whose range is smaller from dominating the resampling, while simultaneously crushing the activity in the finer-grained neurons.
The number of particles, $\mathcal{N}_p$, used in the SMC sweep was $1000$, with $5000$ particles used in the initialization calculation.
In total, the wall-clock time was $800$ seconds, when run on a single node equipped with $48$ Intel Xeon Platinum 2.10GHz 8160F CPUs.
We used the ``default'' parameters of our simulator, which are far too numerous to list even here, although, for consistency with the Parameter Estimation sections, we use $w_{m}=9.0$ and $w_{s}=10.3$, two parameters we identified as highly important.
We place a Rayleigh prior over these parameters, with parameter equal to the true parameter value, although we note that the the likelihood term dominates the prior probability.

We also note in the main text that we observe a small offset in the position and orientation of the worm at the start of the trace.
This is because the initialization of the particles is not perfect, and therefore there is a small amount of ``integration error'' as this imperfection is corrected.
This results in a small error in the initial development of position and orientation of the worm.
This error cannot be corrected under the model since the absolute position of the worm is not conditioned on.
Therefore, we post-hoc centre the reconstruction on the true body shape by applying a rigid body transform using the iterative closest point algorithm.

We show below in Figure \ref{fig:wormsimraw} the unaligned reconstructions for completeness.
\begin{figure}
\centering
	\begin{minipage}[b!]{0.32\textwidth}%
	\centering%
	\includegraphics[width=1.\textwidth]{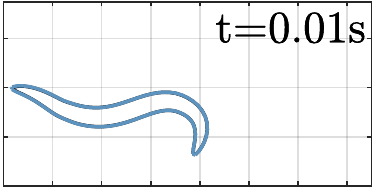}
	\includegraphics[width=1.\textwidth]{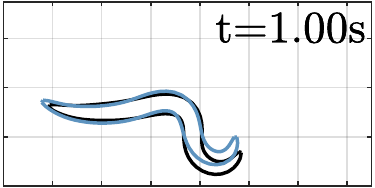}
	\end{minipage}%
	\hfill%
	\begin{minipage}[b!]{0.32\textwidth}%
	\centering%
	\includegraphics[width=1.\textwidth]{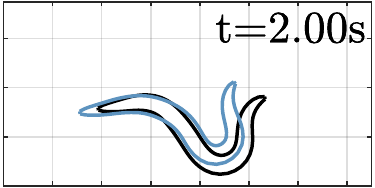}
	\includegraphics[width=1.\textwidth]{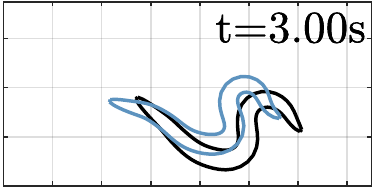}
	\end{minipage}%
	\hfill%
	\begin{minipage}[b!]{0.32\textwidth}%
	\centering%
	\includegraphics[width=1.\textwidth]{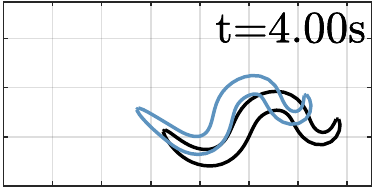}
	\includegraphics[width=1.\textwidth]{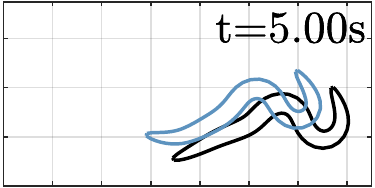}
	\end{minipage}%
	\caption{Unaligned WormSim reconstructions. The shape of the reconstructed worm is correct at later timesteps is correct, but is offset from the true position due to integration error at the start of the trace, and after four seconds of simulation, the reconstruction has ``locked-on'' to the true body shape, offset by this small integration error. }
	\label{fig:wormsimraw}%
	
\end{figure}

\subsection{Autoregressive Parameter Estimation Experiments}
We now give more details on the models and experiments configured for demonstrating the parameter estimation capabilities.

\subsubsection{Generative Model}
To investigate problem domain, we first conduct experiments in a simplified model on synthetic data.
In lieu of the neural simulator, we use a simple autoregressive model (AR), where the model is defined as:
\begin{align}
\mathbf{x}_{t} &\sim p(\mathbf{x}_{t}|\mathbf{x}_{t-1}, \boldsymbol\theta) = \boldsymbol\theta_{\mathbf{W}} \times \mathbf{x}_{t-1} + \mathcal{N}(0, \sigma_p^2\times\mathbb{I}^{N}),\label{equ:ar_plant}\\
\mathbf{y}_{t} &\sim p(\mathbf{y}_t | \mathbf{x}_t, \boldsymbol\theta) = \boldsymbol\theta_F \frac{\mathbf{x}_{t}}{\boldsymbol\theta_K + \mathbf{x}_{t}} + \boldsymbol\theta_d + \mathcal{N}(0, \sigma_m^2\times\mathbb{I}^{N}), \label{equ:ar_obs}
\end{align}
where $\mathbf{x}_t \in \mathbb{R}^N$ is the state vector at time $t$, $\mathbf{y}_t \in \mathbb{R}^M$ is the vector of observations at time $t$, $\boldsymbol\theta_W\in\mathbb{R}_{\geq 0}^{N\times N}$ is a sparse matrix dictating the evolution of the latent state, $\sigma_p^2 \in \mathbb{R}_+$ is a represents the variance of the noise in the plant model, $\sigma_m^2 \in \mathbb{R}_+$ is the variance of the observations and $M$ traces are observed.
We assume that all species are observed in the AR model.
The form of the observation function, $p(\mathbf{y}_t | \mathbf{x}_t, \boldsymbol\theta)$, is chosen to mirror \emph{C. elegans} data as tightly as possible, as proposed by \citet{rahmati2016inferring}.
We assume that the parameters of the additive noise distributions, $\sigma_p^2$ and $\sigma_m^2$, are known.
The observation function uses parameters of $\left\lbrace F,\ K,\ d \right\rbrace\ =\ \left\lbrace 1.0,\ 1.0,\ 10.0 \right\rbrace$.
The process noise and observation noise kernels are then taken to be independent Gaussian  per dimension with diagonal covariance elements of $\sigma_p^2 = \sigma_m^2 = 0.01^2$.
The initial distribution over state is defined as $p(\mathbf{x}_0 | \boldsymbol\theta) = \mathcal{N}(0, 1.0^2 \times \mathbb{I}^N)$.
This is also used at inference time.
For each method, we perform a single importance sampling step at $t=0$ to limit the degeneracy in the first step of the particle filter.
The AR process is then iterated for $200$ timesteps.

We place a prior over each of the parameters, $\boldsymbol\theta$, denoted $p(\boldsymbol\theta)$.
For the prior distribution over parameter values we use a Rayleigh distribution, due to its strictly positive, continuous support, again, mimicking the {\ce} scenario.
The Rayleigh distribution has support $x \in \left[0, \infty \right)$, is parametrized by a single scale parameter, denoted $\sigma^2$, a density function of $\frac{x}{\sigma^2} e^{-x^2 / (2 \sigma^2)}$, and has a form reminiscent of a ``continuous Poisson'' distribution.
We denote draws from the distribution as $x \sim Ray(\sigma^2)$ and the value of the density as $Ray(x; \sigma^2)$.
We generate a sparse, anti-symmetric random transition weight matrix, $W$, by uniformly sampling off-diagonal elements from the strictly upper triangular region and populating these elements samples values from a Rayleigh distribution with known mean, $\boldsymbol\theta_{W_i} \sim Ray(\mu_W),\ i \in \left\lbrace 1, \dots, \mathtt{floor}(\alpha\times N \times (N-1) / 2) \right\rbrace$, where $\alpha$ is the target sparsity. 
The lower triangular is then set to the negation of the upper triangular portion to enforce the resulting matrix to be anti-symmetric.
We denote the full transition matrix as $\boldsymbol\theta_W$, were, fractionally, $\alpha$, entries are fixed at zero for all possible parameter values.
The resulting transform from this matrix to the parameters we have freedom over is simply picking off those values that are not pinned at zero, denote those values we pick off as $\boldsymbol\theta_{W_i}$.
Since the mechanics of the model and the structure of the transition matrix are fixed, we simply refer to the parameters of the transition matrix as the non-zero entries and assume population of the $\boldsymbol\theta_{W}$ matrix is implicit.

Our full generative model is therefore:
\begin{equation}
p(\mathbf{x}_{0:T}, \mathbf{y}_{1:T}, \boldsymbol\theta) \sim p(\boldsymbol\theta) p(\mathbf{x}_0 | \boldsymbol\theta) p(\mathbf{y}_0 | \mathbf{x}_0, \boldsymbol\theta) \Pi_{t=1:T} p(\mathbf{x}_t|\mathbf{x}_{t-1}, \boldsymbol\theta) p(\mathbf{y}_t | \mathbf{x}_t, \boldsymbol\theta).
\end{equation}
We wish to solve for to posterior over the parameter values, $\boldsymbol\theta$, conditioned on the data $\mathbf{y}_{1:T}$, marginalizing over the unobserved states:
\begin{equation}
p(\boldsymbol\theta | \mathbf{y}_{1:T}) = \frac{p(\boldsymbol\theta) \int_{\mathbf{x}_{0:T}\in\mathcal{X}^{T+1}} p(\mathbf{x}_0 | \boldsymbol\theta) \Pi_{t=1:T} p(\mathbf{x}_t|\mathbf{x}_{t-1}, \boldsymbol\theta) p(\mathbf{y}_t | \mathbf{x}_t, \boldsymbol\theta) d\mathbf{x}_{0:T}}{p(\mathbf{y}_{1:T})}.
\end{equation}
Alternatively, optimization entails pointwise maximization of this distribution:
\begin{equation}
\boldsymbol\theta^* = \argmax_{\boldsymbol\theta} p(\boldsymbol\theta | \mathbf{y}_{1:T}).
\end{equation}

A further detail we add is that observations are not available at every timestep. 
For instance, the simulator we build iterates with a timestep of $\delta t = 0.01$ seconds, while the capture interval used in \citet{kato2015global} is $\Delta t = 0.343$ seconds. 
We do not express this in our notation to keep the notation readable, but the effect is that the length $\mathbf{y}_{1:T}$ is actually reduced by a factor of $\Delta t / \delta t$, where intermediate, or ``unobserved'' steps, do not have an associated $\mathbf{y}$. 
For the AR examples contained herein, we set $\delta t=1.0$ and $\Delta t=10.0$, i.e. an observation is received on every tenth step.

\subsubsection{Method Comparison}
Before applying our methods to {\ce}, we compare the methods presented on a small AR process.
For this, we compare ``vanilla'' PMMH, PT and VO on a $30$-dimensional AR process, where the weight matrix has a sparsity of $0.9$, i.e. $N=30$ and $\alpha=0.1$. 
This results in performing parameter inference in a $44$ dimensional parameter space, i.e. $W\in\mathcal{W}=\mathbb{R}_{\geq 0}^{44}=\boldsymbol\theta$.
The elements of $W$ are drawn from a Rayleigh distribution with mean parameter $0.0185$, where this distribution also serves as the prior over parameter values, $p(W)=p(\boldsymbol\theta)= Ray(0.0185)^{44}$.
For this experiment, we assume the observation parameters are known and fixed, since these parameters can be estimated from data and additional experiments~\cite{yasuda2004imaging, kato2015global, rahmati2016inferring}. 
We also present an experiment in the supplementary materials where these parameters are also learned, whereas here we focus on learning the model and comparing methods.

For all methods in this section we use $\mathcal{N}_p=200$.
We now present the details of each of the methods used.

\paragraph{PMMH} As PMMH operates at only a single location, such that $\mathcal{N}_r=1$, we can afford to average over more SMC sweeps and so use $\mathcal{N}_s=5$.
We then take $\mathcal{N}_T=8000$ steps.
This yields a budget $\mathcal{B}=8\times 10^6$.
The Rayleigh distribution over parameters has a mean parameter of $0.0185$.
The proposal distribution we use is a multivariate normal distribution centred on the current parameter value, with covariance $0.00185^2 \mathbb{I}^N$.

\paragraph{PT} We then enhance this approach through the use of parallel chains and likelihood tempering.
For this experiment we use $\mathcal{N}_r = 5$ and $\mathcal{T}_{1:\mathcal{N}_r} = \left[4.0, 2.0, 1.33, 1.14, 1.0 \right]$.
Due to the increase in evaluations at each steps, we reduce $\mathcal{N}_s$ to $2$, and reduce the number of steps $\mathcal{N}_T$ to $4,000$, resulting in an expenditure $\mathcal{B}=8 \times 10^6$.
This inverse of this temperature corresponds to the weighting we apply to the likelihood in the evaluation of the tempered joint density.
We also utilize a tempered (or temperature-dependent) proposal distribution, where higher temperature chain have a wider proposal.
The hottest chain, $\mathcal{T}_1$ proposes from the prior each time to induce the most mixing.
The standard deviation of the proposal is then as follows: $\mathcal{T}_{2:\mathcal{N}_r} = \left[ 0.00180, 0.00156, 0.00147, 0.00138 \right]$.
Swaps are proposed between adjacent chains at every step, where the adjacent pair is drawn from a uniform distribution.
For clarity however, the log-joint density plots shown in Figure 2(c) of the main text are evaluated under the \emph{true density}, i.e., the lowest temperature, for fair comparison. 

\paragraph{PMVO} We finally compare to our PMVO method.
We use $\mathcal{N}_r = 20$ and $\mathcal{N}_s = 2$, and, to match the budget, limit ourselves to use $\mathcal{N}_T = 1,000$, such that our expenditure is $\mathcal{B}=8 \times 10^6$.
We use an isotropic Gaussian proposal variational distribution to draw the samples that are used to estimate the gradient, centred on the current parameter values and with a covariance of $0.000923^2 \times \mathbb{I}^N$.
The width of this proposal is reduced by a factor of $5$ during the optimization.
We utilize the ADAM optimizer~\cite{kingma2014adam} with an initial learning rate of $10^{-3}$, which is logarithmically annealed to $10^{-4}$ during the optimization.
Similarly, we begin estimating the gradient on a tempered joint density, with initial temperature, $1/ \beta = 100$, which is logarithmically annealed to the true density, $1 / \beta = 1$, during the optimization.
We also temper the SMC sweep by relaxing the observation kernels used in the SMC sweep by expanding them by a factor of $2.5$ initially, again, logarithmically annealing this towards the ``true'' kernels specified under the model.
For clarity however, the log-joint density plots shown in Figure 2(c) of the main text are evaluated under the \emph{true density} for fair comparison. 
We found that these relaxations make the gradient-based method more stable.

\paragraph{Comparison}
The results of this experiment are shown in Figure 2 of the main text. 
The true state, $\mathbf{x}_{0:T}$ is shown as a black dashed line in Figure 2(a), where we only image $5$ of the $30$ species for clarity. 
The first thing to note is that we are able to generate roughly stable traces, where different traces exhibit notably different behaviours. 
The blue lines correspond to the filtering distribution returned by the SMC sweep when using the true model and true parameters. 
This reconstruction corresponds to the best we can feasibly hope to reconstruct the true state.
We also show the reconstruction, in red, using parameters drawn from the prior.
As one would expect, the reconstruction using parameters drawn from the prior, remembering that the prior corresponds to our prior belief about the parameter values before seeing any data. 
It the congruence of the blue reconstruction and the black true trace, in contrast to the red reconstruction, suggest that ``finding'' the true parameters strongly permits better reconstructions, and that our approach is sensible.

In Figure 2(b) we show the evolution of the MAP parameter values during the execution. 
The reason for the performance gap between the methods is visible here. 
The PMMH algorithm is taking steps in a $44$ dimensional state using an isotropic normal proposal.
This means that very few steps are going to be accepted.
One could design a more sophisticated transition kernel, but at the expense of greatly increased design effort, and difficulty in designing kernels for more complex systems.
The use of tempering partly alleviates this.
Higher temperature chains can much more easily take larger steps due to the attenuation of the data dependency.
These chains then ``pass down'' their larger steps to cooler chains that then refine the steps more readily. 
This behaviour is observed in the propensity of the system to swap parameters between replicates, and the higher acceptance rate within replicates, particularly at intermediate temperatures.
As is seen in Figure 2(c) parallel tempering reaches higher probability regions of parameter space, but does not approach the true joint density.
Relaxing inference to optimization and using our approximate gradient approach leads to much better performance. 
The parameter values also converge much more strongly to the true parameters.

Figure 2(c) shows the evolution of the MAP joint density estimated at that point during the optimization or inference.
On the $x$ axis we plot against function evaluations, i.e. $\mathcal{N}_s \times \mathcal{N}_r \times \mathcal{N}_t$, where $\mathcal{N}_t \leq \mathcal{N}_T$, represent the computational effort expended.
We normalize the $y$ axis by plotting the discrepancy between the joint density of the MAP parameters and the joint density of the true parameters.
Intuitively, this means that, under the model, any parameters that achieve a value of zero are indistinguishable from, or equally apt as, the true parameters.
To put it even more straightforwardly, we wish to find parameter values which have a joint density at least as high as the joint obtained by the true parameter values, corresponding to a value of zero on the $y$ axis.
One can see straight away that the variational optimization approach outperforms the parallel chains method, which in turn outperforms the vanilla PMMH approach.

Importantly as well, designing the PMVO method was more straightforward, not having to design the temperature ladder and designing proposals for each temperature.
Therefore, we conclude that performing optimization is a sensible objective and herein we discuss only the PMVO method.
However, the one end-goal is still to obtain the posterior density over the parameter values, as this density is of considerable interest to practitioners as it convey more information than a point estimate of the optimal parameter values, and so we are still actively developing these inference methods.


\subsection{Synthetic {\cef} Parameter Estimation}
We now investigate the use of the full simulation pipeline in conjunction with learning parameters. 
We can do this synthetically, much like we did in the AR example, by generating data with known parameters and then attempting to re-learn those parameters, simply replacing \eqref{equ:ar_plant} and \eqref{equ:ar_obs} with the simulation and observation models defined in Section \ref{sec:prior:neural}.
To be more concrete, this task involves performing SMC in a model with latent dimensionality of $N=994$, where, as described earlier, this corresponds to voltage and calcium concentration of the $302$ neurons, and the body state representation used by WormSim.
For this, we use predominantly the same configuration as for the VPC experiment introduced in Section 3 of the main text.
However, for computational speed, we use fewer particles, only using $500$ particles in the initialization, and $90$ particles in the main sweep.
We note that running more particles improves both the quality of the reconstructions, and the parameter learning.
We use the ADAM optimizer~\cite{kingma2014adam} with a learning rate of $0.01$.

We chose to optimize here the two parameters that we introduce as part of meshing SCE and WormSim.
We also chose these parameters predicated on the observation that they are \emph{highly} influential on the resulting traces.
However, our method, and implementation, scales for optimizing other parameters as well. 
For this experiment we place a Rayleigh prior over each parameter, with mean parameter equal to the true parameters, $w_{\text{m}}=9.0$ and $w_{\text{s}}=10.3$.
We note, however, that the likelihood term dominates the prior term, and so the prior, in essence, only guides the initialization.

To initialize the gradient optimization we sample as many parameter values from the prior as we have worker nodes, in this case $9$, and evaluate the joint density at each of these points.
We then select the best scoring parameter values as the initial value of the gradient optimization.
We observe that many of the initially sampled parameter values, as described above, immediately ``crash'' the simulator due to the integrator failing.
These parameter values are immediately rejected as having a probability of zero.
Often, sampled parameter values are too low, leading to insufficient activation and the worm ``dying'' almost immediately.

\begin{figure}[h!]
\centering
\includegraphics[width=\textwidth]{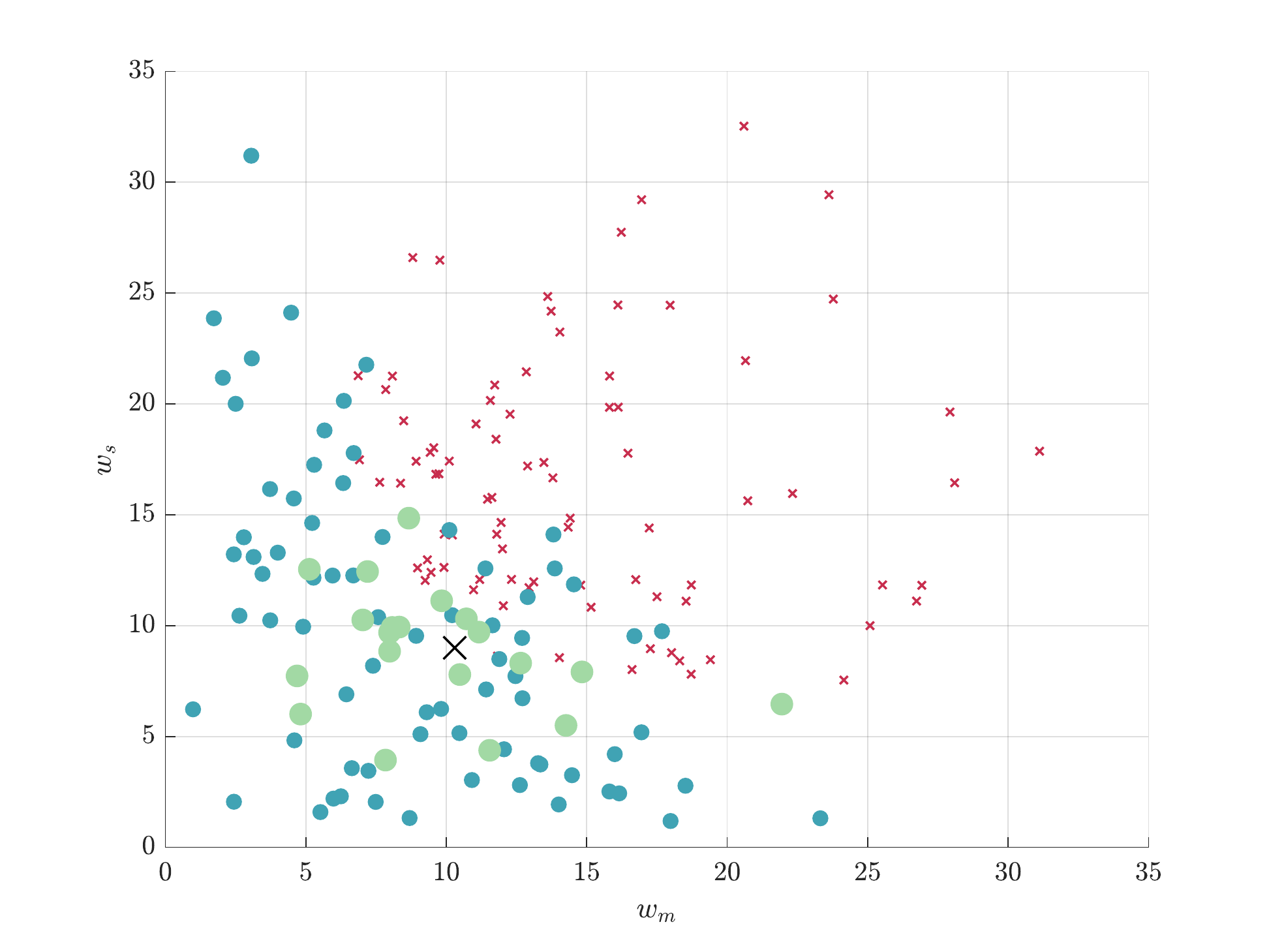}  
\caption{Experiment showing the results of the initialization. Failed initializations are shown as red crosses, initial values that ``passed'' but were not selected are shown as blue dots, and the selected initializations are shown as green dots.
The true parameter value is shown as a black cross.}
\label{fig:sce:init}
\end{figure}

We show a visualization of this procedure in Figure \ref{fig:sce:init}.
Shown as red crosses are parameter values that led to the simulator crashing immediately.
In blue are parameter values that did not crash, but were not optimal among the initial set and so were not executed.
In green are the parameter values that were selected by each of the $20$ experimental repeats.
We see that this initial screening, or random search, dramatically restricts the parameter space to be explored.
In high dimensions, this random search becomes more difficult, but arguably more important as spending a comparatively small amount of effort at the start of the optimization may save considerable computational effort by simply starting from a better place.
To be more concrete, we perform our random initialization procedure, more often than not, in the same time as a single gradient step, representing one five-hundredth of the overall computational cost.

\begin{figure}[t]					

\setcounter{figure}{3}    																								  %

\begin{minipage}[b!]{0.48\textwidth}
	\subfloat[][]{
		\centering
		\includegraphics[width=\textwidth]{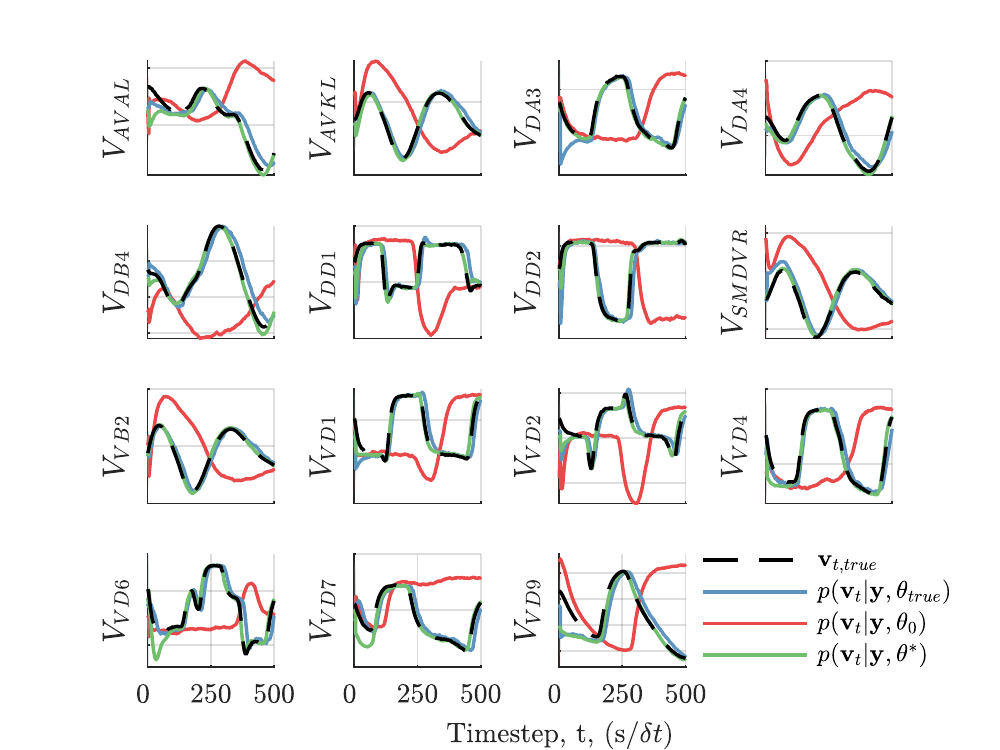}  
		\label{fig:sup:302:reconstruction}
	}
\end{minipage}
\hfill%
\begin{minipage}[b!]{0.48\textwidth}
	\subfloat[][]{
		\centering
		\includegraphics[width=\textwidth]{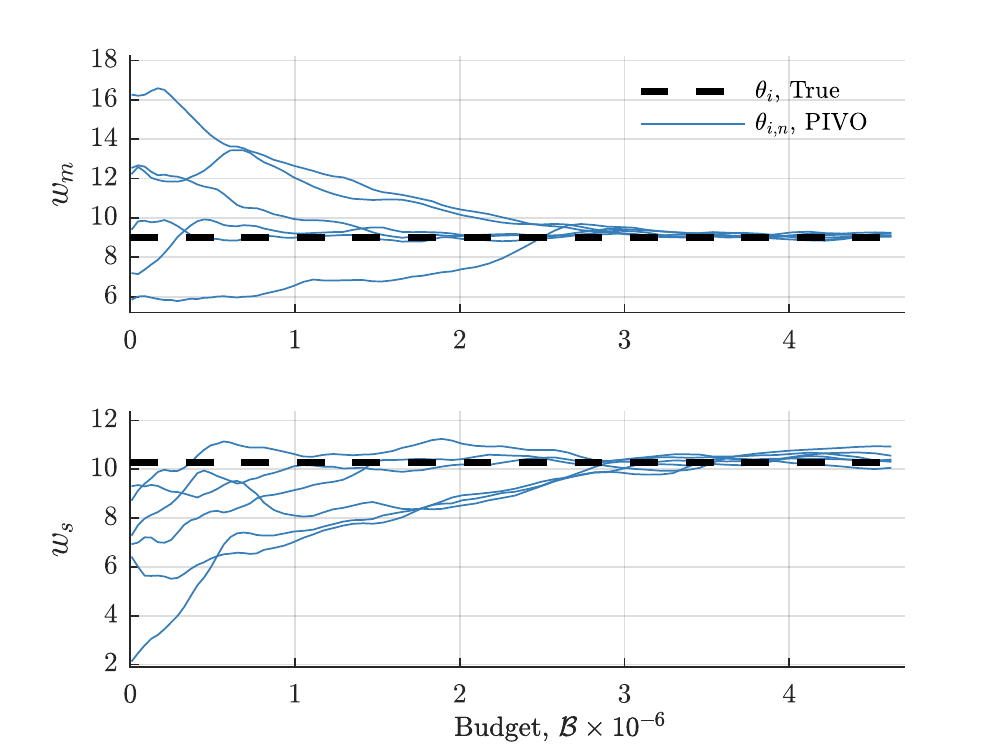}  
		\label{fig:sup:302:parameters}
	}
\end{minipage}

\subfloat[][]{
	\begin{minipage}[b!]{\textwidth}
		\centering
		\begin{minipage}[b!]{0.16\textwidth}
		\includegraphics[width=\textwidth]{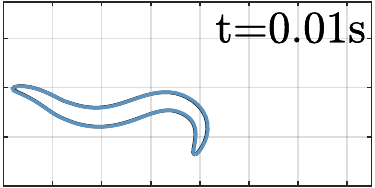} 
		\end{minipage}
		\begin{minipage}[b!]{0.16\textwidth}
		\includegraphics[width=\textwidth]{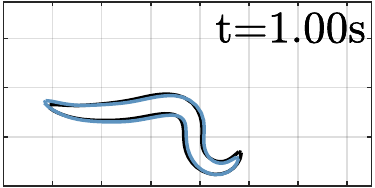} 
		\end{minipage}
		\begin{minipage}[b!]{0.16\textwidth}
		\includegraphics[width=\textwidth]{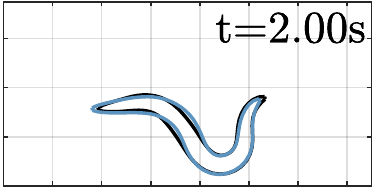} 
		\end{minipage}
		\begin{minipage}[b!]{0.16\textwidth}
		\includegraphics[width=\textwidth]{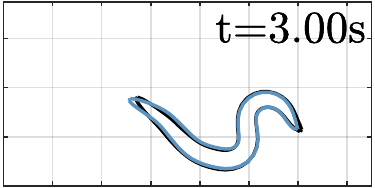} 
		\end{minipage}
		\begin{minipage}[b!]{0.16\textwidth}
		\includegraphics[width=\textwidth]{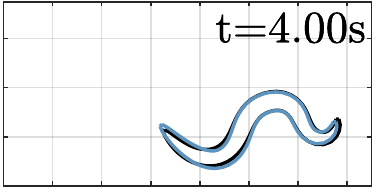} 
		\end{minipage}
		\begin{minipage}[b!]{0.16\textwidth}
		\includegraphics[width=\textwidth]{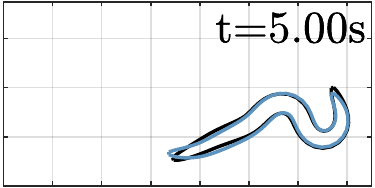} 
		\end{minipage}
		
		\begin{minipage}[b!]{0.16\textwidth}
		\includegraphics[width=\textwidth]{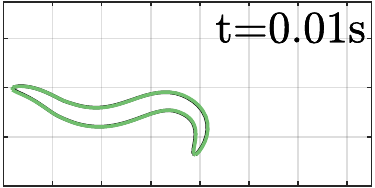} 
		\end{minipage}
		\begin{minipage}[b!]{0.16\textwidth}
		\includegraphics[width=\textwidth]{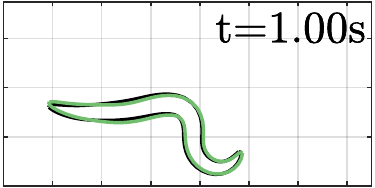} 
		\end{minipage}
		\begin{minipage}[b!]{0.16\textwidth}
		\includegraphics[width=\textwidth]{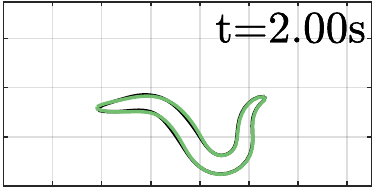} 
		\end{minipage}
		\begin{minipage}[b!]{0.16\textwidth}
		\includegraphics[width=\textwidth]{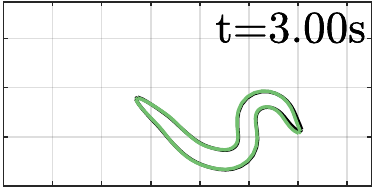} 
		\end{minipage}
		\begin{minipage}[b!]{0.16\textwidth}
		\includegraphics[width=\textwidth]{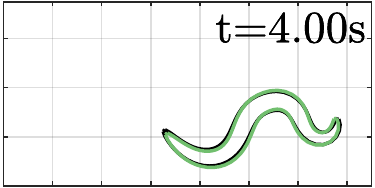} 
		\end{minipage}
		\begin{minipage}[b!]{0.16\textwidth}
		\includegraphics[width=\textwidth]{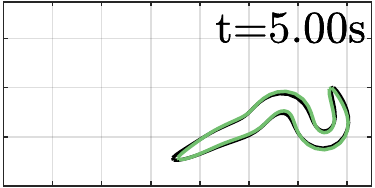} 
		\end{minipage}
		
		\begin{minipage}[b!]{0.16\textwidth}
		\includegraphics[width=\textwidth]{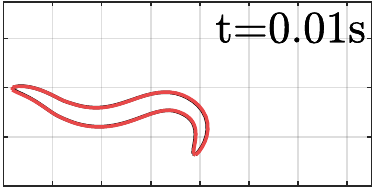} 
		\end{minipage}
		\begin{minipage}[b!]{0.16\textwidth}
		\includegraphics[width=\textwidth]{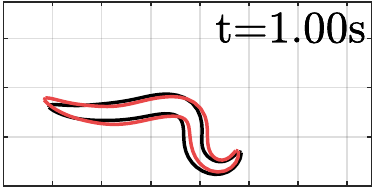} 
		\end{minipage}
		\begin{minipage}[b!]{0.16\textwidth}
		\includegraphics[width=\textwidth]{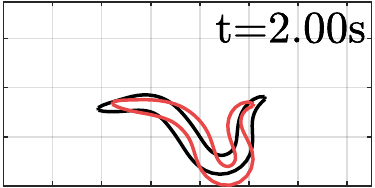} 
		\end{minipage}
		\begin{minipage}[b!]{0.16\textwidth}
		\includegraphics[width=\textwidth]{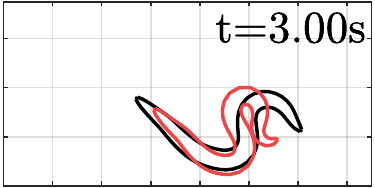} 
		\end{minipage}
		\begin{minipage}[b!]{0.16\textwidth}
		\includegraphics[width=\textwidth]{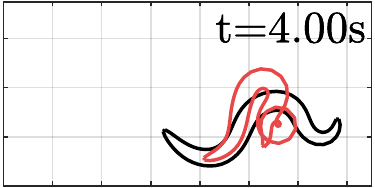} 
		\end{minipage}
		\begin{minipage}[b!]{0.16\textwidth}
		\includegraphics[width=\textwidth]{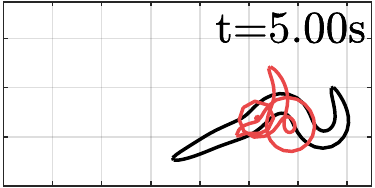} 
		\end{minipage}
		
		\begin{minipage}[b!]{0.16\textwidth}
		\includegraphics[width=\textwidth]{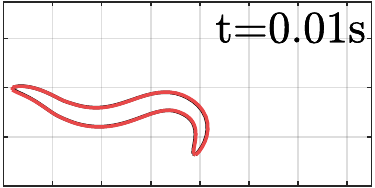} 
		\end{minipage}
		\begin{minipage}[b!]{0.16\textwidth}
		\includegraphics[width=\textwidth]{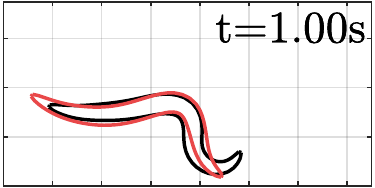} 
		\end{minipage}
		\begin{minipage}[b!]{0.16\textwidth}
		\includegraphics[width=\textwidth]{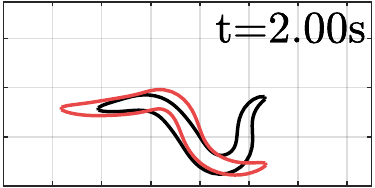} 
		\end{minipage}
		\begin{minipage}[b!]{0.16\textwidth}
		\includegraphics[width=\textwidth]{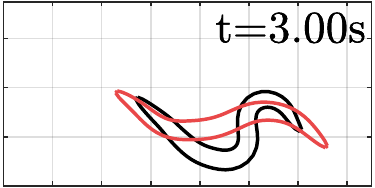} 
		\end{minipage}
		\begin{minipage}[b!]{0.16\textwidth}
		\includegraphics[width=\textwidth]{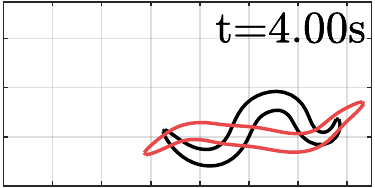} 
		\end{minipage}
		\begin{minipage}[b!]{0.16\textwidth}
		\includegraphics[width=\textwidth]{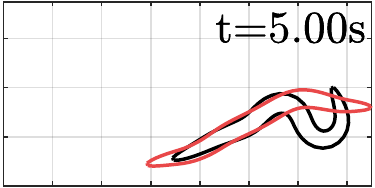} 
		\end{minipage}
	\end{minipage}
	\label{fig:sup:302:wormsim}
}

\setcounter{figure}{2}    					

\caption{
Estimating {\ce} simulator parameters when observing all $302$ neurons.
\protect\subref{fig:sup:302:reconstruction} and \protect\subref{fig:sup:302:wormsim} show the filtering distributions of SMC reconstructions of the membrane potentials of $15$ cells given the true generative parameter (blue), two different optimization algorithm initial parameters (red), and optimized parameters (green). 
\protect\subref{fig:sup:302:parameters} shows the distribution of parameter convergence paths over the course of PMVO optimization, plotted as the mean and variance for $6$ random restarts. 
}
\label{fig:sup:302}
\end{figure}

\subsection{Fully Observed Virtual Patch Clamp Experiment}
For interest, we also include a virtual patch clamp experiment where we observe \emph{all} of the neuron's florescence signal.
Of course, this scenario is infeasible, however, observing and identifying more than $49$ neurons is not infeasible as calcium imaging techniques develop.
One of the difficulties in using calcium imaging data is being unable to associate ``flashes'' with the originating neuron, precluding the use of many of the flashes that are tightly spatially arrange in the head of the worm.
However, neurons, particularly motor and sensory neurons, are spread throughout the length of the worm with a lower spatial density.
Therefore, if the field of view of the calcium imaging could be expanded these neurons could also be unambiguously identified, it would be feasible to condition the simulation on many more neurons.
To confirm the benefit of this supposition, we perform SMC and parameter optimization conditioning on all $302$ neurons.
The results of this are shown in Figure \ref{fig:sup:302}.

We see, unsurprisingly maybe, that the SMC sweep performs very well, practically perfectly reconstructing the latent states and the body pose when all neurons are observed.
Furthermore, the parameters converge quickly and tightly to the true values, suggesting that our PMVO approach is working well, and that the spread over optimized parameter values when only observing $49$ neurons is due to intrinsic uncertainty given the short data traces and the number of unobserved states.
The experimental settings for this are the same as the experiment presented in the main text, apart from that we only use $400$ particles to calculate the initialization and $48$ particles in the SMC sweep itself, as well as reducing the size of the noise kernels used used at inference time in the SMC sweep by a factor of $20$. 
We also re-evaluated the expected model noise and and observation noise by simulating data conditioned on the current model, although we do not believe this greatly impacted the experimental results.
We further modified the prior to have mean parameters $10.8$ and $12.36$ for $w_m$ and $w_sr$ respectively.

\section{Sequential Monte Carlo}
In Section 2 of the main text, we defined the simulation platform that defines the density over possible simulations as $p(\mathbf{x}_{0:T}, \mathbf{y}_{1:T}|\boldsymbol\theta)$.
Here we explicitly include the model parameters for notational consistency, denoted as $\boldsymbol\theta$.
We now give extended details on the sequential Monte Carlo (SMC) technique that underpins our methodological approach.
SMC fulfils both requirements of latent state inference in time series models and approximation of the model evidence, two facets we explore now.
Absolutely exhaustive exposition of the intricacies of SMC, also referred to as sequential importance resampling, is outside the scope even of this supplement and so we refer the reader to \citet{Doucet11atutorial} for more information.

\subsection{Latent State Estimation}
We now give a more detailed introduction into state space estimation using sequential Monte Carlo (SMC) in graphical models, particularly hidden Markov models (HMMs).
SMC provides an incredibly general framework for sampling from \emph{any} distribution, often denoted $\pi(\mathbf{x}|\mathbf{y}, \boldsymbol\theta)$, where $\mathbf{x}\in\mathcal{X}$ are all the latent states we wish to estimate and $\mathbf{y}\in\mathcal{Y}$ is the data we are conditioning on. 
We also include $\boldsymbol\theta \in \boldsymbol\Theta$ defining global parameters which any of the intermediate densities may also be conditioned on.
Of course we may not be conditioning on data and may not have global parameters, in which case the $\mathbf{y}$ and $\boldsymbol\theta$ dependencies may be dropped.

On a high level, SMC operates by defining a series of intermediate target distributions, often denoted as $\pi_t(\mathbf{x}_t|\mathbf{x}_{0:t-1}, \mathbf{y}, \boldsymbol\theta), t\in \left[0, \dots\, T\right],\ T \in \mathcal{T} = \mathbb{Z}_{\geq 0},\ \mathbf{x}_{0:t-1} \in \mathcal{X}_{0:t-1}  \subset \mathcal{X}$, where each intermediate distribution is defined over subsets of the latent state and observed data.
These intermediate distributions are defined such that they are a factorization of the target distribution:
\begin{equation}
\pi(\mathbf{x}|\mathbf{y}, \boldsymbol\theta) = \Pi_{t\in \left[0, \dots, T\right]} \pi(\mathbf{x}_{t}|\mathbf{x}_{0:t-1}\mathbf{y}, \boldsymbol\theta).
\end{equation}
Many different possible factorizations of this expression exist for any given problem, and therefore one of the the design challenges in the deployment of SMC is establishing the most beneficial factorization, in terms of the fidelity of the result traded off against the effort required to determine the form of the required distributions.
In our scenario, we wish to estimate the distribution $p(\mathbf{x}_{0:T} | \mathbf{y}_{1:T}, \boldsymbol\theta)$, where $\mathbf{x}_t \in \mathbb{R}^{994}$ represents the total state of the worm at each time step.

One of the most common deployments of SMC is in state-space estimation in time series models.
Time series models implicitly assume that the value of the state at the current time, $\mathbf{x}_t$ are not conditioned on future states, $\mathbf{x}_{t'>t}$
Another common simplification to make is that the state at the current time step is only conditioned on the state at the previous time step, i.e. the state has no ``memory'' of previous states, that is not directly encoded in the state at the current time.
Similarly, the observation at each time step is only conditioned on the latent state at that time step.
This configuration is referred to as a hidden Markov model (HMM), and is a commonly studied as it describes many physical processes.
To be precise, inclusion of global parameters, $\boldsymbol\theta$, actually defines a parametric HMM, however this distinction does not affect the methods we go onto describe as we will always explicitly condition on a single parameter value, and so we refer to our problem as a standard HMM.

A HMM is described by an initial state distribution, $p(\mathbf{x}_0 | \boldsymbol\theta)$, a transition kernel a transition kernel, $p(\mathbf{x}_{t} | \mathbf{x}_{t-1}, \boldsymbol\theta),\ t \in \left[ 1, \dots, T\right]$ and an observation model, $p(\mathbf{y}_{t} | \mathbf{x}_{t}, \boldsymbol\theta),\ t \in \left[ 1, \dots, T\right]$, where again, each of these distributions can be conditioned on some global parameters $\boldsymbol\theta$.
The initial state distribution describes our prior belief about the state before observing any data.
The transition kernel describes our belief about the evolution of the state with time.
The observation function describes our belief about how the observations are dependent on the latent state.
We assume, as is common, that these kernels are constant and hence we are operating in a homogeneous, parameteric, finite time horizon, discrete time hidden Markov model.
We note that the transition kernel and observation function can be functions of time, leading to an ``inhomogeneous (hidden) Markov model,'' although this is less common and we do not discuss here.

To relate the definitions of the HMM to SMC is straightforward, through the use of the joint density and Bayes' rule:
\begin{align}
\pi(\mathbf{x} | \mathbf{y}, \boldsymbol\theta) &\propto p(\mathbf{x}, \mathbf{y}, \boldsymbol\theta), \\
&= p(\mathbf{y} | \mathbf{x}, \boldsymbol\theta) p(\mathbf{x}, \boldsymbol\theta), \\
&= p(\mathbf{x}_0 | \boldsymbol\theta) \Pi_{t \in \left[ 1, \dots, T \right]}  p(\mathbf{x}_{t} | \mathbf{x}_{0:t-1}, \boldsymbol\theta) p(\mathbf{y}_{t} | \mathbf{x}_{0:t}, \boldsymbol\theta), \label{equ:hmm:2}\\
&= p(\mathbf{x}_0 | \boldsymbol\theta) \Pi_{t \in \left[ 1, \dots, T \right]}  p(\mathbf{x}_{t} | \mathbf{x}_{t-1}, \boldsymbol\theta) p(\mathbf{y}_{t} | \mathbf{x}_{t}, \boldsymbol\theta), \label{equ:hmm:3},
\end{align}
where we have exploited the aforementioned structure in the HMM, via conditional independences, to go from \eqref{equ:hmm:2} to \eqref{equ:hmm:3}.
The intermediate distributions can then be explicitly represented as:
\begin{align}
\pi_0(\mathbf{x}_t | \mathbf{y}, \boldsymbol\theta) &= p(\mathbf{x}_0 | \boldsymbol\theta), \\
\pi_t(\mathbf{x}_t | \mathbf{x}_{0:t-1}, \mathbf{y}, \boldsymbol\theta) &= p(\mathbf{x}_{t} | \mathbf{x}_{t-1}, \boldsymbol\theta) p(\mathbf{y}_{t} | \mathbf{x}_{t}, \boldsymbol\theta)\ \forall\ t \in\left[ 1, \dots, T\right].
\end{align}
Each of these terms we have already described as the initial state distribution, state transition kernel and observation model.
The HMM defines the intermediate distributions to use as the distribution over the latent state at each time conditioned on the observed data up to that time, i.e. $\pi_t(. | .) = p(\mathbf{x}_0 | \boldsymbol\theta) \Pi_{t' \in \left[ 1, \dots, t \right]}  p(\mathbf{x}_{t'} | \mathbf{x}_{t'-1}, \boldsymbol\theta) p(\mathbf{y}_{t'} | \mathbf{x}_{t'}, \boldsymbol\theta)$.
Accordingly we can perform SMC in HMMs through the use of these equations by sequentially estimating each of these distributions, such that their product is the distribution of interest.
This process can be seen as at each intermediate distribution, $\pi_t(. | .)$, propagating that distribution into the next time point and adjusting in light of the new data to available at that time point to yield the next intermediate distribution, $\pi_{t+1}(. | .)$.
This process is referred to as reweighting.

However, in all but the most straightforward of models, analytic calculation of these densities is intractable. 
Therefore, we often resort to numerical approximations of these densities. 
This is referred to as a particle filter.
Particle filters, on a high level, operate by propagating samples from each intermediate distribution through the state transition kernel, retaining those particles that correlate well with the new data, and removing those particles that do not.
Much of the field of state-space estimation and particle filtering is devising more sophisticated and reliable methods for performing this approximation.

To be more concrete, we maintain a set of $\mathcal{N}_p$ particles, denoted $X = \left\lbrace \mathbf{x}^{(n)}_t \right\rbrace_{n\in \left[1, \dots, \mathcal{N}_p \right]}\in \mathcal{X}^\mathcal{N}_p,\ \mathcal{N}_p\in\mathbb{Z}_{\geq 1}$, where we index individual particles by a bracketed superscript at each time step.
The initial particles, $X_0 = \left\lbrace \mathbf{x}^{(n)}_0 \right\rbrace_{n\in \left[1, \dots, \mathcal{N}_p\right]}$, are drawn from the prior $p(\mathbf{x}_0 | \boldsymbol\theta)$, such that, as $\mathcal{N}_p \rightarrow \infty$, the set of particles perfectly approximates $p(\mathbf{x}_0 | \boldsymbol\theta)$.
Each of these particles are then propagated through the state transition kernel, often referred to as a plant model, to yield samples from the distribution $\hat{X}_1^{(n)} \sim p(\mathbf{x}_{1} | \boldsymbol\theta)$, denoted as $\hat{X}_1 \left\lbrace \mathbf{\hat{x}}^{(n)}_1 \right\rbrace_{n\in \left[1, \dots, \mathcal{N}_p\right]}$.
Each of these particles can then be ``scored'' under the observation model by calculating the density $w_1^{(n)} = p(\mathbf{y}_1 ; \mathbf{x}^{(n)}_{1}, \boldsymbol\theta)$.
These weights, referred to as importance weights, are then ``self normalized'' as:
\begin{equation}
W_t^{(n)} = \frac{w_t^{(n)}}{\Sigma_{i=1:\mathcal{N}_p} w_t^{(i)}},
\end{equation}
to yield a multinomial distribution reflecting the relative probability of each particle.
The reweighting is then performed via ``resampling'', where a ``new'' set of particles is created by drawing particles from $\left\lbrace \mathbf{\hat{x}}^{(n)}_1 \right\rbrace_{n\in \left[1, \dots, \mathcal{N}_p\right]}$ proportional to their score under the density induced by the observation.
This resampling may ``kill'' some particles by not resampling them, instead multiply resampling those particles with higher weight.
This process transforms the particles from being distributed according to $p(\mathbf{x}_{1} | \boldsymbol\theta)$ to being distributed according to $p(\mathbf{x}_{1} | \mathbf{y}_1, \boldsymbol\theta)$.
The particle filter then proceeds by exploting the recursive nature of this approach, propagating these resampled particles through the plant model, weighting and resampling until the end of the series of distributions. 
These steps yield a series of distributions (represented via particles) referred to as the ``filtering distributions'', $\left\lbrace p(\mathbf{x}_{t} | \mathbf{y}_{1:t}, \boldsymbol\theta) \right\rbrace_{t \in \left[1, \dots, T\right]}$, i.e. the distribution over the latent state conditioned on all the data that has been seen.

However, this is not yet equal to the target distribution, $\pi(\mathbf{x}_{0:T} | \mathbf{y}_{1:T}, \boldsymbol\theta)$, 
A backwards pass must performed, by where the distributions at previous time steps are only reweighted in light of subsequently seen observations.
This process is implemented in a particle filter by constructing the ancestry of each particle, i.e. for each particle at the current time step, which particle at some time point previously is it descended from.
Those particles without descendants at the final time step are then pruned from the set of particles representing the current distribution.
This can be seen as removing the state estimates that do not explain \emph{future} observations.
Once this backwards pass has been performed, the resulting particle sets at each time step, $\left\lbrace \mathbf{\bar{x}}^{(n)}_{t} \right\rbrace_{n\in \left[1, \dots, \mathcal{N}_p\right]}$, is distributed according to the marginal distribution $p(\mathbf{x}_t | \mathbf{y}_{1:T}, \boldsymbol\theta)$.
Therefore, the product of all of these particle sets corresponds to the product of these marginals which is equal to the target distribution, $\pi(\mathbf{x}_{0:T} | \mathbf{y}_{1:T}, \boldsymbol\theta)$.

\subsection{Evidence Approximation}
Through the mechanics of SMC described above we have developed the necessary terms to approximate the posterior distribution over latent states conditioned on the observed data and a particular set of global parameter values.
The second requirement we identified is the need to estimate the evidence, or the likelihood of the model, $p(\mathbf{y} | \mathcal{M})$, with the latent states marginalized out.
Here, the model represents our belief about the mechanics of the world.
While this framework permits comparing entirely different models, where each model has an associated probability $p(\mathcal{M})$, it is more common, and is what we do here, to fix the model, and optimize the parameters of the model, where we define a prior density over these parameters $p(\boldsymbol\theta)$.
Calculation of $p(\mathbf{y} | \boldsymbol\theta)$, multiplied by the prior probability of the parameter values, yields the joint density of parameter values and observed data under the selected model class.
This joint density is proportional to the posterior density over parameter values.
This observation directly motivates the evidence term, or the likelihood of the parameter values, as a metric of how well our model describes the observed data, where maximization of this metric maximizes the model's capacity to represent the observed data, regularized by our prior beliefs about the parameters themselves.

Somewhat incredibly, particle filters provide us with an estimate of this quantity with little extra effort. 
To begin, we note that the evidence can be factorized as:
\begin{equation}
p(\mathbf{y}|\boldsymbol\theta) = p(\mathbf{y}_1 |\boldsymbol\theta) \Pi_{t \in \left[2, \dots, T\right]} p(\mathbf{y}_t| \mathbf{y}_{1:t-1}, \boldsymbol\theta),\label{equ:evidence:1}
\end{equation}
and as such we can calculate this term at each time step and multiply to get the overall evidence.
Application of the sum and product rule to each term inside the product yields:
\begin{equation}
p(\mathbf{y}_t|\mathbf{y}_{1:t-1}, \boldsymbol\theta) = \int_{\mathbf{x}_t\in\mathcal{X}} p(\mathbf{y}_t | \mathbf{y}_{1:t-1}, \mathbf{x}_t, \boldsymbol\theta) p(\mathbf{x}_t | \mathbf{y}_{1:t-1}, \boldsymbol\theta) \text{d}\mathbf{x}_t.\label{equ:evidence:2}
\end{equation}
The choice of marginalizing over $\mathbf{x}_t$ may seem arbitrary, but \eqref{equ:evidence:2} is true by construction, and our motivation will soon become apparent.
We now exploit the conditional independences induced by the structure of a HMM:
\begin{equation}
p(\mathbf{y}_t|\mathbf{y}_{1:t-1}, \boldsymbol\theta) = \int_{\mathbf{x}_t\in\mathcal{X}} p(\mathbf{y}_t | \mathbf{x}_t, \boldsymbol\theta) p(\mathbf{x}_t | \mathbf{y}_{1:t-1}, \boldsymbol\theta) \text{d}\mathbf{x}_t.\label{equ:evidence:3}
\end{equation}
While this expression may look unhelpful, observe that we have particle approximations for the required distributions produced by the particle filter, or can calculate the required density.
The term $p(\mathbf{x}_t | \mathbf{y}_{1:t-1}, \boldsymbol\theta)$ represents the distribution over latent state conditioned on all data seen until that point.
This is exactly the filtering distribution maintained by the forward pass of the particle filter. 
Therefore, we have a Monte Carlo approximation of the integral in \eqref{equ:evidence:3}:
\begin{equation}
\hat{p}(\mathbf{y}_t|\mathbf{y}_{1:t-1}, \boldsymbol\theta) = \frac{1}{N}\Sigma_{n\in\left[1, \dots, N\right]} p(\mathbf{y}_t | \mathbf{x}^{(n)}_t, \boldsymbol\theta),\label{equ:evidence:4}
\end{equation}
where we denote the fact that this is now an approximation by using $\hat{p}$, although we drop this subsequently for clarity.
The term inside this summation, $p(\mathbf{y}_t | \mathbf{x}^{(n)}_t, \boldsymbol\theta)$, is the probability of the observation induced by each particle under the model.
This is exactly the quantity we use to perform resampling in the particle filters' forward sweep. 
Therefore, we can calculate the evidence at each time step by taking the expectation of the unnormalized importance weights, $w_t^{(n)}$, calculated at each resampling step.
By back substitution of \eqref{equ:evidence:4} into \eqref{equ:evidence:1}, we can then estimate the evidence for the whole dataset by multiplying each of these individual evidence calculations.

\subsection{Summary}
Before we proceed, we just pause to take stock of what the mechanics of SMC, and particularly particle filtering, allow us to perform.
SMC allows us to solve for a numerical approximation of the \emph{exact} posterior distribution over \emph{all} of the latent states conditioned on \emph{all} of the data.
We placed no constraints on the form of the initial state distribution, transition kernel and observation model, and so these could be defined as programs, simulators or complex distributions.
However, these non-trivial distributions make analytic calculation of these intermediate densities almost impossible.
Therefore the use of a particle filter can provide a numerical approximation to the action of SMC, which only requires that we can sample from the initial state distribution and transition kernel, and can evaluate the density under the observation model for any state-observation pair.
These are, often, easily achievable requirements as we have agency in how we design these elements, and we are then guaranteed, under mild conditions, to get a correct estimation of the target density.
We also showed how SMC also produces an estimation of the model evidence, or, the likelihood of the model parameters, marginalizing out the latent states, allowing for objective comparison of models.
Finally, although SMC is easily applied to HMMs which represents a large class of problems and we have explained SMC in terms of HMMs here, we stress that SMC is a very general algorithm that can be manipulated to solve a much wider class of problems than just HMMs.

\subsection{Initialization of Particles}
We observed that the ``quality'' of the initialization of particles dramatically impacts the quality of the reconstruction and evidence approximation.
Due to the computational cost of the simulator and high latent dimensionality simply taking the ``brute-force'' approach of using more particles is not feasible. 
We therefore employed a number of assumptions and methods to improve the quality of the initialization, while limiting any deviating from the formal mechanics of SMC.

We mainly focus here on the details of how we ``refine'' the distribution from which membrane potentials are initialized from.
Importance sampling to improve the initialization of particles is not feasible, as the voltage-calcium dynamics under the model specify that time-derivative calcium is a function of potential, and therefore one cannot simply importance sample to refine values of potential.

We therefore improve the distribution from which membrane potential is initialized.
By sampling state trajectories from the model we estimate a time-invariant, per-neuron distribution over potentials.
We then sample a large number of particles from this distribution, iterate these particles through the model once and perform resampling.
The distribution over initial state, conditioned on the first observation, is then estimated by performing a backward pass.
The particles used in the SMC sweep are then initialized from this distribution, with a small perturbation added to increase particle diversity.

We observe that this improved the quality of the initialization by limiting particle degeneracy.
While this is a deviation from the formal mechanics of SMC, it mainly aims to mitigate deficiencies introduced by using finite particle sets, removing those particles that are immediately discontinued by the SMC sweep, and replicating and diversifying those that have reasonable probability under the model and would be retained.
Importantly, this method incurs little computational cost.

\section{Parameter Learning}
Inference generally aims to solve for a representation, either exact or approximate, of the full form of the target density, i.e. in our scenario quantify the value of $\pi(\boldsymbol\theta | \mathbf{y})$ for all $\boldsymbol\theta$ values.
In contrast, optimization solves the much less onerous task of finding the inputs to a function that maximize the functions output, and additional obtain this value.
We now explore both of these approaches, and describe a methods for both.

\subsection{Posterior Density Calculation}
\label{sec:sub:posterior}
First we concretely define our metric of faithfulness.
For this, the posterior density given the observed data, $p(\boldsymbol\theta|\mathbf{y})$, is the natural choice in the Bayesian framework. 
The posterior density objectively represents the probability that any value of $\boldsymbol\theta$ was capable of generating the observations one is conditioning on.
Using Bayes' rule we can decompose this into the ratio of the joint probability, $p(\boldsymbol\theta, \mathbf{y})$, and the evidence, $p(\mathbf{y})$, which acts as a normalizing constant:
\begin{equation}
p(\boldsymbol\theta|\mathbf{y}) = \frac{p(\boldsymbol\theta, \mathbf{y})}{p(\mathbf{y})}.\label{equ:joint}
\end{equation}
The joint density can then be separated using the product rule:
\begin{equation}
p(\boldsymbol\theta|\mathbf{y}) = \frac{p(\mathbf{y} | \boldsymbol\theta) p(\boldsymbol\theta)}{p(\mathbf{y})}.\label{equ:likelihood}
\end{equation}
One can read the joint density as the product of a prior term, $p(\boldsymbol\theta)$, and a likelihood term, $p(\mathbf{y}|\boldsymbol\theta)$, where the prior term reflects our beliefs about the parameter values before conditioning on any data, and the likelihood term then re-weighting this prior belief according to how well it explains the observed data.

Analytic calculation of the evidence term $p(\mathbf{y})$ is intractable. 
However, we note that this is a normalizing constant that does not depend on $\boldsymbol\theta$ and so the surface defined by $p(\boldsymbol\theta|\mathbf{y})$ over $\boldsymbol\theta$ values is identical to the surface of the joint density $p(\boldsymbol\theta,\mathbf{y})$ up to a multiplicative constant.
Therefore, we can discuss maximizing (or performing inference in) the posterior space by actually working in the joint space, despite not being able to calculate the normalizing constant:
\begin{equation}
p(\boldsymbol\theta|\mathbf{y}) \propto p(\mathbf{y}|\boldsymbol\theta)p(\boldsymbol\theta).\label{equ:posterior}
\end{equation}
The prior distribution is determined by prior knowledge, informed by the beliefs of neuroscientists. 
The likelihood term expresses how well the parameter values describe the observed data. 
Evaluation of this likelihood term is less straightforward, since it requires a `fusion' of the model and the observed data in a time series.
However, we have already described how SMC can provide a numerical approximation of this likelihood, and as such, we can approximate the required joint density.

\subsection{Inference versus Optimization}
As described above, inference aims to quantify a target distribution over its entire support. 
This result, in the particular case of parameter inference under a model, is particularly desirable as it also conveys the strength of the dependence \emph{jointly} under the model and observed data.
As a brief example, suppose the model of interest takes as input two parameters, denoted $\theta_a$ and $\theta_b$, but one of the parameters, $\theta_a$ is not used, while $\theta_b$ is used.
The posterior distribution over the two parameters given the observed data is flat in the $\theta_a$ dimension, but is sharply peaked in the $\theta_b$ dimension. 
This topology suggests that $\theta_a$ does not affect the random variables, but the value of $\theta_b$ strongly influences the values, and, if sharply peaked, only one value for $\theta_b$ is suitable.
In this scenario, quantification of the posterior density for \emph{all} input values has informed us not only of the optimal parameters (the peaked value in $\theta_b$, and any $\theta_a$ value), but also that the value of $\theta_b$ is far more `important' than that of $\theta_a$, a notion quantitatively expressed through the entropy of the distribution.
This result may be as important to practitioners, as it indicates which aspects of their model add little discriminative or descriptive power.
The result of optimization would merely return the values of $\theta_a$ and $\theta_b$ that maximize this function, and indicate nothing about the relative importance of each parameter.

Furthermore, the posterior distribution is also conditioned on the observed data itself. 
A parameter may be observed to have no affect on one dataset, especially if some behaviour or motif is not observed and identified by a flat posterior distribution; and yet be much more highly peaked when conditioned on another dataset, where such behaviours or motifs are observed.
In the same vein, (exact) inference will not be ``over confident'' in asserting the parameter values for the simulator, returning a flat distribution over parameters that are not dependent on the data.

Finally, a ``good'' inference result will quantify any multi-modality in the system.
It may be that there are two (or more) distinct modes that (almost, if not exactly) equally explain the data.
If this is the case, this is almost certainly more interesting than the values themselves because it suggests that the model has two distinct operating regimes which may be indicative of the underlying system.
In this case, returning one of the modes as the ``optimum'' has missed an entire mode and potential operating regime.
The model may appear to not have the capacity to represent the data that led to the second mode being present, but instead, there may be a higher system that switches the behaviour of the system -- and intuition that is not garnered by vanilla optimization.

Optimization on the other hand is a fundamentally easier objective, merely solving for $\boldsymbol\theta^* = \argmax_{\boldsymbol\theta\in\boldsymbol\Theta} p(\mathbf{y}, \boldsymbol\theta)$.
Optimizations can often be accelerated through the use of gradients, and are generally more flexible due to the removal of the constraints that make the result of inference valid.

Therefore, we set out initially to show we can at least optimize the parameters of the simulator conditioned on data, although remaining open to calculating the scientifically more interesting result of the full posterior distribution over parameter values.

\subsection{Variational Optimization}
A major benefit of optimization methods is the ability to use gradient information to direct the optimization algorithm to higher utility regions of parameter space. 
However, it is not trivial, and may not even be possible, to differentiate through an arbitrary simulator, and therefore approximate gradient methods can be used.
An approximate method for obtaining gradients is to use the REINFORCE gradient~\cite{williams1992simple}.
The REINFORCE gradient is inspired from the log-derivative trick:
\begin{equation}
\nabla_{\boldsymbol\theta} f(\boldsymbol\theta) = f(\boldsymbol\theta) \nabla_{\boldsymbol\theta} \log f(\boldsymbol\theta).
\end{equation}
This system requires only the gradient of the logarithm of the objective function, which may be more easily calculated than the gradient of the original objective function.
The REINFORCE gradient has found great utility in reinforcement learning where it permits approximate derivatives to be calculated through non-differentiable elements.
An extension to the REINFORCE gradient is explored in \citet{staines2012variational}, and is referred to as variational optimization (VO).
This method is inspired by the REINFORCE gradients, but uses a variational proposal distribution to ease the problem further by removing the need to differentiate the even the log-objective function.

We first start by defining a variational proposal distribution $q(\boldsymbol\theta | \boldsymbol\phi)$.
This distribution, parametrized by $\phi$, generates sample values for $\boldsymbol\theta$.
The algorithm is predicated on the observation that:
\begin{equation}
\min_{\boldsymbol\theta} f(\boldsymbol\theta) \leq \mathbb{E}_{\boldsymbol\theta \sim q(\boldsymbol\theta | \phi)} \left[ f(\boldsymbol\theta) \right] = U(\boldsymbol\phi),\label{equ:vo_inequality}
\end{equation}
where $U(\boldsymbol\phi)$ is the expectation of the objective function under the proposal distribution.

We now try to build some intuition as to why this assertion is correct, and also because this intuition will help in understanding the VO algorithm more widely.
We first offer a proof of this by logical arguments, before offering a more intuitive set of arguments as to its correctness.
Suppose, for simplicity and without loss of generality, that $\boldsymbol\theta$ and $\boldsymbol\phi$ exist in the same space, and we have a proposal distribution, $q^*(\boldsymbol\theta|\boldsymbol\phi)$, that is a $\delta$-function.
Let us denote the minimum value of the function as $y^*$, occurring at the minimizing input value, $\boldsymbol\theta^*$.
The optimal proposal, with respect to minimizing the expectation of the function under that proposal, is clearly $\delta(\boldsymbol\phi - \boldsymbol\theta^*)$, yielding $U(\phi) = y^*$.
Accordingly, the optimal $\boldsymbol\phi$ value is $\boldsymbol\theta^*$.
Suppose now that we move $\boldsymbol\phi$ away from $\boldsymbol\theta^*$, the value of the expectation in \eqref{equ:vo_inequality} \emph{must} increase, implying that in this $\delta$-proposal schema, the optimal value of $\boldsymbol\phi$ must exist at the global minima of the function.
Let us now extend this by letting our proposal be two equally weighted $\delta$ functions, such that $\boldsymbol\phi$ is now a two-vector denoting the centre of each $\delta$.
The value of $U(\boldsymbol\phi)$ under this proposal is therefore:
\begin{equation}
U(\boldsymbol\phi_1, \boldsymbol\phi_2) = \frac{1}{2} (f(\boldsymbol\phi_1) + f(\boldsymbol\phi_2)),
\end{equation}
which by induction from above must be greater than $y^*$ unless $\boldsymbol\phi_1 = \boldsymbol\phi_2$, or the function $f(x)$ has multiple global minima, of which two are located at $\boldsymbol\phi_1$ and $\boldsymbol\phi_2$.
Since any arbitrary proposal distribution can be built up through such (weighted) summation of $\delta$ functions, this proves equation \eqref{equ:vo_inequality} is generally correct for any proposal distribution.

An alternative, and slightly more intuitive argument is to consider $q(\boldsymbol\theta|\boldsymbol\phi)$ as simply generating a set of values, $\left\lbrace \boldsymbol\theta_1, \boldsymbol\theta_2, \dots, \boldsymbol\theta_N \right\rbrace$, distributed according to the proposal distribution.
These values are in the space of the inputs to the function $f$, resulting in function evaluations of $\left\lbrace y_1, y_2, \dots, y_N \right\rbrace$.
Monte Carlo integration lets $N\rightarrow\infty$ such that we can evaluate integrals exactly using samples.
A (hopefully) trivially simple inequality is that the smallest value in the set must be smaller than or equal to the average value of the set:
\begin{equation}
\min(\left\lbrace y_1, y_2, \dots, y_N \right\rbrace)  \leq \frac{1}{N} \Sigma_{n=1}^N y_n ,
\end{equation}
where the two sides are only equal when all the values are equal.
We now note, that by definition:
\begin{equation}
 y^* \leq \min(\left\lbrace y_1, y_2, \dots, y_N \right\rbrace) ,
\end{equation}
since we defined $y^*$ to be the smallest value of the function $f$; again noting that the equality only holds when one of $y_n$ is equal to $y^*$.
It therefore follows that:
\begin{equation}
y^* \leq \frac{1}{N} \Sigma_{n=1}^N y_n.
\end{equation}
Recasting this inequality back to the original nomenclature, and ``undoing'' the Monte Carlo integration step yields:
\begin{align}
\min_x f(\boldsymbol\theta) &\leq \Sigma_{n=1}^N f(\boldsymbol\theta_n), \boldsymbol\theta_n \sim q(\boldsymbol\theta|\boldsymbol\phi) \\
&= \mathbb{E}_{\boldsymbol\theta\sim q(\boldsymbol\theta|\boldsymbol\phi)} \left[ f(\boldsymbol\theta) \right]. 
\end{align}
The equality condition (naively, see footnote) only holds when all sampled values are equal, and equal to the global function minimizer $\boldsymbol\theta^*$,\footnote{Of course, if the function has multiple global minima, this condition does not preclude $q$ placing non-zero mass on each of these minima, yielding a family of $q$ distributions that are \emph{all} optimal.} corresponding to the first $\delta$ proposal suggested above.

Using the inequality shown in \eqref{equ:vo_inequality} one can re-cast the optimization as instead taking gradient steps in the parameters of the proposal distribution, $\boldsymbol\phi$, such that the optimized proposal samples ``good'' parameter values.
Gradient steps are taken to maximize the function $U(\boldsymbol\phi)$:
\begin{equation}
\nabla_{\boldsymbol\phi} U(\boldsymbol\phi) = \nabla_{\boldsymbol\phi} \mathbb{E}_{\boldsymbol\theta \sim q(\boldsymbol\theta | \boldsymbol\phi)} \left[ f(\boldsymbol\theta) \right].
\end{equation}
Expansion of the expectation and application of the log-derivative trick once more yields:
\begin{equation}
\nabla_{\boldsymbol\phi} U(\boldsymbol\phi) = \mathbb{E}_{\boldsymbol\theta \sim q(\boldsymbol\theta | \boldsymbol\phi)} \left[ f(\boldsymbol\theta) \nabla_{\phi} \log q(\boldsymbol\theta|\boldsymbol\phi) \right].\label{equ:vo_grad}
\end{equation}
Inspection of this equation shows its benefit. 
We need only define a parametrized proposal distribution $q(\boldsymbol\theta |\boldsymbol\phi)$ which is differentiable with respect to its parameters $\boldsymbol\phi$ and we can draw samples from.
We then only have to be able to evaluate the target function at these sampled values.

Evaluation of this expectation is not analytically tractable and therefore we resort to Monte Carlo approximation:
\begin{align}
\hat{\nabla}_{\boldsymbol\phi} U(\boldsymbol\phi) & \approx \Sigma_{i \in \left\lbrace 1, \dots, \mathcal{N}_r \right\rbrace}  f(\boldsymbol\theta^{(i)}) \nabla_{\phi} \log q(\boldsymbol\theta^{(i)} ; \boldsymbol\phi),\label{equ:vo_mc_grad}\\
\boldsymbol\theta^{(i)} & \sim q(\boldsymbol\theta | \boldsymbol\phi),\ i \in \left\lbrace 1, \dots, \mathcal{N}_r \right\rbrace,
\end{align}
where we use $\mathcal{N}_r$ samples in the Monte Carlo estimation, and we differentiate that this is an approximate gradient with a hat on the derivative, $\hat{nabla}$.

The choice of $q(\boldsymbol\theta | \boldsymbol\phi)$ is a degree of freedom we are free to design.
We choose $q(\boldsymbol\theta | \boldsymbol\phi)=\mathcal{N}(\boldsymbol\phi | \boldsymbol\Sigma )$, where $\boldsymbol\Sigma$ is a diagonal covariance matrix.
Implcitly, through this choice, we are setting $\boldsymbol\Theta = \boldsymbol\Phi$ and $\boldsymbol\theta=\boldsymbol\phi$ at each time step.
To elucidate this a little further, we are centring an independent Gaussian proposal over each optimization parameter at the current value of that parameter.
The width of each proposal can then be tuned appropriately, potentially per-parameter if need be.
We set the width of this proposal to in accordance with the expected scale of the parameter value and reduce the width of the proposal during optimization.

Using VO, we are able to calculate an approximate gradient for the parameters of the proposal distribution, without having to differentiate through our simulator.
This is advantageous as it means any simulation platform can be used without requiring differentiability. 
Although the gradient estimator is a Monte Carlo approximation, and hence requires additional computational effort, the calculation in \eqref{equ:vo_grad} is embarrassingly parallelizable and hence we can leverage distributed supercomputer clusters with little overhead.

\subsubsection{Particle Marginal Variational Optimization}
We enhance the variational optimization algorithm by taking inspiration from particle marginal Metropolis Hastings, utilizing a pseudo-marginal estimation of the objective function as the target density.
In this case, we replace the analytic joint density of the parameters, $p(\mathbf{y}, \boldsymbol\theta)$, with the pseudo-marginal joint density estimated by SMC, $\hat{p}(\mathbf{y} | \boldsymbol\theta)p(\boldsymbol\theta)$.
Since this approximation is unbiased, it is a sensible objective to optimize, and corresponds to performing model selection.
We believe this is the first time variational optimization has been combined with a pseudo marginal method to maximize model evidence by optimizing the  parameter values of a black-box simulator.
We describe this as particle marginal variational optimization (PMVO) and suggest that this approach is apt for our problem domain as it is agnostic to the particulars of the simulator, leverages gradient information  which should make it scale to higher dimenional problems compared to, say, Bayesian optimization or simulated annealing, and also facilitates likelihood tempering.

Our particular PMVO algorithm then proceeds as follows: We first initialize the parameter values, $\boldsymbol\phi_0$. 
We chose to implement this by randomly sampling $\mathcal{N}_r$ parameters from the prior and selecting the parameter with the joint density, i.e. $\boldsymbol\theta_0 = \boldsymbol\theta_0^{(i^*)},\ i^* = \argmax_{i\in\left\lbrace 1, \dots, \mathcal{N}_r \right\rbrace} p(\mathbf{y}, \boldsymbol\theta^{(i)}_0),\ \boldsymbol\theta^{(i})_0\sim p(\boldsymbol\theta)$.
The parameters of the proposal distribution are then set to this value, $\boldsymbol\phi_1 \leftarrow \boldsymbol\theta_0$, since $\Phi = \Theta$.
VO then proceeds by sampling from the proposal distribution, $\boldsymbol\theta_1^{(i)} \sim q(\boldsymbol\theta | \boldsymbol\phi_1),\ i\in\left\lbrace 1, \dots, \mathcal{N}_r \right\rbrace$, evaluating the joint distributions at each of these points, $f(\boldsymbol\theta_0^{(i)}) = p(\mathbf{y}, \boldsymbol\theta_0^{(i)})$, and calculating an approximating derivative of the proposal with respect to the parameters of the proposal, $\hat{\nabla}_{\boldsymbol\phi} U(\boldsymbol\phi)$, as specified in \eqref{equ:vo_mc_grad}.
This estimate of the derivative is then used with the stochastic gradient ascent algorithm ADAM~\cite{kingma2014adam} to update the value of $\boldsymbol\phi_t$, denoted as $\boldsymbol\phi_{t+1} \leftarrow \mathtt{ADAM}(\boldsymbol\phi_t, \hat{\nabla}_{\boldsymbol\phi_t}(\boldsymbol\phi_t), \alpha)$, with a learning rate of $\alpha$.
This process iterates, sampling from the proposal, calculating the joint density, estimating the gradient and then taking a gradient step, until some computational budget has been used or a stopping condition is met.
Further augments such as introducing tempering of the joint density, annealing the learning rate and reducing the width of the proposal distribution can all be applied in-between gradient steps to improve performance.

A slight wrinkle in the chosen implementation is for bounded parameters (i.e. strictly positive parameters) the Gaussian proposal places non-zero probability mass on outside of these bounds. 
Therefore, we perform rejection sampling on each bounded parameter value to ensure proposed values are valid.
This is somewhat of a corner case however, as the proposal distribution is normally narrow compared to the bounds of the variable and hence this rejection sampling only comes into effect near the boundaries, which are, under any sensibly specified prior, likely to be low-probability regions, and so the optimization will actively seek to move away from these regions.
In practice, we find that very few samples are rejected.
Similarly, the gradient calculation may take the current parameter value outside these bounds.
If this is the case, we set the parameter to the value of the bound.
Again, we find that if the gradient method hits this constraint, it soon moves away from the constraint into higher probability regions.

\subsubsection{Tempering Gradient-Based Methods}
As described above, tempering of methods can improve their performance, especially when operating in high dimensional and sharply peaked distributions.
Therefore, it also makes sense to temper the objective function when performing gradient-based optimization, in turn, tempering the gradients one receives.
Tempering in the gradient based setting is actually more straightforward, as we are only search for a point estimate of the optimal parameter set, and therefore we can take gradients in the tempered distribution while annealing towards the true objective function, but also evaluate the true objective function at each iteration (since it is just raising the likelihood to a different power) and select the best-performing parameters on the true objective function, regardless of the temperature of the system.

\section{Parameter Inference Methods}
We now present additional information on the inference methods compared to in Section 4 of the main text.
We include these for completeness only, and do not present any experimental details in these sections.

\subsection{Metropolis Hastings}
The canonical asymptotically correct parameter inference algorithm, for distributions that cannot be directly sampled from, is the Markov chain Monte Carlo (MCMC) algorithm Metropolis Hastings (MH)~\cite{hastings1970monte, chib1995mh}.
MH is a method for drawing samples from a distribution that cannot be directly sampled from, but, can be evaluated for any given input value.
MH is an asymptotically correct algorithm that constructs a series of samples which, in the limit of infinite samples, is equal in distribution to the true distribution.
While full elucidation of the ergodic theory that underpins MH is outside the scope here, we provide a high level overview of the algorithm and its implementation, sufficient for the remainder of this paper to be understood.

On a high level, MH generates a series of samples by making ``moves'', or transitions, in the local vicinity of the ``current'' state.
This means that, fundamentally, two successive samples are highly correlated.
However, when many steps are taken, sufficiently distant samples become ``decorrelated,'' i.e. they become independent samples from the target distribution.
By construction MH ensures that some of the more straightforward requirements, such as detailed balance, yielded from the theoretical analysis to ensure samples in the limit of infinite samples approximate samples from the target distribution, are met, reducing the burden on the designer.

More formally, MH constructs a Markov chain over state values, which, under mild conditions, returns samples distributed according to the target distribution, here denoted as $f(\boldsymbol\theta)$.
One first defines a distribution over the initial state of the chain, often taken to be the prior distribution over state, $p(\boldsymbol\theta)$. 
Then, a transition kernel, or update rule, is defined to describe how the Markov chain evolves.
The distribution $q(\boldsymbol\theta_{t+1} | \boldsymbol\theta_t)$, referred to as a proposal distribution, defines a ``mutation'' to the current parameters to create the proposed parameters, denoted using a dash, $\boldsymbol\theta_t' \sim p(\boldsymbol\theta_t' | \boldsymbol\theta_t)$.
The proposed parameters are then accepted with probability $A$, calculated as the ratio of the target densities multiplied by the relevant proposal distribution for the current and proposed parameters:
\begin{equation}
A = \min\Big(1, \frac{q(\boldsymbol\theta_t | \boldsymbol\theta_{t}') f(\boldsymbol\theta_{t}')}{q(\boldsymbol\theta_{t}' | \boldsymbol\theta_t) f(\boldsymbol\theta_t)}\Big), \label{equ:mh:a}
\end{equation}
If the proposed parameters are accepted, the current parameters are updated to be the proposed parameters.
If the proposed parameters are rejected, the current parameters are duplicated. 
In our case $f(.)$ is taken to the be joint density $p(\boldsymbol\theta, \mathbf{y})$, evaluated at the data we observe.

We now have the required components to build a MH sampler.
First, at $t=0$, a sample is drawn from the initial distribution, $\boldsymbol\theta_0 \sim p(\boldsymbol\theta)$.
The Markov chain is set to this value.
This sample is added to the set of samples, $\left\lbrace \boldsymbol\theta_{t'} \right\rbrace_{t'\in\left\lbrace 0 \right\rbrace}$.
A new state is proposed conditioned on the value of the current state of the Markov chain by proposing from $\boldsymbol\theta'_1 \sim q(\boldsymbol\theta_1 | \boldsymbol\theta_0)$.
The acceptance ratio is then calculated as in \eqref{equ:mh:a}, taking into account the target densities' value at current and proposed state values, and also the forward and backwards transition densities, where the $\max$ term ensures the resulting value of $A$ is a valid probability.
The proposed state value, $\boldsymbol\theta'_1$, is then ``accepted'' with probability $A$. 
This means that, with probability $A$, the current state of the Markov chain is set to $\boldsymbol\theta_1 \leftarrow \boldsymbol\theta'_1$.
If the sample is rejected, the state of the Markov chain is set to the value at the previous step, with value $\boldsymbol\theta_{1} \leftarrow \boldsymbol\theta_{0}$.
The current state of the Markov chain, whether the step was accepted or rejected, is appended to the set of samples, $\left\lbrace \boldsymbol\theta_{t'} \right\rbrace_{t'\in\left\lbrace 0, 1\right\rbrace}$.
This process iterated proposing, accepting or rejecting and appending until the desired number of samples, $\mathcal{N}_T$ have been drawn, i.e. one possesses the set $\left\lbrace \boldsymbol\theta_{t'} \right\rbrace_{t'\in\left\lbrace 0, \dots, \mathcal{N}_T \right\rbrace}$.

While formal proof is outside the scope of this paper, it can be shown, that under mild conditions on the target density and proposal distribution, that $\left\lbrace \boldsymbol\theta_{t'} \right\rbrace_{t'\in\left\lbrace 0, \dots, \mathcal{N}_T \right\rbrace}$ exactly represents the target density, i.e. $\lim_{\mathcal{N}_T \rightarrow \infty} \Sigma_{t=\left\lbrace 0, \dots, \mathcal{N}_T \right\rbrace} \delta (\boldsymbol\theta_{t}) = f(\boldsymbol\theta)$.
A common choice of proposal distribution that, in most cases, meets these requirements is to set $q(\boldsymbol\theta'|\boldsymbol\theta) = \mathcal{N}(\boldsymbol\theta'; \boldsymbol\theta, \sigma^2)$, where $\sigma^2$ is some \emph{apriori} designed covariance. 
While the definition of the Metropolis-Hastings algorithm, and the use of a Gaussian kernel means that the theoretical requirements are likely to be met, it does not describe the convergence of the samples to the true distribution.
Designing good proposal distributions is at the heart of MCMC.

\subsection{Particle Marginal Metropolis Hastings}
While MH allows us to draw samples from a target distribution, in our scenario, we also wish to perform inference over the latent state in a time series.
Using the approximation of the likelihood, as estimated by the SMC sweep, constitutes a pseudo-marginal method, i.e. we have (approximately) marginalized over the latent state using a numerical approximation.
A more elegant view of this is to consider particle marginal Metropolis Hastings (PMMH)~\cite{andrieu2010particle, kantas2015particle} as an extension of the theoretical framework used in MH.

What we wish to do is construct a Markov chain sampling in the joint space $\left\lbrace \mathbf{x}_{0:T}, \boldsymbol\theta \right\rbrace$, targeting the posterior distribution $p(\mathbf{x}_{0:T}, \boldsymbol\theta | \mathbf{y})$.
We wish to create a MH acceptance term:
\begin{equation}
A = \min\Big(1, \frac
{q(\mathbf{x}'_{0:T}, \boldsymbol\theta' | \mathbf{x}_{0:T}, \boldsymbol\theta) p(\mathbf{x}'_{0:T}, \boldsymbol\theta' | \mathbf{y})}
{q(\mathbf{x}_{0:T}, \boldsymbol\theta | \mathbf{x}'_{0:T}, \boldsymbol\theta') p(\mathbf{x}_{0:T}, \boldsymbol\theta | \mathbf{y})}
\Big).
\end{equation}
We are free to choose the form of our proposal distribution, but a reasonable choice to use is:
\begin{equation}
q(\mathbf{x}'_{0:T}, \boldsymbol\theta' | \mathbf{x}_{0:T}, \boldsymbol\theta) = q(\boldsymbol\theta' | \boldsymbol\theta) q(\mathbf{x}'_{0:T} | \mathbf{y}, \boldsymbol\theta') = q(\boldsymbol\theta' | \boldsymbol\theta) p(\mathbf{x}'_{0:T} | \mathbf{y}, \boldsymbol\theta').
\end{equation}
This proposal can be viewed of as first proposing a new parameter value, conditioned only on the current parameter value, through the proposal term, $q(\boldsymbol\theta' | \boldsymbol\theta)$.
The second term then proposes the value of the latent variables $\mathbf{x}_{0:T}$ conditioned on both the observed data and the newly sampled parameter values.
Note that this is not conditioned on the value of $\mathbf{x}_{0:T}$ currently held by the Markov chain.
If we were to try and condition on this value, it would introduce a number of intractable likelihood terms, therefore independently proposing new $\mathbf{x}_{0:T}$ values each time is more generally tractable.
To emphasize, this proposal proposes values for \emph{both} $\boldsymbol\theta$ and $\mathbf{x}_{0:T}$.

Substitution of this proposal into the acceptance ratio yields:
\begin{equation}
A = \min\Big(1, \frac
{q(\boldsymbol\theta | \boldsymbol\theta') p(\mathbf{x}_{0:T} | \mathbf{y}, \boldsymbol\theta) p(\mathbf{x}'_{0:T}, \boldsymbol\theta' | \mathbf{y})}
{q(\boldsymbol\theta' | \boldsymbol\theta) p(\mathbf{x}'_{0:T} | \mathbf{y}, \boldsymbol\theta') p(\mathbf{x}_{0:T}, \boldsymbol\theta | \mathbf{y})}
\Big).
\end{equation}
Expansion of the third term then yields:
\begin{equation}
A = \min\Big(1, \frac
{q(\boldsymbol\theta | \boldsymbol\theta') p(\mathbf{x}_{0:T} | \mathbf{y}, \boldsymbol\theta) p(\mathbf{x}'_{0:T} | \mathbf{y}, \boldsymbol\theta') p(\boldsymbol\theta' | \mathbf{y})}
{q(\boldsymbol\theta' | \boldsymbol\theta) p(\mathbf{x}'_{0:T} | \mathbf{y}, \boldsymbol\theta') p(\mathbf{x}_{0:T} | \mathbf{y}, \boldsymbol\theta) p(\boldsymbol\theta | \mathbf{y})}
\Big).
\end{equation}
The second and third terms in numerator and denominator then cancel to yield:
\begin{equation}
A = \min\Big(1, \frac
{q(\boldsymbol\theta | \boldsymbol\theta') p(\boldsymbol\theta' | \mathbf{y})}
{q(\boldsymbol\theta' | \boldsymbol\theta) p(\boldsymbol\theta | \mathbf{y})}
\Big).\label{equ:pmmh:a}
\end{equation}

This rearrangement justifies the somewhat arbitrary choice we made above to propose values for $\mathbf{x}_{0:T}$ independently from the current value.
The SMC sweep returns two things, an estimate of the likelihood of the parameters, $p(\mathbf{y}|\boldsymbol\theta)$, and also the posterior distribution over the latent state, $p(\mathbf{x}_{0:T} | \mathbf{y}, \boldsymbol\theta)$, where this latter term is nothing more than the particles retained after the backwards pass. 
Therefore, the SMC sweep provides us with all the required samples and values to perform this update, there the evidence approximation is used in \eqref{equ:pmmh:a}, and the particles (after the backwards pass) are the samples of $p(\mathbf{x}_{0:T} | \mathbf{y}, \boldsymbol\theta)$.
PMMH then proceeds as an MH sampler, proposing, accepting or rejecting and appending samples until the required number of samples have been reached.

Most importantly, using PMMH is practically identical to independently using ``standard'' MH and SMC.
One only has to define the required kernels for the SMC sweep, in state-space, and the proposal distributions for the parameter updates in MH.
The conjunction of these two is the PMMH algorithm, with no heed needing to be paid to the interaction of the two, greatly reducing the burden on the designer.

\subsection{Coordinate-wise Ascent}
One drawback of MH-based algorithms is the dependence on the proposal distribution. 
Often one can only define an ``independent'' proposal distribution, by where individual states are perturbed with no knowledge of how other parameters are mutated.
However, in high dimensions one must jointly propose perturbations to avoid low acceptance rates, and poor performance.
Jointly proposing ``good'' perturbations is often as difficult as the original inference problem, requiring quantification of the local surface. 
Therefore, a commonly used modification is to perform block-MH, or coordinate wise ascent~\cite{haario2005componentwise}. 
In these schemes MH steps are taken on subsets of the parameters or individual parameters respectively.
This is observed to increase acceptance rates in high dimensions over naive, independent proposals.
However, if there is strong correlation between the states, i.e. there is a sharp ``ridge'' in the objective function, coordinate-wise or block method may also have low acceptance rates if these states are not perturbed together.
Furthermore, if the dimensionality of the parameters to be estimated is large, permuting individual coordinates, or even small groups of parameters, can dramatically increase the runtime cost of inference.
We find that coordinate-wise kernels often yield a better final result, albeit at greatly increased computational cost. 

\subsection{Tempering}
An augment to the above methods is the use of tempering~\cite{van2014tempering, maclaurin2015firefly, angelopoulos2008bayesian}.
Tempering is the process of modifying the objective function such that it has more favourable properties, and then annealing the function during execution to restore it to the original function of interest.
A common form of tempering in posterior inference is raising the likelihood term to an ``inverse temperature,'' $\beta$:
\begin{align}
p^{\beta}(\boldsymbol\theta | \mathbf{y}) \propto& \ p(\mathbf{y} | \boldsymbol\theta) ^{\beta} p(\boldsymbol\theta), \\ 
\beta =&\ \frac{1}{T},\\
T \in & \mathcal{T} = \mathbb{R}_+,
\end{align}
where $T$ is referred to as a temperature.
For high temperature systems (low $\beta$ values) the contribution of the likelihood is attenuated relative to the contribution of the prior.   
For infinite temperature systems, one can see that the tempered objective function becomes the prior, while unity temperature systems reflect the original objective function.
This tempering effect is observed to ``flatten'' the ridges in the likelihood function, often making inference and optimization methods perform better~\cite{gupta2018evaluation}.
Of course, one must fully cool the system before using any of the results as the intermediate tempered functions have limited use.

\subsection{Parallel Tempering}
For inference however the tempering scheme described above is of limited use, since samples from the true objective function are only generated once the system is fully cooled, once again exposing the issue related to correlated samples.
An additional augmentation to mitigate this is the use of parallel tempered (PT) chains~\cite{swendsen1986replica, Altekar2003PT}.
In this system, a number of independent MCMC chains, $N_r$, referred to as replicates, are run in parallel, each operating at a different temperature, $T_n \in \mathcal{T}, n \in \left\{ 1, \dots, \mathcal{N}_r \right\}$.
It is customary to organize these chains in descending order of temperature, such that $T_{n-1} \leq T_{n} \leq T_{n+1}$, although this ordering is arbitrary and is for simplicities sake.
With this ordering in mind, and supposing $T_1 = \inf$ and $T_{\mathcal{N}_r}=1$, these chains then define a series of target functions that slowly morph from the prior distribution into the joint distribution.

Mixing between the chains is then permitted by ``swapping'' the state of the Markov chain between replicates, according to an MH acceptance ratio.
The scheme for defining these swaps needs only to obey the same conditions as conventional MH, although it is most common to only propose swaps between adjacent chains with some fixed probability.
Therefore, the swap between two temperatures, $T_{i}$ and $T_{j}$, is accepted according to the MH probability:
\begin{equation}
A = \min\Big(1, \frac
{p^{T_{i}}(\boldsymbol\theta_{j} | \mathbf{y}) p^{T_{j}}(\boldsymbol\theta_{i} | \mathbf{y})}
{p^{T_{i}}(\boldsymbol\theta_{i} | \mathbf{y}) p^{T_{j}}(\boldsymbol\theta_{j} | \mathbf{y})}
\Big).
\end{equation}
This expression does not feature a proposal term as we are used to, as once a swap is proposed, the replicates being swapped have \emph{both} been selected, and so the proposal is symmetric.
The numerator is then the posterior density of the two replicates' (at their own temperatures) if the swap were to be accepted, while the denominator is the posterior density if the swap is rejected.
This means if a swap yields a higher probability configuration it is accepted with certainty, while if it yields a lower probability configuration it is accepted with probability determined by the ratio between these two probabilities. 

The intuitition behind this method is that the high temperature chains can easily roam around the very smooth, high temperature surface.
``Good'' parameter values are then swapped into successively lower temperature chains where they are perturbed over smaller length scales (to yield a reasonable acceptance rate).
At the end of inference, the higher-temperature chains are discarded and the unit temperature chain is retained as the true samples.
This method can be seen as defining a $\mathcal{N}_r - 1$ auxiliary variables, in the form of the higher temperature chains, to aid inference, and then marginalizing over these auxiliary variables at the end of inference, where marginalization is tantamount to discarding. 

Why this is valid algorithm, from the perspective of MCMC, relies on the observation that we can add additional auxiliary variables to our state, so long as we marginalize them at the end of inference. 
Therefore, PT constructs a Markov chain on the extended space $\boldsymbol\Theta ^{\mathcal{N}_r}$, where $\boldsymbol\Theta$ is the space of the state we are performing inference over (in our case, parameters), and $\mathcal{N}_r$ is the number of replicates (interacting chains, most likely at different temperatures).
There are then two distinct MCMC moves made as part of a ``single'' update to the extended space.
The first treats chains as independent, performing a normal MCMC step on the state of each chain.
The second move then involves proposing one or more pairs of replicates to swap and accepting or rejecting those swaps.
Therefore, the overall transition kernel for the state of the extended space is the composition of these two steps.

Well tuned parallel tempering algorithms are observed to improve inference performance, facilitating improved exploitation and exploration characteristics when compared with just running $\mathcal{N}_r$ independent chains.
Of course, using PT requires more computational effort compared to running a single chain, and the higher temperature chains must be evaluated and are then discarded (although often the improvement in performance outweighs this cost).
Finally tuning PT algorithms is difficult, exacerbated in high dimensions and with many temperatures, leading to poor performance and wasted computational effort.
Research into adaptive tempering aim to alleviate this~\cite{miasojedow2013adaptive, lkacki2016state}.

\subsection{Computation Allocation}
The evidence approximation (in our case, the likelihood of the parameters, $p(\mathbf{y} | \boldsymbol\theta)$) produced by the SMC sweep is, in essence a Monte Carlo approximation of the true evidence.
Therefore one must chose how to allocate computation: running more particles to get more accurate individual samples, or, run more cheap SMC sweeps (with fewer particles) and average over the more noisy evidence approximations, but from which we can have a higher throughput.
Separate SMC sweeps are computationally independent from one another and therefore their calculation can be embarrassingly parallelized across distributed compute systems.
We opt to run one SMC sweep per node available, running as many particles on that node as is tractable to in a reasonable time frame.

We denote the number of particles we use in each SMC sweep as $\mathcal{N}_p$, and the number of sweeps as $\mathcal{N}_s$.
We also denote the number of samples, or replicates, used within our VO gradient calculation, as well as the number of temperatures used in the temperature ladder for parallel tempering, as $\mathcal{N}_r$, where the temperatures used are then denoted as $\mathcal{T}_{1:\mathcal{N}_r}$.
We denote the number of steps taken either in inference (MH steps) or optimization (gradient steps) as $\mathcal{N}_T$. 
The total computational expenditure of the learning process is therefore $\mathcal{N}_p \times \mathcal{N}_s \times \mathcal{N}_r \times \mathcal{N}_T = \mathcal{B}$, where $\mathcal{B}$ denotes our total computational budget.
We normalize experiments by this quantity to allow for meaningful comparison of methods, and, most importantly, to ensure that our proposed methodologies remain computationally tractable.

\end{document}